\documentclass[smallextended]{svjour3} 
\smartqed

\RequirePackage{fix-cm}
\usepackage{natbib}
\usepackage{amsmath,amssymb,amsfonts}
\usepackage{algorithmic}
\usepackage{graphicx}
\usepackage{textcomp}
\usepackage{xcolor}
\usepackage{booktabs}
\usepackage{tikz}
\usepackage{hyperref}
\usepackage{float}
\usepackage{fancybox}  

\usepackage{multirow}
\usepackage{array}
\usepackage{threeparttable}
\usepackage{threeparttablex}
\usepackage{booktabs}
\usepackage{tabularx}

\usepackage{subcaption}

\usepackage[most]{tcolorbox}
\newtcolorbox{takeaway}{%
  colback=black!3,
  colframe=black!40,
  boxrule=0pt,
  leftrule=2.5pt,
  arc=2pt,
  sharp corners=west,
  fontupper=\small,
  left=10pt, right=10pt, top=6pt, bottom=6pt,
  before skip=10pt, after skip=10pt,
  breakable
}

\journalname{Empirical Software Engineering}

\def\BibTeX{{\rm B\kern-.05em{\sc i\kern-.025em b}\kern-.08em
    T\kern-.1667em\lower.7ex\hbox{E}\kern-.125emX}}

\newcommand{\fabio}[1]{\textcolor{teal}{\textbf{Fabio}: #1}}

\newcommand{\alberto}[1]{\textcolor{cyan}{\textbf{Alberto}: #1}}

\newcommand{\rev}[1]{\textcolor{black}{#1}}
\newcommand{\stageTwo}[1]{{\color{blue}#1}}
\newcommand{\boxedtext}[1]{\fbox{\scriptsize\bfseries\textsf{#1}}}
\newcommand{\nota}[2]{\boxedtext{#1}{\small$\blacktriangleright$\emph{\textsl{#2}}$\blacktriangleleft$}}
\newcommand{\todo}[1]{{\color{red}\nota{TODO}{#1}}}

\begin{document}

\title{Using Biometrics to Understand AI-Assisted Coding Performance and its Perception}

\titlerunning{Using Biometrics to Understand AI-Assisted Coding}

\author{Paolo Burelli \and
        Fabio Calefato \and
        Daniela Grassi \and
        Mihaela Yurieva Hristova \and
        Nicole Novielli \and
        Alberto Antonio Romano \and
        Paolo Tell
}

\authorrunning{Burelli et al.}

\institute{P. Burelli \at
              IT University of Copenhagen,  Denmark \\
              \email{pabu@itu.dk}
           \and
           F. Calefato \at
              University of Bari, Italy \\
              \email{fabio.calefato@uniba.it} 
           \and
           D. Grassi \at
              University of Bari, Italy \\
              \email{daniela.grassi@uniba.it}
           \and
           M.~Y. Hristova \at
              IT University of Copenhagen, Denmark \\
              \email{mihr@itu.dk}
           \and
           N. Novielli \at
              University of Bari, Italy \\
              \email{nicole.novielli@uniba.it}
           \and
           A.~A. Romano \at
              University of Bari, Italy \\
              \email{a.romano93@studenti.uniba.it}
           \and
           P. Tell \at
              IT University of Copenhagen, Denmark \\
              \email{pate@itu.dk}
            \\ \\ \emph{Authors listed in alphabetical order} 
}

\date{Received: date / Accepted: date}

\maketitle

\begin{abstract}
\stageTwo{AI-based code assistants are transforming software development, yet we lack empirical evidence on how they affect developers' cognitive processes.
We present a multisite study investigating the neurophysiological correlates of AI-assisted programming through a within-subjects crossover design. 
We recruited participants at two universities (Bari, Italy, and Copenhagen, Denmark) and collected electroencephalography, eye-tracking, electrodermal activity, and heart rate variability data alongside a rubric-based performance score and self-reported workload across six dimensions using the NASA Task Load Index (NASA-TLX).
We tested four hypotheses addressing physiological differences between AI-assisted and non-assisted conditions, the moderating role of developer experience, the association between physiology and performance, and the alignment between subjective perceptions and objective measures.}
\stageTwo{Under AI assistance, the EEG $\theta/\alpha$ ratio was lower during the first task and the gaze blink rate was higher during the second, both consistent with reduced cognitive engagement when developers offload generative effort to the model. This pattern did not differ between undergraduate and graduate students. Electrodermal activity correlated with performance under the non-AI condition but not under AI. Among the six NASA-TLX dimensions of self-reported workload, only Physical demand was associated with performance under the non-AI condition but not under AI.
These findings suggest that AI-assisted programming is not a faster version of solo coding but a cognitively distinct activity, with implications for the design of AI assistants and for biometric monitoring in AI-augmented development.}
\end{abstract}

\keywords{AI-assisted programming \and LLM \and Biometrics \and Developer cognition \and Cognitive debt \and Software engineering}

\section{Introduction}
Large Language Models (LLMs) and AI-based code assistants such as ChatGPT and Copilot are rapidly transforming code development, promising increased productivity~\citep{BirdEtAl:copilot,peng2023impactaideveloperproductivity}. 
Despite widespread adoption, there is a significant gap in our understanding of how these tools affect cognitive processes during programming tasks. 
Although subjective reports and productivity metrics provide some insights \citet{vaithilingam2022expectation, Prather:2024, liang2024large} 
they fail to capture the underlying neurophysiological changes in the transition from traditional programming to AI-assisted development.

The cognitive processes involved in programming encompass two distinct mental activities: analytical thinking, where developers break down and understand problems, and synthetic thinking, where they create and implement solutions. Developers continuously shift between these two modes while programming~\citep{pennington1987stimulus,detienne2001}.
With AI assistance, this established workflow is fundamentally altered as developers now engage in a third phase in which they evaluate, validate, and refine AI-generated code. 
This shift potentially reorganizes cognitive load and changes how developers approach coding problems. 
However, we still lack empirical evidence on how these changes affect physiology and \stageTwo{performance}.

Understanding these cognitive shifts is crucial not only for theoretical advancements in software engineering but also for practical applications in tool design and developer training. 
By combining biometric measurements---i.e., electroencephalography (EEG), eye tracking, electrodermal activity (EDA), and heart rate variability (HRV)---with \stageTwo{a rubric-based performance score} and subjective perceptions, we can better understand AI-assisted programming's impact on developer cognition.

To systematically investigate these cognitive shifts, we formulate our research questions as follows. First, although initial subjective reports suggest that AI assistance affects developers, we lack objective evidence of underlying cognitive changes. Physiological measures offer unbiased insight into cognitive processes that would otherwise be invisible. 
Establishing whether physiological differences exist is fundamental to enabling triangulation with self-report on cognitive load, thus informing evidence-based approaches to designing development environments capable of enhancing human-AI collaboration.  
Therefore, we ask:

\textbf{RQ1}: \textit{To what extent do physiological measures differ between developers programming with and without AI assistance?}

\rev{In addition, we aim to study how developer experience moderates AI's cognitive impact, as experienced and novice developers possess different mental models and problem-solving strategies. 
These insights from this analysis would inform the development of experience-adaptive AI systems and personalized training protocols that account for individual differences in AI tool effectiveness.
Therefore, we ask:}

\textbf{RQ\stageTwo{2}}: \textit{How does developer experience moderate the relationship between AI assistance and physiological measures?}

Understanding whether cognitive changes translate into practical outcomes is essential to validate the significance of physiological differences. Hence, we ask:

\textbf{RQ\stageTwo{3}}: \textit{How do physiological measures correlate with performance metrics in both conditions?}

Finally, when using a tool, developers' perceptions may not accurately reflect their actual cognitive states or performance, potentially leading to suboptimal AI tool usage. 
This knowledge is crucial for designing AI transparency mechanisms and metacognitive training systems that help developers accurately assess tool benefits and optimize their interaction strategies.
Therefore, our last research question is:

\textbf{RQ\stageTwo{4}}: \textit{To what extent does the alignment between subjective perceptions and objective measures differ between AI-assisted and non-AI-assisted conditions?}

We conduct a multisite study at the University of Bari, Italy (Uniba) and the IT University of Copenhagen, Denmark (ITU). The multisite design improves external validity through diverse student populations, programming cultures, and instrumentation.
Beyond increasing our sample size, this approach enabled us to recruit participants with linguistic, cultural, and educational contexts, thus improving result generalizability.
\stageTwo{The study was pre-registered as a Stage~1 Registered Report and received In-Principle Acceptance at ESEM~2025~\citep{stage1report}; the experimental design, hypotheses, and confirmatory analyses reported here follow that registered protocol, with deviations introduced during execution documented in~\ref{appendix:deviations}.} 

\stageTwo{Our findings show that the EEG $\theta/\alpha$ ratio is lower under AI assistance during the first task and the gaze blink rate is higher under AI during the second, both consistent with reduced cognitive engagement when developers offload generative effort to the model. This pattern does not differ between undergraduate and graduate students. Electrodermal activity correlates with performance under the non-AI condition but not under AI, and among the six dimensions of the NASA Task Load Index (NASA-TLX), only Physical demand shows the same asymmetric pattern.

This work makes three contributions to the empirical study of human--AI collaboration in SE. First, to our knowledge, this is the first study to combine EEG, eye tracking, electrodermal activity, and heart rate variability with a rubric-based performance score and self-reported workload to compare AI-assisted and non-AI-assisted development. Second, we find that the EEG $\theta/\alpha$ ratio and gaze blink rate differentiate the two programming conditions, unlike electrodermal activity and heart rate variability; this finding provides empirical guidance on which signals to instrument in AI-assisted development. Third, by showing that subjective workload aligns with performance differently across the AI/non-AI conditions only for the Physical demand dimension of the NASA-TLX, it refines the broader perception--reality gap reported in prior work on AI-assisted coding.}

\stageTwo{The remainder of the paper is structured as follows. Section~\ref{sec:background} reviews the theoretical background and related work on cognitive processes in programming, biometric measurement in software engineering, and human--AI collaboration. Section~\ref{sec:hypotheses} states the four pre-registered hypotheses. Section~\ref{sec:methodology} describes the study design, participants, instrumentation, and protocol. Section~\ref{sec:analysis} details the analysis plan, including preprocessing, variables, and the test of each hypothesis. Section~\ref{sec:results} reports the results, organized by research question. Section~\ref{sec:discussion} discusses the findings and their theoretical and practical implications. Section~\ref{sec:limitations} addresses limitations and threats to validity. We draw conclusions in Section~\ref{sec:conclusion}.}

\section{Background and Related Work}\label{sec:background}

\subsection{Cognitive and Neural Mechanisms in Programming}
Cognitive load theory suggests that the limited capacity of working memory creates constraints on learning and problem-solving~\citep{sweller1994cognitive}. \stageTwo{Dual-process theory further distinguishes between fast, automatic (Type~1) and slow, deliberative (Type~2) cognitive processes~\citep{kahneman2011thinking}; in programming, developers rely on Type~1 processes for familiar patterns and syntax, and on Type~2 processes for novel problem-solving and code analysis.} In the context of programming, these constraints manifest as developers face complex challenges while simultaneously maintaining mental models of code structure, program behavior, and problem requirements~\citep{HEINONEN2023107300}. The cognitive processes involved in programming articulate into two distinct phases~\citep{pennington1987stimulus,detienne2001}, with developers moving from analytical thinking required for problem understanding, to synthetic thinking, where they construct and implement solutions. The introduction of AI-assisted development introduces a third cognitive task---the evaluation and refinement of AI-generated code---thus potentially reorganizing cognitive load by shifting the focus from generating solutions to critically assessing and integrating machine-generated ones~\citep{Prather:2024}. \stageTwo{Distributed cognition theory~\citep{hutchins1995cognition} provides a complementary lens, framing cognitive processes as extending beyond the individual mind to encompass interactions with tools and environments; in AI-assisted programming, cognition is distributed between the developer and the AI system, which may fundamentally alter the physiological signatures of programming activity.}

With the increasingly widespread availability of neuroimaging hardware, researchers started investigating the neural mechanisms underlying computer programming, providing unprecedented insights into how developers process and understand code. 
\citet{siegmund2014, siegmund2017} conducted pioneering work applying functional magnetic resonance imaging (fMRI) to study programming, identifying a network of brain regions involved in bottom-up program comprehension. 
Their study revealed activation in left-lateralized brain regions associated with working memory, attention, and language processing.
\citet{liu2020} found code comprehension recruits a left-lateralized fronto-parietal network overlapping with formal logic and mathematical reasoning regions. \citet{krueger2020} showed that code writing engages bilateral regions associated with attention, working memory, and planning. In contrast, prose writing primarily activates the regions of the language of the left hemisphere.
Research on code complexity's impact on cognitive load demonstrated that size-based and vocabulary metrics correlate with anticipatory cognitive load, while data-flow metrics correlate with program comprehension network activation. Traditional metrics like McCabe's cyclomatic complexity showed no significant correlation with neural patterns~\citep{peitek2021}.
Beyond fMRI, functional Near-Infrared Spectroscopy (fNIRS) studies revealed increased prefrontal cortex activation corresponding to task complexity and during variable memorization~\citep{nakagawa2014,ikutani2014}. These findings collectively indicate that program comprehension engages neural circuits distinct from language processing and mathematical reasoning, with different programming activities utilizing specialized cognitive processes that vary with expertise~\citep{floyd2017}.

\subsection{Biometric Measurement in Software Engineering}
\todo{@Daniela add some recent paper on biometric}
The use of biometrics to understand cognitive processes in software engineering represents a growing research area. \citet{fritz2014icse} pioneered the use of biometrics in software engineering by using eye-tracking, electroencephalography (EEG), and electrodermal activity (EDA) to predict task difficulty during programming. Their work demonstrated that these physiological measures can help distinguish between easy and difficult programming tasks.
\stageTwo{Building on this foundation, \cite{muller2015icse} and \cite{Girardi:2020:ICSE} collected EEG, EDA, and heart rate measurements to detect developers' emotions and progress during programming tasks. They found that biometric features can be used to train a classifier to identify developers' positive or negative emotions.} 
\citet{fucci2019icpc} propose a machine learning model to automatically identify
the type of tasks developers are working on, leveraging EEG, EDA, and heart-related signals collected with lightweight biometric sensors. \citet{LaudatoEtAl:predictingBugs} introduced a developer-based bug prediction model using behavioral (keystrokes, mouse movements), biometric (heart rate, EDA, attention), and control factors (experience, task time). Biometrics were also used in combination with computer-related metrics to predict developers' interruptibility in a field study~\citep{Zueger:interruptability}. Overall, these findings suggest that lightweight, non-invasive biometric sensors can be a viable alternative to neuroimaging techniques when ecological validity and naturalness of the experimental environment have to be prioritized over accuracy, thus allowing researchers to study cognitive processes as they unfold during real programming sessions, including in professional settings or laboratory studies involving extended development tasks.

\subsection{Human-AI Collaboration in Programming}

\stageTwo{\subsubsection{Programming with AI Assistants}}
LLMs have significantly impacted software development, particularly pair programming. \citet{ma2023ai} found mixed results comparing human-AI and human-human pair programming, reporting that human pairs excel at knowledge sharing while AI partners offer advantages in code generation. Along the same line, \citet{takai2023classification} demonstrated how AI pair programming enhances code readability and development speed, while human-human strengthens knowledge sharing.
\citet{friedmann2024pair} observed experienced programmers use AI for acceleration of familiar tasks, while novices use it more exploratively for design discussions. In education, \citet{chen2024pair} explored ChatGPT's potential as a replacement for traditional pair programming in introductory computer science courses and found that ChatGPT offers rapid responses and error-free code generation, though understanding AI-generated code remains challenging.
\citet{stray2025generative} examined how tools like ChatGPT influence development efficiency. \citet{ganeshan2024impact} discovered that developers use AI to speed up coding in solo sessions but may view them as workflow interruptions during pair programming. \citet{friedmann2024pair} emphasized balanced integration strategies promoting critical thinking, while \citet{ganeshan2024impact} highlighted the importance of considering the impact of AI usage on knowledge sharing and team dynamics.
\stageTwo{\citet{welter2025developer} extended these comparisons to knowledge transfer, finding that developers accept AI suggestions with less scrutiny than those from human partners, suggesting qualitative differences in cognitive engagement across collaboration modes.}

\stageTwo{Recent work has also examined how developers actually interact with AI code assistants at a finer granularity. \citet{barke2023grounded} identified two distinct interaction modes in a grounded theory study of 20 programmers using Copilot: an \emph{acceleration} mode, in which the developer knows what to write and uses the tool for speed, and an \emph{exploration} mode, in which the developer is uncertain and uses suggestions as a starting point. The distinction implies that cognitive demands vary substantially depending on the developer's current goal. \citet{mozannar2024reading} quantified this further by developing a taxonomy of 12 programmer activities, revealing that developers spend 22.4\% of their time verifying suggestions and up to 34.3\% of total session time on double-checking and editing AI-generated code. These verification activities are largely invisible to traditional acceptance-rate metrics. \citet{vaithilingam2022expectation} provided early evidence of an expectation-experience gap: in a within-subjects study, Copilot did not improve task completion time or success rate, yet 79\% of participants preferred it, perceiving it as a useful starting point. \citet{tang2024developer} combined eye tracking with IDE interaction logging and NASA-TLX assessments to study how developers validate LLM-generated code, finding that awareness of code provenance led to improved performance but also higher cognitive workload.}

\stageTwo{\subsubsection{Productivity Perception, Cognitive Debt, and Skill Erosion}}
\stageTwo{A growing body of evidence suggests that the perceived productivity benefits of AI assistance may not align with objective outcomes. In a randomized controlled trial, \citet{becker2025metr} found that allowing experienced open-source developers to use frontier AI tools actually increased task completion time by 19\%, contradicting both developers' pre-study forecasts of a 24\% speedup and their post-hoc estimates of a 20\% speedup. This perception-reality gap persisted even after task completion, indicating that subjective impressions of AI-assisted productivity are systematically miscalibrated.}

\stageTwo{Several theoretical frameworks help explain why this gap arises. \citet{tankelevitch2024metacognitive} argued that generative AI imposes substantial \emph{metacognitive} demands on users, who must formulate prompts, evaluate outputs, and continuously calibrate their reliance on the tool. \citet{lee2025critical} provided empirical evidence that higher confidence in GenAI is associated with reduced critical thinking effort, while higher task-specific self-confidence is associated with more critical thinking. These findings suggest that the subjective ease of AI-assisted work may partly reflect reduced cognitive engagement rather than genuine efficiency gains.}

\stageTwo{The cognitive science literature offers further grounding. \citet{grinschgl2021offloading} demonstrated experimentally that cognitive offloading improves immediate task performance but diminishes subsequent memory for the offloaded information, establishing a fundamental trade-off between short-term efficiency and long-term retention. In the context of AI-assisted development, this trade-off has been conceptualized as \emph{cognitive debt}---the accumulated comprehension deficit that arises when code is produced faster than developers can understand what the program does, how their intentions were implemented, and how the software can be changed~\citep{storey2026cognitive}. Empirical evidence supports these concerns: \citet{shen2026skill} conducted a randomized experiment showing that AI assistance during unfamiliar coding tasks reduced subsequent skill assessment scores by 17\%, with full delegation to AI producing the largest learning deficits. \citet{sankaranarayanan2026epistemic} reported that unrestricted AI users matched the productivity of scaffolded users during initial construction but suffered a 77\% failure rate in subsequent maintenance tasks without AI, compared to 39\% for the scaffolded group. \citet{starr2025troubleshooting} developed a theory of troubleshooting grounded in developer interviews, finding that AI tools initially reduce friction but then increase difficulty when developers must debug code they did not write, contributing to cognitive fatigue. \citet{kosmyna2025brain} provided neurophysiological evidence using EEG, showing that LLM assistance systematically reduces neural connectivity compared to unassisted work, with accumulated effects over repeated sessions.}

\stageTwo{\subsubsection{Physiological Measurement in AI-Assisted Development}}
\stageTwo{Despite the growing behavioral and self-report evidence on the cognitive costs of AI-assisted development, physiological measurement in this context remains nascent. \citet{alhaque2025decoding} proposed a study protocol combining EEG, eye tracking, and NASA-TLX to study developer cognition with and without AI assistance, but no completed study yet combines multimodal biometrics (EEG, EDA, HRV, eye tracking) in AI-assisted programming contexts. \citet{elfares2025gazecopilot} demonstrated that real-time gaze data reflects developers' cognitive load and attention during code comprehension, further supporting the use of physiological signals in this domain. Our study addresses this gap by combining multiple biometric signals with performance metrics and subjective assessments to provide an integrated view of how AI assistance affects developer cognition.}

\bigskip
\section{Hypotheses}\label{sec:hypotheses}

According to the aforementioned theoretical background and related work, we formulate our hypotheses as follows.
To address RQ1 (\textit{extent to which developers' physiological measures differ when programming with and without AI}), we expect AI assistance to cause a shift in developers' cognitive load from solution generation to evaluation and verification, potentially changing physiological patterns as mental demands transition from creating code to assessing AI suggestions. Therefore, we hypothesize:


\textbf{H1}: \textit{Developers will exhibit distinct physiological patterns when programming with vs. without AI assistance}.

To address RQ\stageTwo{2} (\textit{moderating role of developer experience}), we note that experience likely plays a key role in how developers interact with and benefit from AI assistance. Drawing on the observation by \citet{friedmann2024pair} that experienced programmers use AI tools differently than novices, we \rev{speculate} that cognitive adaptations to AI assistance \rev{may} vary based on experience. Experienced developers have automated many programming processes and developed efficient mental models, potentially reducing the cognitive impact of AI assistance. \stageTwo{In our setting, we operationalize developer experience as academic seniority (undergraduate vs.\ graduate students), adopted as an accessible proxy within a student sample; we acknowledge that this contrast does not span the full junior--senior range studied by \citet{friedmann2024pair} and revisit this caveat in the Limitations.} Therefore, we hypothesize:

\textbf{H\stageTwo{2}}: \textit{Developer experience will moderate the relationship between AI assistance and physiological measures}.
            

To address RQ\stageTwo{3} (\textit{physiological measures correlation with performance metrics across conditions}), we note that the relationship between physiological states and performance outcomes is well-established in cognitive psychology but unexplored in AI-assisted programming. Building on previous findings that biometric measurements can predict task difficulty~\citep{fritz2014icse} and outcome~\citep{LaudatoEtAl:predictingBugs}, we expect physiological measures associated with efficient cognitive processing to correlate with improved performance outcomes regardless of whether AI assistance is used. Therefore, we hypothesize:

 \textbf{H\stageTwo{3}}: \textit{Physiological measures will correlate with performance metrics in both conditions}.
            
            
            
            

Finally, to address RQ\stageTwo{4} (\textit{alignment between subjective perceptions and objective measures across conditions}), we consider that the novel human-AI interaction paradigm may disrupt developers' metacognitive awareness---their ability to assess their performance accurately~\citep{Prather:2024}. In traditional programming, developers have developed calibrated mental models of their capabilities through years of experience. However, the introduction of AI assistance represents a significant shift in the development process that may challenge these established metacognitive frameworks. Therefore, we hypothesize:

\textbf{H\stageTwo{4}}: \textit{The alignment between subjective perceptions and objective measures (both physiological and performance-based) will differ between AI-assisted and non-AI-assisted conditions}.
            

\section{Methodology}\label{sec:methodology}

\subsection{Study Design}
We employed a within-subjects, 2x2 factorial \rev{crossover} design where each participant completed two programming tasks: one with AI assistance and one without. 
This design counterbalances both treatment order (AI-assisted vs. non-AI-assisted) and task order (Task A vs. Task B), creating four experimental sequences to ensure that each treatment appears equally often in each temporal position and in combination with each task. 
Participants were randomly assigned to one of these four sequences to control for learning effects, fatigue effects, and task-specific effects. 
This approach allowed us to directly compare the same individual's physiological responses, performance, and perceptions across both conditions while controlling for individual differences in programming ability and physiological baselines, as well as potential interactions between AI assistance effectiveness and specific task characteristics.


\subsection{Participants and Sampling Strategy}
\begin{table*}[t]
  \centering
  \caption{Participant demographics and self-reported questionnaire responses.
    Continuous variables are reported as mean\,$\pm$\,SD; categorical variables as $n$\,(\%).}
  \label{tab:participants}
  \resizebox{\textwidth}{!}{
  \begin{threeparttable}
  \footnotesize
  \setlength{\tabcolsep}{4pt}
  \renewcommand{\arraystretch}{0.92}
  \begin{tabular}{@{}l rrr@{}}
    \toprule
    \textbf{Characteristic} & \textbf{Uniba} ($n=34$) & \textbf{ITU} ($n=26$) & \textbf{Combined} ($N=60$) \\
    \midrule
    Age, mean ± SD & 23.3 ± 2.2 & 25.8 ± 3.2 & 24.4 ± 3.0 \\
    \quad Range & [20–27] & [21–35] & [20–35] \\
    \multicolumn{4}{@{}l}{\textit{Gender}} \\
    \quad Female & 4 (11.8\%) & 5 (19.2\%) & 9 (15.0\%) \\
    \quad Male & 29 (85.3\%) & 19 (73.1\%) & 48 (80.0\%) \\
    \quad Other / Non-binary & 1 (2.9\%) & 2 (7.7\%) & 3 (5.0\%) \\
    \multicolumn{4}{@{}l}{\textit{Education}} \\
    \quad Bachelor's degree (BSc) & 17 (50.0\%) & 11 (42.3\%) & 28 (46.7\%) \\
    \quad Master's degree (MSc) & 8 (23.5\%) & 13 (50.0\%) & 21 (35.0\%) \\
    \quad PhD & 9 (26.5\%) & 2 (7.7\%) & 11 (18.3\%) \\
    \multicolumn{4}{@{}l}{\quad\textit{Aggregated (Academic seniority)}} \\
    \quad\quad Undergraduate (BSc) & 17 (50.0\%) & 11 (42.3\%) & 28 (46.7\%) \\
    \quad\quad Graduate (MSc + PhD) & 17 (50.0\%) & 15 (57.7\%) & 32 (53.3\%) \\
    \multicolumn{4}{@{}l}{\textit{Native language}} \\
    \quad Danish & — & 21 (80.8\%) & 21 (35.0\%) \\
    \quad Italian & 34 (100.0\%) & — & 34 (56.7\%) \\
    \quad Other\tnote{a} & — & 5 (19.2\%) & 5 (8.3\%) \\
    \multicolumn{4}{@{}l}{\textit{AI usage frequency}} \\
    \quad Never & 0 (0.0\%) & 0 (0.0\%) & 0 (0.0\%) \\
    \quad Rarely & 1 (2.9\%) & 2 (7.7\%) & 3 (5.0\%) \\
    \quad Weekly & 6 (17.6\%) & 2 (7.7\%) & 8 (13.3\%) \\
    \quad Daily & 27 (79.4\%) & 22 (84.6\%) & 49 (81.7\%) \\
    \multicolumn{4}{@{}l}{\textit{AI literacy}} \\
    \quad No idea how to interact with AI tools & 0 (0.0\%) & 0 (0.0\%) & 0 (0.0\%) \\
    \quad Knows, but rarely has useful results & 4 (11.8\%) & 2 (7.7\%) & 6 (10.0\%) \\
    \quad Knows and has useful results & 18 (52.9\%) & 16 (61.5\%) & 34 (56.7\%) \\
    \quad Can build prompts and obtain desired results & 12 (35.3\%) & 8 (30.8\%) & 20 (33.3\%) \\
    \multicolumn{4}{@{}l}{\textit{Prior experience w/ AI coding tools}} \\
    \quad In the last month & 0 (0.0\%) & 0 (0.0\%) & 0 (0.0\%) \\
    \quad In the last 6 months & 1 (2.9\%) & 2 (7.7\%) & 3 (5.0\%) \\
    \quad In the last year & 6 (17.6\%) & 2 (7.7\%) & 8 (13.3\%) \\
    \quad Before this year & 27 (79.4\%) & 22 (84.6\%) & 49 (81.7\%) \\
    \bottomrule
  \end{tabular}
  \begin{tablenotes}
    \footnotesize
    \item[a] Greek ($n=1$), Nepali ($n=1$), Romanian ($n=2$), Spanish ($n=1$).
    \item SD\,=\,Standard Deviation.
  \end{tablenotes}
  \end{threeparttable}
  }
\end{table*}

We recruited 60 computer science students, with 34 from Uniba and 26 from ITU, exceeding the minimum target of at least 40 participants (20+ per site) set in the protocol registration~\citep{stage1report}. Their demographic, AI familiarity, and self-reported characteristics are summarized in Table~\ref{tab:participants}. \stageTwo{Participants were predominantly male (80.0\%), with a mean age of 24.4\,$\pm$\,3.0 years (range 20--35); on average, participants at ITU were slightly older than those at Uniba (25.8\,$\pm$\,3.2 vs.\ 23.3\,$\pm$\,2.2 years). Their educational background spanned undergraduate (BSc, 46.7\%), graduate (MSc, 35.0\%), and doctoral (PhD, 18.3\%) levels, with a higher concentration of MSc students at ITU (50.0\%) and of PhD students at Uniba (26.5\%). Participants' native languages reflected the recruitment sites: all Uniba participants were native Italian speakers (100.0\%), whereas at ITU the majority were native Danish speakers (80.8\%), with the remaining 19.2\% covering other backgrounds. Across the combined sample, this corresponded to 56.7\% Italian, 35.0\% Danish, and 8.3\% other.}

For inclusion, candidates had to be enrolled in a computer science or related program at either site, possess at least one year of programming experience, and demonstrate familiarity with Java programming languages. Additionally, they had to have prior experience with AI coding assistants (ChatGPT) and integrated development environments (VS Code).
Additionally, as part of our screening process, we used a technology familiarity scale to comprehensively assess participants' experience with AI coding assistants. 
This included measuring the frequency of use, types of tasks performed with AI assistance, self-reported proficiency levels, and specific exposure to the ChatGPT platform used in our study. 
At each site, participants were stratified by developer experience, operationalized as academic seniority and contrasting undergraduate (BSc) with graduate (MSc and PhD) students.
This stratification ensured representation of both groups and enabled the test of hypothesis H\stageTwo{2} on how developer experience moderates the effects of AI assistance on cognitive processes and performance.
We acknowledge that the two institutions may have different academic curricula and student backgrounds, which we documented and considered during the analysis.
This diversity in educational contexts enhances the generalizability of our findings rather than limiting them.

Recruitment was conducted through university mailing lists, department bulletin boards, and direct outreach to CS courses at both institutions. 
Potential participants were required to complete a screening questionnaire to determine their eligibility based on the inclusion and exclusion criteria. 
Recruitment materials and screening questionnaires were provided in the language of instruction at each site (see \textit{Language Considerations}).
No monetary compensation, gifts, or grade bonuses were provided to selected participants, who received their individual biometric data plots and access to study results upon request.

\stageTwo{To characterize participants' prior exposure to AI coding tools, the screening questionnaire included a  technology familiarity scale focused on: (1) frequency of use (never/rarely/weekly/daily), (2) AI literacy, ranging from no prior interaction to the ability to craft prompts that obtain the desired results, and (3) recency of first use (within the last month, the last six months, the last year, or before this year). The resulting distributions are reported in Table~\ref{tab:participants}. The registered protocol committed to using a composite of these items as a covariate in the models, but ceiling effects in the sample prevented this (see~\ref{appendix:deviations}).}

Finally, to enable a brief environment familiarization, upon arrival, participants received a 5-minute orientation on the specific VS Code configuration used in our study. 
This allowed them to verify their comfort with the environment setup and confirm their basic familiarity with ChatGPT functionality.


\subsection{Experimental Protocol}
\begin{figure*}
     \centering
     \includegraphics[width=\textwidth]{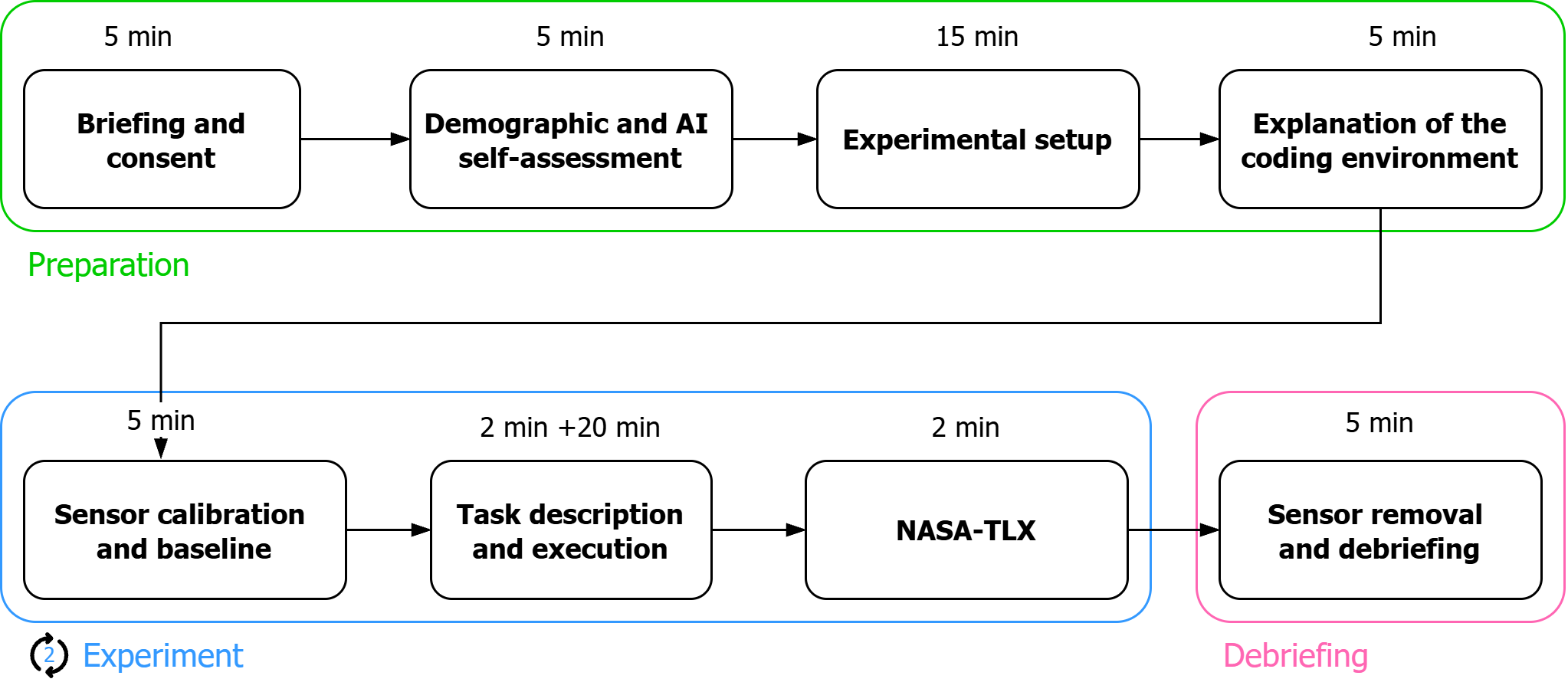}
     \vspace{-6mm}
     \caption{Experimental protocol phases.}
    \label{fig:exp-timeline}
 \end{figure*}

At each site, the experiment was conducted in a controlled laboratory environment to minimize external distractions and ensure consistent measurement conditions. 
The overall timeline required about 90 minutes to complete (see Figure~\ref{fig:exp-timeline}).

\textit{Preparation Phase}.
Upon arrival, participants received a briefing about the purpose of the study without revealing specific hypotheses and provided informed consent. 
This initial step took approximately 5 minutes. 
Then, participants completed a background questionnaire covering demographics, programming experience, prior experience with AI programming tools, and self-assessed proficiency (5 min.). 
Then, participants received a 5-minute introduction to the split-screen environment, where they were guided through the interactions with VS Code to ensure comfort with the configuration before the experimental phase.
Finally, biometric equipment was set up, with variations by site (see Sect.~\ref{sec:biometric-equipment}). A brief calibration phase ensured the proper functioning of all sensors, with standardized calibration protocols across both sites (5 min.). Prior to accessing the task description and beginning the task execution phase, participants were asked to remain still for 30 seconds while baseline physiological measurements were collected. During this period, the screen displayed a white page instructing participants to wait.
%


\textit{Experimental Phase}.
The experimental phase began with a presentation of the first programming task. 
Participants then had 20 minutes to complete the task, either with or without AI assistance, based on their assigned condition order. 
Throughout the task, physiological measurements were continuously recorded, along with screen activity, keystrokes, and mouse events to capture development behavior. 
Upon completion or after 20 min., participants filled out the NASA-TLX post-task questionnaire (5 min.) for self-assessment of workload. 
Before starting the second task, we recalibrated the sensors (2 min.). 
Participants then proceeded to the second 20-minute programming task in the alternate condition, followed by another 5-minute NASA-TLX questionnaire.

\textit{Debriefing Phase}.
After the biometric equipment was removed, participants took part in a semi-structured interview (5 min.), in which open-ended questions invited them to compare their experiences across the two conditions, reflecting on task complexity, their perception of the development environment, and the use of AI. Finally, they were debriefed on the complete purpose and hypotheses of the study.

\subsection{Instrumentation}\label{sec:materials-instrument}
\textit{Programming Tasks}.
We developed two programming tasks of equivalent difficulty, validated through pilot testing. Each task was presented primarily via UML class and activity diagrams, with minimal textual description to discourage direct copying into AI prompts. The class diagrams depicted 4–6 classes, including their attributes, methods, and relationships, while the activity diagrams illustrated the main workflow and control flow.
Both tasks were designed to be completable within 20 minutes, yet complex enough to require significant cognitive effort. 
Task A involved implementing a simplification of the connection logic of a mobile device. Given a UML class diagram showing \texttt{Device}, \texttt{Connection}, \texttt{Wi-Fi}, and \texttt{MobileData} classes with their relationships, participants were required to implement the logic of connecting to the preferred connection type. The implementation required using the provided \texttt{DeviceAPI} and \texttt{ConnectionAPI} interfaces, which described the functions to determine which connections are available.
Task B involved implementing a simplified online shopping application. Given a UML class diagram showing \texttt{ShopItem}, \texttt{ShoppingCart}, \texttt{PaymentMethod}, \texttt{MemberCard} and \texttt{CreditCard} classes with their relationships, participants had to implement the logic of adding shop items to the shopping cart, purchasing the items in the shopping cart if they were all in stock, and the payment method had the necessary funds available. The implementation required using the provided \texttt{StockAPI} interface, which describes the functions for determining whether shop items are in stock.

\begin{figure}[t!]
  \centering
  \begin{subfigure}{\textwidth}
    \centering
    \fcolorbox{gray!30}{white}{%
      \includegraphics[width=\linewidth]{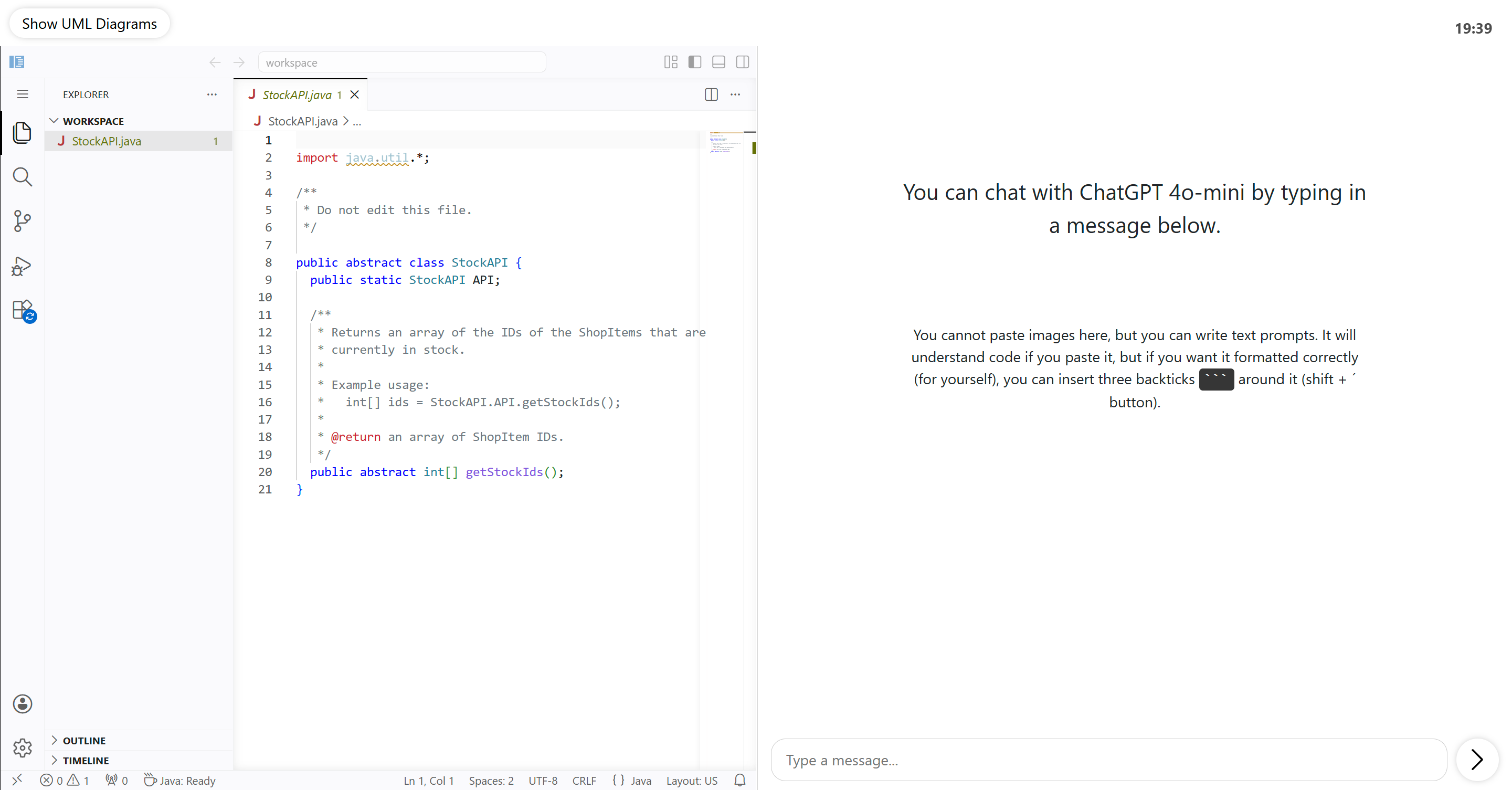}%
    }
    \caption{AI-assisted condition.}
    \label{fig:ai}
  \end{subfigure}

  \vspace{1em}

  \begin{subfigure}{\textwidth}
    \centering
    \fcolorbox{gray!30}{white}{%
      \includegraphics[width=\linewidth]{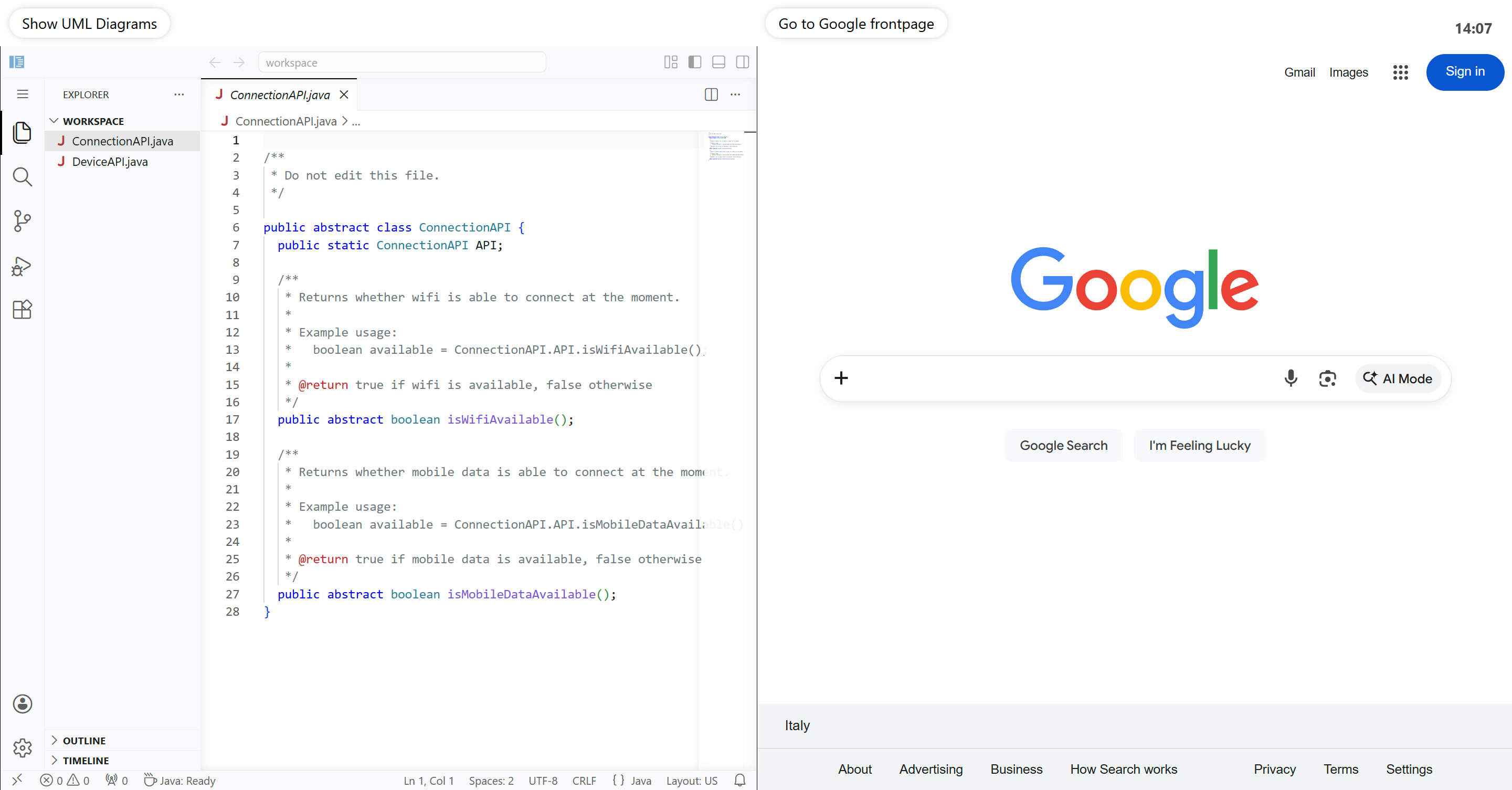}%
    }
    \caption{NON-AI-assisted condition.}
    \label{fig:nonai}
  \end{subfigure}
  \caption{Development environment screenshots for AI-assisted vs. non-assisted conditions.}
  \label{fig:dev-env-screenshots}
\end{figure}

\textit{Development Environment}. 
All participants used an identical development setup at both sites, consisting of a workstation equipped with an external monitor, keyboard, and mouse. Development activity was recorded using OBS Studio for screen capture. In the AI-assisted condition, ChatGPT (version 4o-mini) served as the AI assistant. The integrated development environment (IDE) was the web-based version of Visual Studio Code, configured with Java extensions and embedded within a custom \texttt{oTree} \rev{framework~\citep{chen2016otree}}, which controls the sequence and timing of the experimental stages. 
oTree is an open-source, Python-based web framework that enables researchers to design, implement, and administer interactive behavioral experiments.
In our study, oTree served as a browser-based platform for presenting programming tasks to participants, collecting their responses, and managing the experimental flow across different conditions (AI-assisted vs. non-assisted), as shown in
Figure~\ref{fig:dev-env-screenshots}. The framework provides session management, automated data collection, and synchronized timing mechanisms, while offering a standardized interface that ensures consistent task presentation across both research sites.


Participants interacted exclusively with this application. Each stage of the experiment, from sensor calibration and baseline acquisition phases, to development tasks and questionnaire completion, was implemented as a distinct view within \texttt{oTree}, each governed by either a preset timeout or a manual continuation mechanism to ensure temporal consistency across sessions.
The core of the experiment is the \textit{task view}, also presented as a split screen. The left pane contains the IDE, while the right pane either provides a chat interface for interaction with ChatGPT (for the AI-assisted condition) or displays a static webpage, such as the Google homepage (for the control condition). 
Participants accessed the task description 
via dedicated buttons. 

The \texttt{oTree} application communicated with a local server that managed the interaction with the ChatGPT API and handles the setup and teardown of source files in the IDE before and after each task. Additionally, a local Lab Streaming Layer (LSL) server was used to stream data from the \texttt{oTree} application, including event markers with timestamps and phase descriptions, as well as mouse and keyboard input data.

\textit{Biometric Equipment}.\label{sec:biometric-equipment}
\begin{figure}[t!]
  \centering
  \begin{subfigure}[t]{0.48\textwidth}
    \centering
    \includegraphics[width=\linewidth]{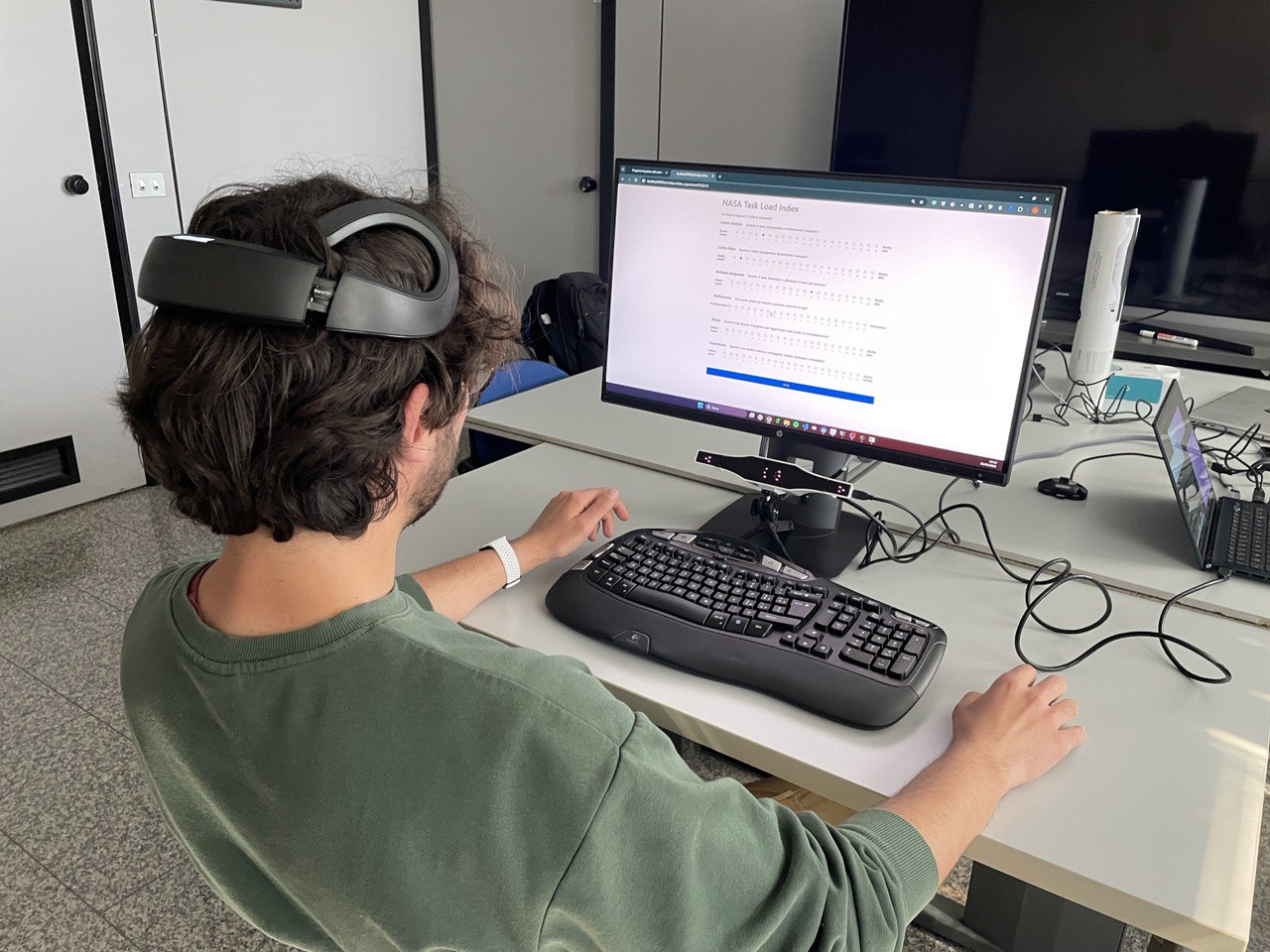}
    \caption{Uniba site.}
    \label{fig:uniba}
  \end{subfigure}
  \hfill
  \begin{subfigure}[t]{0.48\textwidth}
    \centering
    \includegraphics[width=\linewidth]{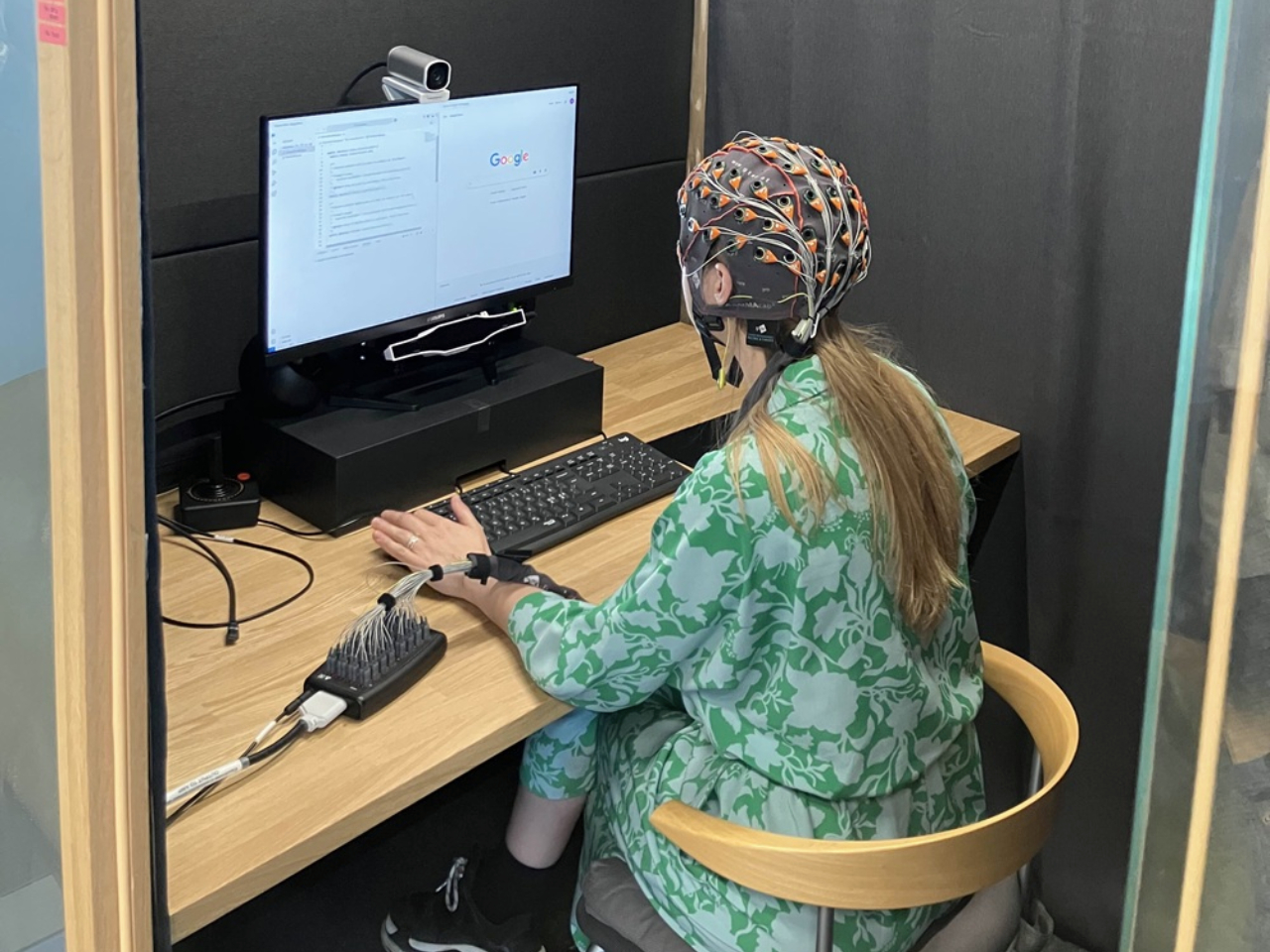}
    \caption{ITU site.} 
    \label{fig:itu}
  \end{subfigure}
  \caption{Experimental setup for participants in both Uniba and ITU sites.}
  \label{fig:participant-setup}
\end{figure}
One of the strengths of our multisite approach is the use of different but comparable biometric equipment at each site (see Figure~\ref{fig:participant-setup}), allowing us to assess the robustness of our findings across different measurement systems.
At Uniba (Figure~\ref{fig:uniba}), we used the Neurosity Crown\footnote{\url{https://neurosity.co}}, a wireless 8-channel EEG headset, and the Empatica EmbracePlus wristband\footnote{\url{https://www.empatica.com/en-eu/embraceplus}} for continuous measurements of electrodermal activity (EDA) and heart rate variability (HRV).
At ITU (Figure~\ref{fig:itu}), EEG measurements were collected using a g.HIamp amplifier with 64 channels and dry g.SAHARA electrodes\footnote{\url{https://www.gtec.at/product/g-hiamp-256-channel-biosignal-amplifier}}.
\stageTwo{Unlike the registered protocol, no wristband device was available at ITU; therefore, EDA and HRV data were collected only at the Uniba site (see~\ref{appendix:deviations}).}
Both teams used the Gazepoint GP3 HD\footnote{\url{https://www.gazept.com}} high-precision eye tracker with a 150Hz sampling rate.


\subsection{Physiological Signals}\label{sec:physio-signal}

\stageTwo{The theoretical motivation for our multimodal physiological approach rests on the hypothesis that AI assistance reorganizes the cognitive architecture of programming---reducing self-generated solution effort while increasing evaluation and validation load. No single physiological signal is sufficient to capture this reorganization; evidence 
consistently points to the need for a multimodal approach combining brain activity,heart and electrodermal responses, and eye movements and pupillary response~\citep{PUMA:theta:alpha:cognitiveLoad, kosmyna2025brain,fritz2014icse, castaldo2015acute}.} 

\stageTwo{In what follows, we detail each signal family, the specific metrics derived from it, and the theoretical justification for expecting differences between the AI-assisted and non-AI-assisted conditions.
Table~\ref{tab:signal-summary} summarizes the key metrics, the cognitive constructs they index, and the expected direction of differences between the non-AI and AI conditions.}
 
\subsubsection{Electroencephalography (EEG)}

\stageTwo{EEG provides the most direct window into cognitive load and is the most established signal in this context. We analyzed Power Spectral Density (PSD) in the theta (4--8\,Hz), 
alpha (8--12\,Hz), and beta (13--30\,Hz) bands. We adopt the canonical 
topographic division established in the cognitive workload 
literature e.g. frontal-midline electrodes for theta and parietal electrodes for alpha~\citep{borghini2014measuring, 
cavanagh2014frontal}.\footnote{Frontal-midline 
ROI: \texttt{F5, F6} (Uniba) and \texttt{Fp1, Fpz, Fp2, AF7, AF3, AF4, AF8, F5, F3, F1, Fz, F2, F4, F6, AFz} (ITU). Parietal ROI: \texttt{PO3, PO4, CP3, CP4} (Uniba) and \texttt{P7, P5, P3, P1, Pz, P2, P4, P6, P8} (ITU). Beta is averaged across all available electrodes.}}

\stageTwo{Our primary index is the \textbf{theta/alpha ratio} 
($\theta/\alpha$), which has been validated as a workload indicator in applied multitasking and human factors settings~\citep{borghini2014measuring, PUMA:theta:alpha:cognitiveLoad}. Since theta rises and alpha falls as cognitive demand increases, a \emph{higher} ratio reflects \emph{greater} cognitive engagement \cite{borghini2014measuring}. 
Frontal theta in particular has been shown to scale with the number of concurrently processed subtasks in EEG workload 
paradigms~\citep{PUMA:theta:alpha:cognitiveLoad}. \textbf{Beta-band PSD} is additionally extracted as a complementary marker of mental arousal: a Microsoft EEG study of office workers reported that beta activity rises during stressful, demanding activities~\citep{microsoft2021brain}}
 
\stageTwo{\cite{kosmyna2025brain} applied EEG connectivity analysis in an essay-writing paradigm and found that participants using an LLM exhibited systematically \emph{lower overall brain connectivity} across multiple frequency bands (including alpha and beta) compared to those using a search engine or no tool (brain-only condition). The brain-only group exhibited the strongest, most distributed connectivity networks, suggesting that LLM assistance reduces the level of active neural engagement, consistent with cognitive offloading reducing the brain's need to self-organize information~\citep{grinschgl2021offloading}. Based on these findings, and considering that frontal-midline theta and parietal alpha already track cognitive effort~\citep{fritz2014icse, PUMA:theta:alpha:cognitiveLoad}, we hypothesize that the non-AI condition will elicit higher theta PSD and lower alpha PSD than the AI-assisted condition, yielding a higher $\theta/\alpha$ ratio.
}

\begin{table*}[t]
\caption{Summary of physiological signals, metrics, expected direction of differences between non-AI-assisted and AI-assisted conditions, and supporting references. AI assistance is generally expected to produce lower cognitive load signatures; for EDA metrics and blink rate, the expected direction is uncertain.}
\label{tab:signal-summary}
\centering
\scriptsize
\renewcommand{\arraystretch}{1.25}
\setlength{\tabcolsep}{3pt}
\begin{tabular*}{\textwidth}{@{\extracolsep{\fill}}
  l
  >{\raggedright\arraybackslash}p{0.12\textwidth}
  >{\raggedright\arraybackslash}p{0.20\textwidth}
  c
  c
  >{\raggedright\arraybackslash}p{0.26\textwidth}
  @{}}
\toprule
\textbf{Signal} & \textbf{Metric} & \textbf{Cognitive Construct} & \textbf{Non-AI} & \textbf{AI} & \textbf{References} \\
\midrule
\multirow{2}{*}{\shortstack[l]{EEG\\(PSD)}}
  & $\theta/\alpha$ ratio & Cognitive load & higher & lower & \citet{PUMA:theta:alpha:cognitiveLoad,cavanagh2014frontal} \\
  & Beta power            & Stress         & higher & lower & \cite{microsoft2021brain} \\
\midrule
\multirow{2}{*}{HRV}
  & LF/HF ratio &  Arousal, cognitive effort           & higher & lower  & \citet{castaldo2015acute,Wood:HRV:cognitiveChallenge} \\
  & RMSSD       & Parasympathetic activity 
  & lower  & higher & \citet{castaldo2015acute,Wood:HRV:cognitiveChallenge} \\
\midrule
\multirow{3}{*}{EDA}
  & Mean EDA (tonic)         & Baseline sympathetic arousal           & higher           & lower     & \citet{fritz2014icse,fucci2019icpc,boucsein2012electrodermal} \\
  & EDA peak count (phasic)  & Discrete arousal/stress responses      & uncertain & uncertain & \citet{westerink2020deriving} \\
  & SCR amplitude (phasic)   & Magnitude of arousal/stress responses  & uncertain & uncertain & \citet{westerink2020deriving} \\
\midrule
\multirow{3}{*}{\shortstack[l]{GAZE}}
  & Fixation duration & Effortful visual comprehension                     & longer & shorter & \citet{Zagermann:et:Al:eye-tracker:cogLoad,tang2024developer,elfares2025gazecopilot} \\
  & Fixation count    & Breadth of visual exploration                      & uncertain   & uncertain   & \citet{Zagermann:et:Al:eye-tracker:cogLoad,tang2024developer} \\
  & Blink rate        & Cognitive load -- blink suppression during reading & lower  & higher  & \citet{lenskiy2016blink,rosenfield2015cognitive} \\
\bottomrule

\end{tabular*}
\end{table*}
 
\subsubsection{Heart Rate Variability (HRV)}
 
\stageTwo{We characterized HRV using two complementary metrics: \textbf{RMSSD} (root mean square of successive RR-interval differences), a time-domain index of parasympathetic activity, and the \textbf{LF/HF} ratio (low-to-high frequency power ratio), reflecting sympathetic--parasympathetic balance \cite{castaldo2015acute}. Together, these two indices capture the autonomic nervous system's response to mental effort: under elevated sympathetic engagement, parasympathetic activity decreases, producing lower RMSSD and a higher LF/HF ratio~\citep{castaldo2015acute, kim2018stress}.} \stageTwo{ \citet{charles2019measuring} reports that cardiovascular measures, including time-domain and frequency-domain HRV indices, are sensitive to changes in mental workload across applied and laboratory tasks. The effect is detectable even on short recordings: \citet{Wood:HRV:cognitiveChallenge} showed that HRV indices react measurably to attention-demanding tasks within minutes, although the magnitude and direction of the response can vary across populations.} 

\stageTwo{Building on this evidence, we expect non-AI coding---which requires sustained generative effort---to suppress RMSSD and elevate LF/HF more strongly than AI-assisted coding, where part of the cognitive effort is offloaded to the model. We note that this prediction assumes that reduced cognitive engagement under AI assistance also translates into reduced sympathetic arousal, an assumption that may not hold uniformly~\citep{kosmyna2025brain}.}

 
\subsubsection{Electrodermal Activity (EDA)}
 
\stageTwo{EDA reflects sympathetic nervous system arousal and is widely used as a proxy for stress and emotional engagement during cognitively demanding tasks~\citep{boucsein2012electrodermal}. We extracted three components: (1) \textbf{mean skin conductance level (tonic EDA)}, reflecting baseline sympathetic arousal and sustained engagement~\citep{boucsein2012electrodermal}; (2) the \textbf{count of skin conductance peaks (phasic EDA)}, used as a proxy for the frequency of discrete arousal events triggered by salient or surprising 
stimuli~\citep{dawson2007electrodermal, westerink2020deriving}; and (3) the \textbf{mean amplitude of skin conductance responses (SCR amplitude)}, which captures the intensity of each arousal event and provides a fine-grained index of stress 
reactivity~\citep{dawson2007electrodermal}.}
 
\stageTwo{\citet{fritz2014icse} used EDA alongside EEG and eye tracking to distinguish easy from difficult programming tasks, demonstrating that EDA responds to changes in task difficulty. \citet{westerink2020deriving} validated EDA peak counts as a cortisol-related stress indicator using wearable sensors, and \citet{fucci2019icpc} replicated this in a code comprehension context, combining EDA with EEG and heart rate to classify developer task types.}
 
\stageTwo{The expected direction for tonic EDA follows from the cognitive load hypothesis: non-AI coding demands more independent problem-solving and sustained effort, likely producing higher mean tonic EDA. However, the expected direction for phasic EDA peak count is less certain. While sustained cognitive load is expected to be lower in the AI condition on average, individual moments of uncertainty (such as discovering that AI-generated code is incorrect or poorly matched to requirements) can trigger discrete arousal spikes. The net peak count therefore depends on how frequently such surprises occur. Given the well-documented inter-individual variability in phasic EDA responses~\citep{boucsein2012electrodermal}, this variability may be further amplified by differences in developer or AI experience across participants.}
 
\subsubsection{Eye Tracking}
 
\stageTwo{Eye tracking provides oculomotor indices of visual attention and cognitive processing. We derived three metrics: \textbf{fixation count}, \textbf{mean fixation duration}, and \textbf{blink rate}; this set differs from the eye-tracking indicators committed to in the registered protocol (see~\ref{appendix:deviations} for the rationale). 

\citet{Zagermann:et:Al:eye-tracker:cogLoad} proposed a 
descriptive model linking eye-tracking metrics to cognitive load in visual computing, arguing on the basis of prior empirical work~\citep{chen2011eye} that longer fixation durations reflect greater strain on working memory and indicate increased processing difficulty. In programming, \citet{tang2024developer} found that awareness of code provenance altered cognitive workload during the validation of LLM-generated code, and \citet{elfares2025gazecopilot} demonstrated that gaze data captures developers' cognitive load during the comprehension of AI-generated code.
The interpretation of blink rate is more nuanced. While higher blink rates accompany cognitive load in physical or auditory tasks, the relationship is reversed in reading paradigms: \citet{lenskiy2016blink} reported lower number of blinks during reading, and \citet{rosenfield2015cognitive} found that greater cognitive demand on digital screens decreases blink rate, as individuals suppress blinking to sustain visual intake. }

\stageTwo{Building on these results, we derive condition-specific predictions as follows. For \textbf{fixation duration}, we expect non-AI coding to elicit longer fixations: developers must actively construct the solution by repeatedly inspecting the UML diagrams and writing code from scratch, requiring sustained visual processing of dense information sources. For \textbf{blink rate}, the inverse blink--load relationship documented in reading paradigms leads us to expect a \emph{lower} blink rate under non-AI than under AI assistance, as developers suppress blinking to maintain visual intake while constructing the solution. For \textbf{fixation count}, the directional prediction is less straightforward: while a higher count could reflect a broader visual exploration of the UML diagrams and the code under construction in non-AI, the literature also documents that increased cognitive load may instead reduce fixation rate as attention focuses on fewer information sources~\citep{Zagermann:et:Al:eye-tracker:cogLoad, chen2011eye}. We therefore treat the direction of fixation count as exploratory.}

\subsection{Quality Assurance} 
To ensure the validity and reliability of our study, we implemented a comprehensive set of quality assurance procedures. Before the main data collection, we conducted a pilot study at each site to validate the equivalence and difficulty of the programming tasks, assess the reliability of biometric measurements, and consolidate the experimental procedures.

Throughout the study, we maintained high data quality standards. EEG signal quality was monitored in real time, and post-processing included established artifact detection and removal procedures. EDA and HRV signals were evaluated for continuity and plausibility, with recordings exhibiting noise or irregularities flagged for verification. 

To ensure environmental consistency across sessions, we controlled for ambient conditions such as lighting and noise levels, and scheduled sessions during comparable times of day across sites. This reduced the risk of uncontrolled variables that may affect physiological or cognitive responses.

Task equivalence was established through 
pilot testing to ensure similar programming concepts and cognitive demands.
Participants were also asked to rate the understandability of each task and their confidence in completing it within 20 minutes.



\textit{Ethical Considerations}.
The ethical approval was secured from the IT University of Copenhagen  and the University of Bari.
To ensure ethical conduct, we obtained informed consent from all participants before their participation, ensuring that they would have been able to withdraw at any time during the study. All data collected were stored securely in compliance with GDPR and anonymized during analysis and reporting at each site independently. 
Screen recordings focused solely on the oTree window, excluding the participants. Finally, participants were debriefed about the true nature of the experiment.

\textit{Cross-Site Standardization}.  
We ensured consistency across sites through regular cross-site meetings, shared protocols for recruitment, procedures, and data collection, and conducted pilot testing at both locations. 
Physiological measurements followed standardized calibration, and site-specific factors were coded as variables in the analysis.  

\textit{Language Considerations.}
At ITU, researchers conducted all verbal interactions with participants in Danish (the majority of participants' native language) and English, while experimental materials, consent forms, and questionnaires were in English, consistent with the English-language master's program curriculum.
At Uniba, all materials were translated into Italian, including consent forms, the NASA-TLX questionnaire, experimental instructions, and debriefing materials, with researchers conducting all interactions in Italian.
Programming-related content was standardized in English across both sites, including UML class diagrams, variable names, function names, and code-related terminology. This approach aligned with standard programming education practices at both institutions.
The consistency of programming language elements across sites ensured that task difficulty and cognitive demands remained equivalent, whereas the use of the participants' native language in procedural communication reduced potential comprehension barriers that could confound our measurements. We acknowledge that language differences may have introduced site-specific effects and, therefore, included site as a factor in our statistical models to account for potential linguistic influences on cognitive load and performance metrics.


\subsection{Pilot Study}\label{sec:pilot}
We employed a multi-stage piloting strategy to validate the experimental setup, calibrate sensors, test data collection pipelines, and ensure task equivalence.

At ITU, the piloting process comprised four exploratory interviews with students meeting our participation criteria to validate the task design, followed by two formal pilot sessions and five secondary pilot sessions. 
These stages enabled us to refine procedural and technical aspects of the study, including the timing of stimuli, synchronization of sensor data (e.g., eye tracking, keystrokes, and screen recordings), and the integrity of logged behavioral metrics. 
To assess task equivalence specifically, we analyzed both self-reported measures and behavioral outcomes from the pilot sessions. While individual task preferences varied among participants, suggesting broadly balanced perceived difficulty, the NASA-TLX cognitive load ratings and debriefing interviews revealed that the Connection task was generally more cognitively demanding than the Shopping Cart task. 
However, the Connection task also had higher completion rates despite its increased difficulty. 
Participants' feedback on abstraction levels and domain familiarity further informed our understanding of task characteristics. These findings support our conclusion that the tasks are sufficiently equivalent for experimental purposes, with observed difficulty differences appearing to reflect individual skillsets and programming preferences rather than systematic imbalance.

At Uniba, we conducted a focused validation pilot with three participants to confirm that the technical infrastructure, experimental setup, and behavioral observations established at ITU could be successfully replicated in the local laboratory environment.
This pilot specifically verified the compatibility of data collection systems, sensor calibration procedures, and task administration protocols within the Uniba setting, demonstrating consistent technical performance and participant responses across sites.

Overall, these piloting efforts confirmed the reliability of our experimental pipeline across both research sites and validated the suitability of both programming tasks for eliciting comparable cognitive and behavioral responses in our target population.

\subsection{Project Execution Timeline}
\stageTwo{The study was executed over approximately 12 months, as illustrated in Figure~\ref{fig:project-timeline}.
Experimental environments at both sites were set up in Jun 2025.
Pilot testing was conducted between Jul and Aug 2025 and used to refine the protocol, calibrate sensors, and validate task equivalence (see Section~\ref{sec:pilot}).
Participant recruitment and data collection ran in parallel at the two sites on staggered schedules between Sep 2025 and Mar 2026.
Data preprocessing, analysis, and Stage~2 manuscript preparation were carried out between Mar and May 2026.}

\stageTwo{
\begin{figure}[t]
\centering
\resizebox{\columnwidth}{!}{
%
%
\begin{tikzpicture}[
    x=1.05cm,
    bar/.style={rounded corners=1pt, minimum height=0.3cm},
    lblact/.style={anchor=east, font=\scriptsize\bfseries},
    lblsite/.style={anchor=east, font=\scriptsize},
    monthfont/.style={font=\scriptsize},
    yearfont/.style={font=\scriptsize\bfseries}
]

\def\rowh{0.5}   
\def\bh{0.15}    

\def\maxdepth{10}

\foreach \i in {0,...,11} {
    \draw[gray!25] (\i, 0.4) -- (\i, -\maxdepth*\rowh - 0.1);
}
\draw[gray!25] (12, 0.4) -- (12, -\maxdepth*\rowh - 0.1);

\node[yearfont] at (3.5, 0.7) {2025};
\node[yearfont] at (9.5, 0.7) {2026};

\foreach \m [count=\i from 0] in {Jun,Jul,Aug,Sep,Oct,Nov,Dec,Jan,Feb,Mar,Apr,May} {
    \node[monthfont] at (\i + 0.5, 0.25) {\m};
}

\def\rA{0}       
\def\rAu{1}      
\def\rAi{2}      
\node[lblact] at (-0.15, -\rA*\rowh) {Setup};
\node[lblsite] at (-0.15, -\rAu*\rowh) {Uniba};
\fill[orange!60, bar] (0, -\rAu*\rowh - \bh) rectangle (1, -\rAu*\rowh + \bh);
\node[lblsite] at (-0.15, -\rAi*\rowh) {ITU};
\fill[teal!50, bar] (0, -\rAi*\rowh - \bh) rectangle (1, -\rAi*\rowh + \bh);

\def\rB{3}
\def\rBu{4}
\def\rBi{5}
\node[lblact] at (-0.15, -\rB*\rowh) {Pilot testing};
\node[lblsite] at (-0.15, -\rBu*\rowh) {Uniba};
\fill[orange!60, bar] (1, -\rBu*\rowh - \bh) rectangle (3, -\rBu*\rowh + \bh);
\node[lblsite] at (-0.15, -\rBi*\rowh) {ITU};
\fill[teal!50, bar] (1, -\rBi*\rowh - \bh) rectangle (3, -\rBi*\rowh + \bh);

\def\rC{6}
\def\rCu{7}
\def\rCi{8}
\node[lblact] at (-0.15, -\rC*\rowh) {Data collection};
\node[lblsite] at (-0.15, -\rCu*\rowh) {Uniba};
\fill[orange!60, bar] (3, -\rCu*\rowh - \bh) rectangle (9, -\rCu*\rowh + \bh);
\node[lblsite] at (-0.15, -\rCi*\rowh) {ITU};
\fill[teal!50, bar] (6, -\rCi*\rowh - \bh) rectangle (10, -\rCi*\rowh + \bh);

\def\rD{9}
\def\rDj{10}
\node[lblact] at (-0.15, -\rD*\rowh) {Analysis \& writing};
\node[lblsite] at (-0.15, -\rDj*\rowh) {Uniba + ITU};
\fill[blue!50, bar] (9, -\rDj*\rowh - \bh) rectangle (12, -\rDj*\rowh + \bh);

\end{tikzpicture}}
\caption{Project execution timeline showing the parallel activities at the two sites.}
\label{fig:project-timeline}
\end{figure}
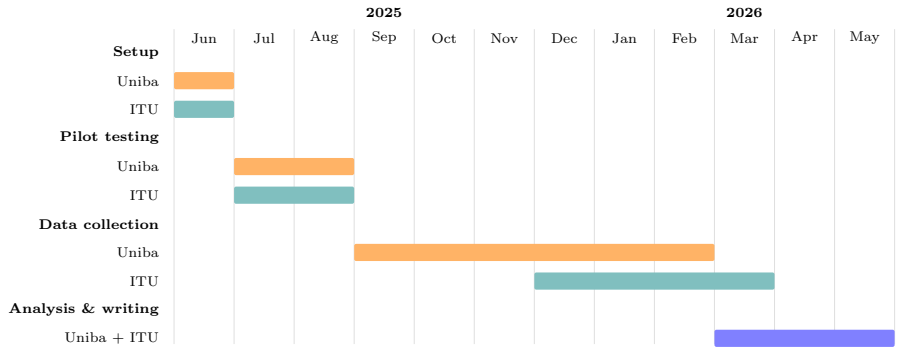
}

\section{Analysis}\label{sec:analysis}

In this section we describe the analyses performed and executed as computational Jupyter notebooks. All the notebooks are part of the replication package described in the \hyperref[sec:data-availability]{Data Availability} statement, which also includes anonymized data, to ensure full reproducibility.

\subsection{Physiological Data Preprocessing}
\label{sec:preprocessing}
For the preprocessing of \textit{physiological data}, we followed consolidated practices in this research field. For EEG data, we filtered the raw recordings and apply Independent Component Analysis to remove artifacts from muscle activity, eye movements, and electrical noise. Finally, the data were normalized relative to each participant's baseline measurements. These steps are necessary to ensure good signal quality in naturalistic interaction scenarios~\citep{hegedues2023investigating}. 
\stageTwo{The raw EDA signal was first low-pass filtered using a Butterworth filter with a 1 Hz cutoff frequency \cite{Girardi:2020:ICSE} 
The filtered signal was then decomposed into tonic (skin conductance level, SCL) and phasic (skin conductance response, SCR) components using the \texttt{cvxEDA} algorithm \cite{greco}. SCR peaks were identified with the \texttt{kim2004} method (minimum amplitude threshold = 0.1$\mu). S$ 
Prior to feature extraction, the EDA signal was z-score normalized against the baseline, in line with consolidated literature \cite{Girardi:2020:ICSE, fritz2014icse}.
For HRV analysis, we used the \texttt{hrvanalysis} package\footnote{https://pypi.org/project/hrv-analysis/
} to process the BVP signal. Processing involved computing RR intervals from the raw systolic peaks, cleaning them via range filtering (300–2000 ms) and Malik-rule ectopic beat removal \cite{Malik}. From the cleaned RR intervals, \texttt{hrvanalysis} provided both time-domain and frequency-domain metrics.
For gaze data, blinks were identified following the methodology 
implemented in PyGaze Analyser~\cite{PyGaze_Dalmaijer2014}, which detects them as consecutive signal losses reported by the eye-tracking device. A valid blink was defined as an eye closure lasting between 200 and 400ms, in accordance with established physiological thresholds~\cite{Alsaeedi2019,blink-rate_Al-gawwam}. Given the nominal sampling rate of 150Hz of the device used, these temporal bounds correspond to 30 and 60 missing samples, respectively. Validity flags for fixations provided by the device, together with the maximum duration of each fixation, were employed to compute fixation count and mean fixation duration.
All EEG, eye-gaze, EDA, and HRV features were extracted using 60-second sliding windows with 80\% overlap, following \cite{Anders2024}. For each window, we computed the metrics described in Table \ref{tab:signal-summary}.
}

\subsection{Temporal Alignment of Physiological Data Streams} 

To ensure a reproducible analysis workflow, we implemented a structured pipeline based on the Lab-Streaming Layer framework~\citep{kothe2025lab} to integrate the multiple physiological data streams collected during the study. 
Specifically, we synchronized all collected data streams, EEG, EDA, HRV and eye tracking, using timestamp alignment. Each stream was automatically annotated with data derived from key events logged by the oTree platform and participant-level experimental information (e.g., anonymized ID), capturing the task being performed, its order, and whether AI assistance was provided. At both sites, key behavioral events, such as reading UML diagrams or interacting with ChatGPT, were identified based on the analysis of annotated data. 
In addition, the event marker function of the Empatica Embrace Plus was used by participants in Bari to flag essential moments in their activity (e.g., start of each coding task).

All this information was combined with additional annotations derived from screen recordings, enabling a fine-grained, event-related analysis of cognitive processes at specific phases of the coding tasks. This additional annotation layer provided essential context for interpreting the statistical patterns observed across the extracted metrics.

\subsection{Variables and Measurements}\label{sec:variables-measurements}

The independent variable of our study is the programming condition, a categorical variable with two levels: AI-assisted vs. non AI-assisted. \stageTwo{In the statistical models, we refer to this as \texttt{Modality}}. We further include developer experience \stageTwo{ as a between-participants factor, operationalized as academic seniority and used as the moderator in H\stageTwo{2}}. It is a categorical variable with two levels (undergraduate vs.\ graduate students), aligned with the participant stratification criteria\stageTwo{; in the statistical models we refer to this factor as \texttt{Education}. Finally, we include \texttt{Task\_order} as a within-participants factor with two levels (first vs.\ second), encoding the temporal position of each task in a participant's session. Modeling this factor (and its interaction with \texttt{Modality}) is needed to account for potential carryover, sequence, and treatment\,$\times$\,period effects, which are inherently difficult to disentangle in crossover designs in programming contexts~\citep{vegas2016tse}. We further discuss these threats in Section~\ref{sec:limitations}}.

Furthermore, to address our research questions, we model variables spanning three families: physiological signals, self-reported workload, and task performance. 
Physiological variables operationalize the cognitive constructs targeted by each physiological signal, the metrics derived from it, and the expected direction of differences between the AI- and non-AI-assisted conditions are detailed in Section~\ref{sec:physio-signal} and summarized in Table~\ref{tab:signal-summary}. 

\stageTwo{The variable types, role of each variable in each confirmatory test, and the corresponding regression model are laid out in Table~\ref{tab:analytic-design}. All H1--H4 models include a participant-level random intercept ($1 \mid \texttt{Participant}$) to absorb within-subject correlation. Reference levels are first task for \texttt{Task\_order}, graduate for \texttt{Education}, and \emph{low} for both NASA-TLX sub-dimensions and \texttt{Performance}. Note that the physiological metrics play a dual role across the confirmatory tests: they are the dependent variables of H1 and H2 (each metric in a separate LMM) and they re-enter H3 as continuous predictors of \texttt{Performance} (each metric in a separate BinGLMM, jointly with the \texttt{Modality} interaction).}

\begin{table*}[t]
  \centering
  \caption{Role of each variable in the four hypothesis tests (H1--H4), with variable types and the fitted regression model. Within each H column, IV marks the independent variable, DV the dependent variable, \checkmark a variable included in the model, and -- a variable not in the model.}
  \label{tab:analytic-design}
  \begin{threeparttable}
  \footnotesize
  \setlength{\tabcolsep}{6pt}
  \renewcommand{\arraystretch}{1.15}
  \begin{tabular}{@{}>{\raggedright\arraybackslash}p{3.1cm} >{\raggedright\arraybackslash}p{2.3cm} cccc@{}}
    \toprule
    \textbf{Variable} & \textbf{Type} & \textbf{H1} & \textbf{H2} & \textbf{H3} & \textbf{H4} \\
    \midrule
    \texttt{Modality} & Categorical: AI, NON-AI & IV & IV & \checkmark & \checkmark \\
    \texttt{Task\_order} & Categorical: first, second & \checkmark & \checkmark & \checkmark & \checkmark \\
    \texttt{Education} & Categorical: undergraduate, graduate & -- & \checkmark & -- & -- \\
    \texttt{NASA-TLX}\tnote{a} & Binary: high, low & -- & -- & -- & IV \\
    EEG: \texttt{$\theta/\alpha$ ratio},\newline \texttt{Beta power} & Continuous & DV & DV & IV & -- \\
    HRV: \texttt{LF/HF ratio},\newline \texttt{RMSSD} & Continuous & DV & DV & IV & -- \\
    EDA: \texttt{Mean EDA},\newline \texttt{EDA peak count}, \texttt{SCR amplitude} & Continuous & DV & DV & IV & -- \\
    Eye-tracking: \texttt{Fixation duration}, \texttt{Fixation count}, \texttt{Blink rate} & Continuous & DV & DV & IV & -- \\
    \texttt{Performance}\tnote{b} & Binary: high, low & -- & -- & DV & DV \\
    \midrule
    \textit{Model fitted} &   & LMM & LMM & BinGLMM & BinGLMM \\
    \bottomrule
  \end{tabular}
  \begin{tablenotes}
    \footnotesize
    \item[a] Six dimensions plus the averaged score; $z$-scored across participants, then median-split.
    \item[b] Median-split of the rubric-based task-correctness score in $[0,1]$.
  \end{tablenotes}
  \end{threeparttable}
\end{table*}

\stageTwo{As a deviation from the registered protocol (see~\ref{appendix:deviations}), \textit{performance} was operationalized through a single metric: a task-correctness score derived from a structured assessment rubric, jointly designed by members of both research groups and grounded in previous work~\citep{BRABRAND2024111887,Nahdi_Fast-Fleeting-GPT_rubric}. The rubric comprises a series of task-specific items, each scored on an ordinal three-level scale: 0 (incorrect implementation), 0.5 (partially correct implementation), and 1 (correct implementation). The rubric follows a cumulative scoring approach: the higher the number of items implemented correctly, the higher the final score for each task, thus reflecting the degree of completeness of the solution provided. The cumulative score is normalized to the $[0,1]$ interval. Because the item-level scale is ordinal, disagreements between two raters are not all equally severe: a \emph{strong} disagreement (a distance of 1 on the scale, e.g., one rater scoring 0 and the other 1) reflects a qualitatively different judgment than a \emph{light} disagreement (a distance of 0.5, e.g., 0 vs.\ 0.5 or 0.5 vs.\ 1). To account for this graded structure, inter-rater agreement was computed using weighted Cohen's $\kappa$, which penalizes large discrepancies more than small ones.

The rubric was applied through multiple independent rating rounds. Each task solution was first evaluated independently by one researcher at ITU and one researcher at Uniba, with a third team member serving as arbiter for residual disagreements. In an initial calibration round, each of the two raters applied the rubric to the code produced by 5 participants per site (20\% of the sample). Weighted Cohen's $\kappa$ values were $\kappa = 0.60$ for Task~A (Connection) and $\kappa = 0.69$ for Task~B (Shopping), indicating substantial agreement; no rubric modifications were deemed necessary, and disagreements were resolved by the arbiter. 
The same two raters then independently annotated the code submitted by 10 additional participants from both sites; inter-rater reliability rose to $\kappa = 0.88$ for Task~A and $\kappa = 0.89$ for Task~B, indicating almost perfect agreement. The few remaining discrepancies were again resolved by the arbiter. Given the consistently high agreement, the code submitted by the remaining participants was annotated independently by a single researcher.

Neither the test-pass rate nor the time-to-completion indicators originally registered were used in the analyses; the rationale for both omissions is detailed in~\ref{appendix:deviations}.}


Finally, \textit{subjective experiences} were captured through two instruments.
The NASA-TLX questionnaire, administered after each task, provided scores (0--20) for mental demand, physical demand, temporal demand, performance, effort, and frustration.
\stageTwo{In addition, during the debriefing phase, participants were briefly interviewed about the difficulty encountered in each task and their perception of the development setup (split-screen layout with AI chat and VS Code).}

\subsection{\rev{Test of Hypothesis and Decision Criteria}}
\label{sec:hypothesis-test}

\stageTwo{For all confirmatory tests, we controlled the false discovery rate using the Benjamini--Hochberg (BH) procedure at level $\alpha = 0.05$. The family of comparisons over which BH was applied differs by hypothesis according to the structure of the analysis (signal family, temporal period, effect type, NASA-TLX dimension) and is specified below for each test.}

\paragraph{H1: Developers will exhibit distinct physiological patterns when programming with vs.\ without AI assistance.}
\stageTwo{
To test \textbf{H1}, we examined whether physiological responses differ between the AI-assisted and non-AI-assisted conditions, while accounting for potential order-related effects.
}

\stageTwo{The registered protocol planned a mixed ANOVA; however, the windowed feature extraction described in Section~\ref{sec:preprocessing} produced a dataset in which each participant contributed hundreds of overlapping 60-second windows. To properly account for this non-independence, we relied on linear mixed-effects models (LMMs), which are well-suited for repeated measures and provide robust estimates in the presence of within-subject correlation~\citep{gueorguieva2004move}.}

\stageTwo{
Before proceeding with the analysis, we verified that the two programming tasks imposed comparable cognitive demands: a paired $t$-test on the NASA-TLX overall workload scores between the Connection and Shopping tasks revealed no significant difference ($t = 0.21$, $p = .83$), confirming that the tasks were perceived as cognitively equivalent by participants.
}

\stageTwo{For each physiological metric (represented below as \texttt{DV}, see Table~\ref{tab:signal-summary}), we fitted a separate LMM with the physiological metric as the dependent variable and the following fixed effects: \texttt{Modality} (AI vs.\ NON-AI, with NON-AI as the reference level), \texttt{Task\_order} (first vs.\ second task, with first as the reference level), and their interaction \texttt{Modality\,$\times$\,Task\_order}. A random intercept was included for each participant to account for between-subject variability in baseline physiological levels:}
\begin{equation}\label{eq:lmm-h1-full}
\begin{aligned}
\texttt{DV} \sim {} &  \texttt{Modality} + {} \texttt{Task\_order} {} + \texttt{Modality}\!\times\!\texttt{Task\_order} + \\ & (1\,|\,\texttt{Participant}).
\end{aligned}
\end{equation}

\stageTwo{To probe potential order-dependent effects of the task, we complemented the full model with two stratified analyses, fitted separately on the data from the first task only and on the data from the second task only. In these stratified models, we retained \texttt{Modality} as the sole fixed effect of interest, with the same random structure:}
\begin{equation}\label{eq:lmm-h1-strat}
\texttt{DV} \sim \texttt{Modality} + (1\,|\,\texttt{Participant}).
\end{equation}

\stageTwo{For H1, BH correction was applied within each combination of signal family (EEG, HRV, EDA, GAZE) and effect type.}

\paragraph{H\stageTwo{2}: Developer experience will moderate the relationship between AI assistance and physiological measures.}

\stageTwo{To test \textbf{H2}, we extended the LMM specification used for H1 by introducing developer experience as a moderator. Specifically, we included the factor \texttt{Education}, operationalized as a binary variable (undergraduate vs.\ graduate, with graduate as the reference level).}

\stageTwo{For each physiological metric, we fitted an LMM that included an additional \texttt{Modality\,$\times$\,Education} interaction term:}
\begin{equation}\label{eq:lmm-h2}
\begin{aligned}
\texttt{DV} \sim {} & \texttt{Modality} + \texttt{Task\_order} + \texttt{Modality}\!\times\!\texttt{Task\_order} \\
& + \texttt{Modality}\!\times\!\texttt{Education} + (1\,|\,\texttt{Participant}).
\end{aligned}
\end{equation}

\stageTwo{This formulation allowed us to test whether the effect of AI assistance on physiological responses differs between undergraduate and graduate students. As in the RQ1 analysis, we complemented the full model with two stratified models fitted on the first-task and second-task subsets, each retaining the \texttt{Modality\,$\times$\,Education} interaction. BH correction was applied within the same family as for H1 (i.e., signal family $\times$ effect type).}

\paragraph{H\stageTwo{3}: Physiological measures will correlate with performance metrics in both conditions.}

\stageTwo{To test \textbf{H3}, we examined whether physiological measures correlate with task performance and whether this relationship differs between the AI-assisted and non-AI-assisted conditions. }

\stageTwo{The performance score---a rubric-based task-correctness measure described in~\ref{appendix:deviations} and bounded in the $[0, 1]$ interval---was discretised into a binary variable via a median split: scores strictly above the sample median were coded as \emph{high} performance, while scores at or below the median were coded as \emph{low} performance. This procedure produced a balanced binary outcome with 30 high-performance and 30 low-performance task-level observations.}

\stageTwo{Because cognitive demands plausibly evolve over the 20 minutes of each task coding session~\citep{sharafi2020practical}, we structured the analysis along three temporal periods rather than using session-averaged features~\citep{bednarik2012expertise}.}

\stageTwo{We defined three 3-minute periods within each task, in line with consolidated practices~\citep{Zueger:interruptability}. Specifically, we selected: (i) the first 3 minutes after an initial 2-minute interval, which we excluded because participants were still reading the task description; (ii) the 3 minutes at the center of the task; and (iii) the final 3 minutes. For each period, the windowed values of the physiological features were averaged into a single value per participant, which served as the input to the regression analysis.}

\stageTwo{For each combination of period (early, mid, late) and physiological metric, we fitted a separate Bayesian binomial mixed-effects model with a logistic link, using the \texttt{BinomialBayesMixedGLM} implementation and variational inference. The fixed-effects structure included \texttt{Modality}, \texttt{Task\_order}, the physiological predictor (\texttt{Physiological}), and the \texttt{Modality}\,$\times$\,\texttt{Physiological} interaction; a participant-level random intercept absorbed the within-subject correlation between the two tasks:}
\begin{equation}\label{eq:glmm-h3-full}
\begin{aligned}
\text{logit}\bigl(P(\texttt{Performance}=\textit{high})\bigr) \sim {} 
& \texttt{Modality}\!\times\!\texttt{Physiological} \\
& {} + \texttt{Task\_order} + (1\,|\,\texttt{Participant}).
\end{aligned}
\end{equation}
\stageTwo{Reference levels were NON-AI for \texttt{Modality}, first task for \texttt{Task\_order}, and \emph{low} for \texttt{Performance}.}

\stageTwo{For H3, BH correction was applied within each combination of temporal period, signal family, and effect type, reflecting the additional partition introduced by the early/mid/late stratification.}

\stageTwo{For each model in which the \texttt{Modality}\,$\times$\,\texttt{Physiological} interaction survived BH correction, we refitted the model separately on the AI and NON-AI subsets in order to recover the condition-specific slope of the physiological predictor:}
\begin{equation}\label{eq:glmm-h3-strat}
\begin{aligned}
\text{logit}\bigl(P(\texttt{Performance}=\textit{high})\bigr) \sim {} & \texttt{Task\_order} {} + \texttt{Physiological}\\
 & + (1\,|\,\texttt{Participant}).
\end{aligned}
\end{equation}
\stageTwo{These condition-specific estimates are reported in the indented rows labeled \emph{Physiological$\mid$AI} and \emph{Physiological$\mid$NON-AI} of Tables~\ref{tab:glmm_early}--\ref{tab:glmm_late}, and provide a direct interpretation of the interaction in terms of two distinct slopes rather than a single average effect modulated by modality.}

\paragraph{H\stageTwo{4}: The alignment between subjective perceptions and objective measures (both physiological and performance-based) will differ between AI-assisted and non-AI-assisted conditions.}

\stageTwo{To test \textbf{H4}, we examined whether the alignment between self-reported workload (NASA-TLX) and objective task performance differs between the AI-assisted and non-AI-assisted conditions.}

\stageTwo{For the NASA-TLX, each of the six sub-dimensions (mental, physical, temporal, performance, effort, frustration) was first $z$-score normalized across all participants to remove differences in scale usage, in line with common practices~\citep{Zueger:interruptability}, and then median-split into a binary variable (\emph{high} vs.\ \emph{low}). The resulting class distributions were:
\emph{mental} (low: 74, high: 46),
\emph{physical} (low: 62, high: 58),
\emph{temporal} (low: 71, high: 49),
\emph{performance} (low: 60, high: 60),
\emph{effort} (low: 66, high: 54), and
\emph{frustration} (low: 63, high: 57). We additionally computed an overall NASA-TLX score by averaging the six sub-dimensions.}

\stageTwo{For each of the six NASA-TLX sub-dimensions, plus the averaged score, we fitted a binomial generalized mixed-effects model with a logistic link, predicting performance from \texttt{Modality}, the median-split NASA-TLX sub-dimension, \texttt{Task\_order}, and the \texttt{Modality\,$\times$\,NASA-TLX-Dimension} interaction. A random intercept per participant captured the within-subject correlation arising from the crossover design:}
\begin{equation}\label{eq:glmm-h4}
\begin{aligned}
\text{logit}\bigl(P(\texttt{Performance}=\text{High})\bigr) \sim {} 
& \texttt{Modality}\!\times\!\texttt{NASA-TLX-Dimension} \\
& {} + \texttt{Task\_order} + (1\,|\,\texttt{Participant}).
\end{aligned}
\end{equation}
\stageTwo{Reference levels were set to NON-AI for \texttt{Modality}, \emph{low} for \texttt{NASA-TLX-Dimension}, and the first task for \texttt{Task\_order}.}

\stageTwo{We estimated the models using the \texttt{BinomialBayesMixedGLM} implementation in \texttt{statsmodels}, which fits a Bayesian mixed-effects logistic regression via variational inference. When the interaction between modality and the NASA-TLX dimension reached significance, we refitted the model separately within each level of \texttt{Modality} (AI and NON-AI), as done for H2. These condition-specific estimates are reported in the indented rows of Table~\ref{tab:glmm_workload_interaction}.}

\stageTwo{For H4, BH correction was applied within each effect type across the seven NASA-TLX models (six sub-dimensions plus the averaged score).}

\section{Results}
\label{sec:results}


\subsection{H1: Physiological patterns differ between AI- and non-AI-assisted conditions}
\label{sec:res:h1}

\stageTwo{
Among all the physiological metrics considered, only the \texttt{$\theta/\alpha$ ratio} exhibits significant differences after  correction (Table~\ref{tab:lmm-full-combined}). Specifically, we observe a significant difference between modalities ($\beta = -1.11$, $p_{\text{BH}} = .01$), with a lower \texttt{$\theta/\alpha$ ratio} in the AI condition compared to the NON-AI condition. Differences in \texttt{$\theta/\alpha$ ratio} were also significant with respect to task order ($\beta = -1.03$, $p_{\text{BH}} = .02$), with lower values observed during the second task. The interaction \texttt{Modality} $\times$ \texttt{Task\_order} does not reach significance.
No significant difference emerges for any EDA, HRV, or gaze-based metric.
}

\begin{table}[t]
\centering
\caption{Fixed-effect estimates for H1 (full sample); model in Eq.~\ref{eq:lmm-h1-full}.}
\label{tab:lmm-full-combined}
\resizebox{\textwidth}{!}{%
\begin{threeparttable}
\footnotesize
\setlength{\tabcolsep}{4pt}
\renewcommand{\arraystretch}{0.92}
\begin{tabular}{ll l rrrrr r}
\toprule
\textbf{Family} & \textbf{Metric} & \textbf{Effect} & \textbf{Coef.} & \textbf{SE} & \textbf{$z$} & \textbf{$p$} & \textbf{$p_{\text{BH}}$} & \textbf{$R^2_m / R^2_c$} \\
\midrule
\multirow{9}{*}{\shortstack[l]{EDA\\\scriptsize($N{=}34$\\\scriptsize UNIBA)}}
  & \multirow{3}{*}{Tonic Mean}
    & Modality (AI)                       &     0.09 &  4.34 &  0.02 & .98 & .98 & \multirow{3}{*}{.03\,/\,.42} \\
  & & Task\_order (2)                    &     1.83 &  4.35 &  0.42 & .67 & .88 \\
  & & Modality (AI) $\times$ Task\_order (2)    &     6.21 &  8.65 &  0.72 & .47 & .71 \\
  \cmidrule{2-9}
  & \multirow{3}{*}{SCR Amplitude}
    & Modality (AI)                       &     3.21 &  5.76 &  0.56 & .58 & .87 & \multirow{3}{*}{.03\,/\,.60} \\
  & & Task\_order (2)                    &  $-$0.85 &  5.76 & $-$0.15 & .88 & .88 \\
  & & Modality (AI) $\times$ Task\_order (2)    &  $-$9.43 & 11.50 & $-$0.82 & .41 & .71 \\
  \cmidrule{2-9}
  & \multirow{3}{*}{SCR Peaks}
    & Modality (AI)                       &  $-$2.48 &  3.46 & $-$0.72 & .47 & .87 & \multirow{3}{*}{.05\,/\,.71} \\
  & & Task\_order (2)                    &  $-$4.87 &  3.46 & $-$1.41 & .16 & .48 \\
  & & Modality (AI) $\times$ Task\_order (2)    &  $-$0.46 &  6.91 & $-$0.07 & .95 & .95 \\
\midrule
\multirow{10}{*}{\shortstack[l]{HRV\\\scriptsize($N{=}34$\\\scriptsize UNIBA)}}
  & \multirow{3}{*}{RMSSD}
    & Modality (AI)                       &    16.12 & 11.51 &  1.40 & .16 & .32 & \multirow{3}{*}{.09\,/\,.49} \\
  & & Task\_order (2)                    & $-$23.71 & 11.51 & $-$2.06 & .04 & .08 \\
  & & Modality (AI) $\times$ Task\_order (2)    &  $-$9.87 & 22.93 & $-$0.43 & .67 & .67 \\
  \cmidrule{2-9}
  & \multirow{3}{*}{LF/HF Ratio}
    & Modality (AI)                       &  $-$0.03 &  0.17 & $-$0.19 & .85 & .85 & \multirow{3}{*}{$<$.01\,/\,.10} \\
  & & Task\_order (2)                    &     0.02 &  0.17 &  0.14 & .89 & .89 \\
  & & Modality (AI) $\times$ Task\_order (2)    &     0.18 &  0.33 &  0.56 & .58 & .67 \\
\midrule
\multirow{6}{*}{\shortstack[l]{EEG\\\scriptsize($N{=}60$\\\scriptsize UNIBA+ITU)}}
  & \multirow{3}{*}{Beta Power}
    & Modality (AI)                       &     3.04 &  2.15 &  1.41 & .16 & .16 & \multirow{3}{*}{.01\,/\,.65} \\
  & & Task\_order (2)                    &     2.29 &  2.15 &  1.06 & .29 & .29 \\
  & & Modality (AI) $\times$ Task\_order (2)    &  $-$4.94 &  4.29 & $-$1.15 & .25 & .25 \\
  \cmidrule{2-9}
  & \multirow{3}{*}{\textbf{$\theta/\alpha$ Ratio}}
    & \textbf{Modality}              &  \textbf{$-$1.11} &  \textbf{0.40} & \textbf{$-$2.78} & \textbf{$<$.01} & \textbf{.01}$^*$ & \multirow{3}{*}{\textbf{.02\,/\,.24}} \\
  & & \textbf{Task\_order}           &  \textbf{$-$1.03} &  \textbf{0.40} & \textbf{$-$2.59} & \textbf{$<$.01} & \textbf{.02}$^*$ \\
  & & Modality (AI) $\times$ Task\_order (2)    &     1.27 &  0.79 &  1.61 & .11 & .22 \\
\midrule
\multirow{9}{*}{\shortstack[l]{Gaze\\\scriptsize($N{=}60$\\\scriptsize UNIBA+ITU)}}
  & \multirow{3}{*}{Fixation Count}
    & Modality (AI)                       &     4.93 &  9.10 &  0.54 & .59 & .94 & \multirow{3}{*}{.02\,/\,.69} \\
  & & Task\_order (2)                    &    11.60 &  9.10 &  1.27 & .20 & .30 \\
  & & Modality (AI) $\times$ Task\_order (2)    & $-$18.71 & 18.17 & $-$1.03 & .30 & .45 \\
  \cmidrule{2-9}
  & \multirow{3}{*}{Mean Fix.\ Duration}
    & Modality (AI)                       &     0.00 &  0.01 &  0.35 & .73 & .94 & \multirow{3}{*}{$<$.01\,/\,.55} \\
  & & Task\_order (2)                    &     0.00 &  0.01 &  0.03 & .97 & .97 \\
  & & Modality (AI) $\times$ Task\_order (2)    &     0.00 &  0.03 & $-$0.10 & .92 & .92 \\
  \cmidrule{2-9}
  & \multirow{3}{*}{Blink Rate}
    & Modality (AI)                       &     0.04 &  0.48 &  0.07 & .94 & .94 & \multirow{3}{*}{.04\,/\,.40} \\
  & & Task\_order (2)                    &  $-$0.81 &  0.48 & $-$1.68 & .09 & .28 \\
  & & Modality (AI) $\times$ Task\_order (2)    &     1.68 &  0.96 &  1.76 & .08 & .24 \\
\bottomrule
\end{tabular}
\begin{tablenotes}\footnotesize
\item[$^*$] Statistically significant after BH correction at $\alpha = 0.05$.
\item Reference levels: Modality = NON-AI; Task\_order = first.
\end{tablenotes}
\end{threeparttable}
}
\end{table}

\stageTwo{
When restricting the analysis to the first task only (see Table~\ref{tab:lmm-task1}), the \texttt{$\theta/\alpha$ ratio} shows a significant difference after correction ($\beta = -1.16$, $p_{\text{BH}} = .04$) with respect to modality, again indicating a lower ratio in the AI condition relative to NON-AI.
None of the other physiological metrics reach significance.}

\begin{table}[t]
\centering
\caption{Fixed-effect estimates for H1 (first task only); model in Eq.~\ref{eq:lmm-h1-strat}.}
\label{tab:lmm-task1}
\resizebox{\textwidth}{!}{%
\begin{threeparttable}
\begin{tabular}{ll rrrrr r}
\toprule
\textbf{Family} & \textbf{Metric} & \textbf{Coef.} & \textbf{SE} & \textbf{$z$} & \textbf{$p$} & \textbf{$p_{\text{BH}}$} & \textbf{$R^2_m / R^2_c$} \\
\midrule
\multirow{3}{*}{\shortstack[l]{EDA\\\scriptsize($N{=}34$\\\scriptsize UNIBA)}}
  & Tonic Mean        &     0.17 &  4.31 &  0.04 & .97 & .97 & $<$.01\,/\,.77 \\
  \cmidrule{2-8}
  & SCR Amplitude     &     1.01 &  4.47 &  0.23 & .82 & .97 & $<$.01\,/\,.80 \\
  \cmidrule{2-8}
  & SCR Peaks         &  $-$2.39 &  3.69 & $-$0.65 & .52 & .97 & .01\,/\,.86 \\
\midrule
\multirow{2}{*}{\shortstack[l]{HRV\\\scriptsize($N{=}34$\\\scriptsize UNIBA)}}
  & RMSSD             &    15.75 & 14.64 &  1.08 & .28 & .56 & .02\,/\,.72 \\
  \cmidrule{2-8}
  & LF/HF Ratio       &  $-$0.04 &  0.17 & $-$0.23 & .82 & .82 & $<$.01\,/\,.11 \\
\midrule
\multirow{2}{*}{\shortstack[l]{EEG\\\scriptsize($N{=}60$\\\scriptsize UNIBA+ITU)}}
  & Beta Power        &     3.19 &  2.21 &  1.44 & .15 & .15 & .03\,/\,.79 \\
  \cmidrule{2-8}
  & \textbf{$\theta/\alpha$ Ratio}
                      & \textbf{$-$1.16} & \textbf{0.49} & \textbf{$-$2.36} & \textbf{.02} & \textbf{.04}$^*$ & \textbf{.04\,/\,.42} \\
\midrule
\multirow{3}{*}{\shortstack[l]{Gaze\\\scriptsize($N{=}60$\\\scriptsize UNIBA+ITU)}}
  & Fixation Count    &     4.50 &  9.75 &  0.46 & .64 & .87 & $<$.01\,/\,.76 \\
  \cmidrule{2-8}
  & Mean Fix.\ Duration &   0.00 &  0.01 &  0.31 & .75 & .87 & $<$.01\,/\,.64 \\
  \cmidrule{2-8}
  & Blink Rate        &     0.09 &  0.51 &  0.17 & .87 & .87 & $<$.01\,/\,.46 \\
\bottomrule
\end{tabular}
\begin{tablenotes}\footnotesize
\item[$^*$] Statistically significant after BH correction at $\alpha = 0.05$.
\end{tablenotes}
\end{threeparttable}
}
\end{table}

\stageTwo{
Regarding the second task (see Table~\ref{tab:lmm-task2}), the \texttt{$\theta/\alpha$} ratio no longer shows a significant difference between conditions. Instead, the \texttt{blink rate} emerges as the only metric with a significant modality effect ($\beta = 1.76$, $p_{\text{BH}} < .01$), with higher values observed in the AI condition compared to NON-AI. This pattern is consistent with the inverse blink-rate/cognitive-load relationship documented in reading paradigms~\citep{lenskiy2016blink,rosenfield2015cognitive}: greater cognitive demand suppresses blinking to sustain visual intake, so the lower blink rate observed under NON-AI is interpretable as a higher cognitive load when the developer must construct the solution unaided. The modality distinction observed on the first task therefore re-emerges on the second through a different oculomotor channel rather than dissolving.
}

\begin{table}[b]
\centering
\caption{Fixed-effect estimates for H1 (second task only); model in Eq.~\ref{eq:lmm-h1-strat}.}
\label{tab:lmm-task2}
\resizebox{\textwidth}{!}{%
\begin{threeparttable}
\begin{tabular}{ll rrrrr r}
\toprule
\textbf{Family} & \textbf{Metric} & \textbf{Coef.} & \textbf{SE} & \textbf{$z$} & \textbf{$p$} & \textbf{$p_{\text{BH}}$} & \textbf{$R^2_m / R^2_c$} \\
\midrule
\multirow{3}{*}{\shortstack[l]{EDA\\\scriptsize($N{=}34$\\\scriptsize UNIBA)}}
  & Tonic Mean        &     6.20 &  7.31 &  0.85 & .40 & .41 & .02\,/\,.79 \\
  \cmidrule{2-8}
  & SCR Amplitude     &  $-$6.19 &  7.48 & $-$0.83 & .41 & .41 & .02\,/\,.79 \\
  \cmidrule{2-8}
  & SCR Peaks         &  $-$3.40 &  3.78 & $-$0.90 & .37 & .41 & .02\,/\,.81 \\
\midrule
\multirow{2}{*}{\shortstack[l]{HRV\\\scriptsize($N{=}34$\\\scriptsize UNIBA)}}
  & RMSSD             &     4.26 & 15.35 &  0.28 & .78 & .78 & $<$.01\,/\,.78 \\
  \cmidrule{2-8}
  & LF/HF Ratio       &     0.17 &  0.20 &  0.83 & .41 & .78 & $<$.01\,/\,.13 \\
\midrule
\multirow{2}{*}{\shortstack[l]{EEG\\\scriptsize($N{=}60$\\\scriptsize UNIBA+ITU)}}
  & Beta Power        &  $-$1.96 &  2.55 & $-$0.77 & .44 & .79 & .01\,/\,.80 \\
  \cmidrule{2-8}
  & $\theta/\alpha$ Ratio
                      &     0.15 &  0.55 &  0.27 & .79 & .79 & $<$.01\,/\,.39 \\
\midrule
\multirow{3}{*}{\shortstack[l]{Gaze\\\scriptsize($N{=}60$\\\scriptsize UNIBA+ITU)}}
  & Fixation Count    & $-$13.99 &  9.37 & $-$1.49 & .14 & .20 & .03\,/\,.76 \\
  \cmidrule{2-8}
  & Mean Fix.\ Duration &   0.00 &  0.01 &  0.14 & .89 & .89 & $<$.01\,/\,.66 \\
  \cmidrule{2-8}
  & \textbf{Blink Rate}
                      & \textbf{1.76} & \textbf{0.56} & \textbf{3.17} & \textbf{$<$.01} & \textbf{$<$.01}$^*$ & \textbf{.07\,/\,.50} \\
\bottomrule
\end{tabular}
\begin{tablenotes}\footnotesize
\item[$^*$] Statistically significant after BH correction at $\alpha = 0.05$.
\end{tablenotes}
\end{threeparttable}
}
\end{table}

\stageTwo{Across the H1 models (Tables~\ref{tab:lmm-full-combined}--\ref{tab:lmm-task2}), marginal $R^2$ values are low (typically below $0.10$): the fixed effects of \texttt{Modality}, \texttt{Task\_order}, and their interaction jointly explain only a small share of the variance in the physiological signals. Conditional $R^2$ values are markedly higher (often $0.40$--$0.80$), because most of the variance is between-participant and is absorbed by the random intercept. Although the differences between modalities reported above reach statistical significance, they explain only a small share of the variability, most of which is between participants.}

\begin{takeaway}
\textbf{H1: Partially supported.} Of all physiological metrics tested, only the $\theta/\alpha$ ratio differs significantly between conditions, with lower values under AI than under NON-AI assistance, consistent with reduced working-memory engagement when the model takes on part of the generative effort. This difference is robust on the first task; on the second task, the gaze blink rate emerges as the differentiating metric, with higher values under AI than NON-AI, consistent with the inverse blink-rate/cognitive-load relationship documented in reading paradigms (Table~\ref{tab:signal-summary}) and pointing to the same interpretation: a higher cognitive load under NON-AI. EDA, HRV, and the remaining gaze metrics show no significant difference between modalities.
\end{takeaway}

\subsection{H\stageTwo{2}: Developer experience moderates the AI--physiology relationship}
\label{sec:res:h2}

\stageTwo{
The extended model with \texttt{Education} as a moderator confirms the H1 finding: only the \texttt{$\theta/\alpha$ ratio} shows significant effects after BH correction. As shown in Table~\ref{tab:lmm-full-education}, the $\theta/\alpha$ ratio remains significantly lower with respect to modality ($\beta = -1.16$, $p_{\text{BH}} < .01$) and task order ($\beta = -1.03$, $p_{\text{BH}} = .02$). None of the \texttt{Modality} $\times$ \texttt{Education} interaction terms reach significance for any of the physiological metrics considered. 
}

\begin{table}[t]
\centering
\caption{Fixed-effect estimates for H2 (full sample), with Modality$\times$Education interaction; model in Eq.~\ref{eq:lmm-h2}.}
\label{tab:lmm-full-education}
\resizebox{\textwidth}{!}{%
\begin{threeparttable}
\footnotesize
\setlength{\tabcolsep}{4pt}
\renewcommand{\arraystretch}{0.92}
\begin{tabular}{ll l rrrrr r}
\toprule
\textbf{Family} & \textbf{Metric} & \textbf{Effect} & \textbf{Coef.} & \textbf{SE} & \textbf{$z$} & \textbf{$p$} & \textbf{$p_{\text{BH}}$} & \textbf{$R^2_m / R^2_c$} \\
\midrule
\multirow{15}{*}{\shortstack[l]{EDA\\\scriptsize($N{=}34$\\\scriptsize UNIBA)}}
  & \multirow{5}{*}{Tonic Mean}
    & Modality (AI)                                  &     5.99 &  4.36 &  1.37 & .17 & .51 & \multirow{5}{*}{.05\,/\,.44} \\
  & & Task\_order (2)                               &     2.10 &  4.35 &  0.48 & .63 & .83 \\
  & & Modality (AI) $\times$ Task\_order (2)               &     6.06 &  8.66 &  0.70 & .48 & .73 \\
  & & Mod$\times$Edu (AI:grad)                  &  $-$2.07 &  4.35 & $-$0.48 & .63 & .76 \\
  & & Mod$\times$Edu (NON-AI:grad)              &     9.26 &  4.35 &  2.13 & .03 & .20 \\
  \cmidrule{2-9}
  & \multirow{5}{*}{SCR Amplitude}
    & Modality (AI)                                  &  $-$1.35 &  5.77 & $-$0.23 & .82 & .82 & \multirow{5}{*}{.04\,/\,.61} \\
  & & Task\_order (2)                               &  $-$1.25 &  5.76 & $-$0.22 & .83 & .83 \\
  & & Modality (AI) $\times$ Task\_order (2)               &  $-$8.80 & 11.49 & $-$0.77 & .44 & .73 \\
  & & Mod$\times$Edu (AI:grad)                  &  $-$0.02 &  5.76 & $-$0.00 & $>$.99 & $>$.99 \\
  & & Mod$\times$Edu (NON-AI:grad)              &  $-$8.42 &  5.76 & $-$1.46 & .14 & .43 \\
  \cmidrule{2-9}
  & \multirow{5}{*}{SCR Peaks}
    & Modality (AI)                                  &  $-$2.58 &  3.51 & $-$0.74 & .46 & .69 & \multirow{5}{*}{.06\,/\,.72} \\
  & & Task\_order (2)                               &  $-$4.98 &  3.51 & $-$1.42 & .16 & .47 \\
  & & Modality (AI) $\times$ Task\_order (2)               &  $-$0.25 &  7.00 & $-$0.04 & .97 & .97 \\
  & & Mod$\times$Edu (AI:grad)                  &  $-$1.80 &  3.51 & $-$0.51 & .61 & .76 \\
  & & Mod$\times$Edu (NON-AI:grad)              &  $-$1.79 &  3.51 & $-$0.51 & .61 & .76 \\
\midrule
\multirow{10}{*}{\shortstack[l]{HRV\\\scriptsize($N{=}34$\\\scriptsize UNIBA)}}
  & \multirow{5}{*}{RMSSD}
    & Modality (AI)                                  &    24.85 & 11.62 &  2.14 & .03 & .06 & \multirow{5}{*}{.10\,/\,.51} \\
  & & Task\_order (2)                               & $-$24.15 & 11.58 & $-$2.08 & .04 & .07 \\
  & & Modality (AI) $\times$ Task\_order (2)               &  $-$8.29 & 23.09 & $-$0.36 & .72 & .72 \\
  & & Mod$\times$Edu (AI:grad)                  & $-$19.04 & 11.58 & $-$1.64 & .10 & .40 \\
  & & Mod$\times$Edu (NON-AI:grad)              &  $-$0.22 & 11.58 & $-$0.02 & .98 & .98 \\
  \cmidrule{2-9}
  & \multirow{5}{*}{LF/HF Ratio}
    & Modality (AI)                                  &  $-$0.10 &  0.17 & $-$0.60 & .55 & .55 & \multirow{5}{*}{$<$.01\,/\,.10} \\
  & & Task\_order (2)                               &     0.02 &  0.17 &  0.10 & .92 & .92 \\
  & & Modality (AI) $\times$ Task\_order (2)               &     0.19 &  0.33 &  0.58 & .56 & .72 \\
  & & Mod$\times$Edu (AI:grad)                  &  $-$0.04 &  0.17 & $-$0.25 & .80 & .98 \\
  & & Mod$\times$Edu (NON-AI:grad)              &  $-$0.17 &  0.17 & $-$1.02 & .31 & .62 \\
\midrule
\multirow{10}{*}{\shortstack[l]{EEG\\\scriptsize($N{=}60$\\\scriptsize UNIBA+ITU)}}
  & \multirow{5}{*}{Beta Power}
    & Modality (AI)                                  &     2.11 &  2.18 &  0.97 & .33 & .33 & \multirow{5}{*}{.02\,/\,.65} \\
  & & Task\_order (2)                               &     2.12 &  2.17 &  0.98 & .33 & .33 \\
  & & Modality (AI) $\times$ Task\_order (2)               &  $-$4.74 &  4.34 & $-$1.09 & .27 & .27 \\
  & & Mod$\times$Edu (AI:grad)                  &  $-$0.34 &  2.18 & $-$0.16 & .87 & $>$.99 \\
  & & Mod$\times$Edu (NON-AI:grad)              &  $-$1.88 &  2.18 & $-$0.86 & .39 & $>$.99 \\
  \cmidrule{2-9}
  & \multirow{5}{*}{\textbf{$\theta/\alpha$ Ratio}}
    & \textbf{Modality}                         & \textbf{$-$1.16} & \textbf{0.41} & \textbf{$-$2.84} & \textbf{$<$.01} & \textbf{$<$.01}$^*$ & \multirow{5}{*}{\textbf{.02\,/\,.24}} \\
  & & \textbf{Task\_order}                      & \textbf{$-$1.03} & \textbf{0.41} & \textbf{$-$2.55} & \textbf{.01} & \textbf{.02}$^*$ \\
  & & Modality (AI) $\times$ Task\_order (2)               &     1.26 &  0.80 &  1.58 & .12 & .23 \\
  & & Mod$\times$Edu (AI:grad)                  &     0.10 &  0.41 &  0.26 & .80 & $>$.99 \\
  & & Mod$\times$Edu (NON-AI:grad)              &     0.00 &  0.41 & $-$0.01 & $>$.99 & $>$.99 \\
\midrule
\multirow{15}{*}{\shortstack[l]{Gaze\\\scriptsize($N{=}60$\\\scriptsize UNIBA+ITU)}}
  & \multirow{5}{*}{Fixation Count}
    & Modality (AI)                                  &     5.57 &  9.19 &  0.61 & .54 & .90 & \multirow{5}{*}{.02\,/\,.69} \\
  & & Task\_order (2)                               &    11.04 &  9.18 &  1.20 & .23 & .34 \\
  & & Modality (AI) $\times$ Task\_order (2)               & $-$17.33 & 18.33 & $-$0.95 & .34 & .52 \\
  & & Mod$\times$Edu (AI:grad)                  &  $-$8.24 &  9.19 & $-$0.90 & .37 & .63 \\
  & & Mod$\times$Edu (NON-AI:grad)              &  $-$5.69 &  9.19 & $-$0.62 & .54 & .63 \\
  \cmidrule{2-9}
  & \multirow{5}{*}{Mean Fix.\ Duration}
    & Modality (AI)                                  &     0.00 &  0.01 &  0.27 & .79 & .90 & \multirow{5}{*}{$<$.01\,/\,.55} \\
  & & Task\_order (2)                               &     0.00 &  0.01 &  0.08 & .94 & .94 \\
  & & Modality (AI) $\times$ Task\_order (2)               &     0.00 &  0.03 & $-$0.16 & .87 & .87 \\
  & & Mod$\times$Edu (AI:grad)                  &     0.01 &  0.01 &  0.76 & .45 & .63 \\
  & & Mod$\times$Edu (NON-AI:grad)              &     0.01 &  0.01 &  0.49 & .63 & .63 \\
  \cmidrule{2-9}
  & \multirow{5}{*}{Blink Rate}
    & Modality (AI)                                  &     0.06 &  0.48 &  0.13 & .90 & .90 & \multirow{5}{*}{.05\,/\,.40} \\
  & & Task\_order (2)                               &  $-$0.75 &  0.48 & $-$1.57 & .12 & .34 \\
  & & Modality (AI) $\times$ Task\_order (2)               &     1.56 &  0.96 &  1.63 & .10 & .31 \\
  & & Mod$\times$Edu (AI:grad)                  &     0.63 &  0.48 &  1.31 & .19 & .63 \\
  & & Mod$\times$Edu (NON-AI:grad)              &     0.57 &  0.48 &  1.18 & .24 & .63 \\
\bottomrule
\end{tabular}
\begin{tablenotes}\footnotesize
\item[$^*$] BH-FDR correction applied within each signal family (EDA, HRV, EEG, Gaze) and effect type (Modality, Task\_order, Modality$\times$Task\_order, Mod$\times$Edu).
\item Reference levels: Modality = NON-AI; Task\_order = first; Education = undergrad.
\item Mod$\times$Edu (AI:grad) and Mod$\times$Edu (NON-AI:grad) denote the effect of graduate (vs.\ undergrad) education within the AI and NON-AI modality, respectively.
\end{tablenotes}
\end{threeparttable}
}
\vspace{-5mm}
\end{table}

\stageTwo{
When restricting the analysis to the first task (see Table~\ref{tab:lmm-task1-education}), the \texttt{$\theta/\alpha$ ratio} retains significance after BH correction with respect to modality ($\beta = -2.30$, $p_{\text{BH}} < .01$). No \texttt{Modality} $\times$ \texttt{Education} interaction reaches significance for any metric, providing no evidence of a moderating role of developer experience on the first task.
}

\stageTwo{
Regarding the second task (see Table~\ref{tab:lmm-task2-education}), no variables reaches statistical significance after correction, neither \texttt{Modality} nor the \texttt{Modality} $\times$ \texttt{Education} interaction terms, providing no evidence of modality differences or of moderation by developer experience on the second task.}

\stageTwo{The same $R^2$ pattern persists once developer experience is included (Tables~\ref{tab:lmm-full-education}--\ref{tab:lmm-task2-education}): marginal $R^2$ remains low and conditional $R^2$ high, with most of the variance again attributable to between-participant differences.}

\begin{table}[tb]
\centering
\caption{Fixed-effect estimates for H2 (first task only); stratified version of Eq.~\ref{eq:lmm-h2}, with \texttt{Task\_order} terms omitted.}
\label{tab:lmm-task1-education}
\resizebox{\textwidth}{!}{%
\begin{threeparttable}
\footnotesize
\setlength{\tabcolsep}{4pt}
\renewcommand{\arraystretch}{0.92}
\begin{tabular}{ll l rrrrr r}
\toprule
\textbf{Family} & \textbf{Metric} & \textbf{Effect} & \textbf{Coef.} & \textbf{SE} & \textbf{$z$} & \textbf{$p$} & \textbf{$p_{\text{BH}}$} & \textbf{$R^2_m / R^2_c$} \\
\midrule
\multirow{9}{*}{\shortstack[l]{EDA\\\scriptsize($N{=}34$\\\scriptsize UNIBA)}}
  & \multirow{3}{*}{Tonic Mean}
    & Modality (AI)                       &     6.15 &  6.10 &  1.01 & .31 & .58 & \multirow{3}{*}{.05\,/\,.78} \\
  & & Mod$\times$Edu (AI:grad)       &  $-$7.72 &  6.10 & $-$1.27 & .21 & .64 \\
  & & Mod$\times$Edu (NON-AI:grad)   &     4.43 &  6.10 &  0.73 & .47 & .70 \\
  \cmidrule{2-9}
  & \multirow{3}{*}{SCR Amplitude}
    & Modality (AI)                       &  $-$5.47 &  6.31 & $-$0.87 & .39 & .58 & \multirow{3}{*}{.05\,/\,.80} \\
  & & Mod$\times$Edu (AI:grad)       &     7.87 &  6.31 &  1.25 & .21 & .64 \\
  & & Mod$\times$Edu (NON-AI:grad)   &  $-$5.24 &  6.31 & $-$0.83 & .41 & .70 \\
  \cmidrule{2-9}
  & \multirow{3}{*}{SCR Peaks}
    & Modality (AI)                       &  $-$0.87 &  5.39 & $-$0.16 & .87 & .87 & \multirow{3}{*}{.01\,/\,.87} \\
  & & Mod$\times$Edu (AI:grad)       &  $-$0.75 &  5.39 & $-$0.14 & .89 & .89 \\
  & & Mod$\times$Edu (NON-AI:grad)   &     2.21 &  5.39 &  0.41 & .68 & .82 \\
\midrule
\multirow{10}{*}{\shortstack[l]{HRV\\\scriptsize($N{=}34$\\\scriptsize UNIBA)}}
  & \multirow{3}{*}{RMSSD}
    & Modality (AI)                       &    24.01 & 20.71 &  1.16 & .25 & .47 & \multirow{3}{*}{.06\,/\,.73} \\
  & & Mod$\times$Edu (AI:grad)       & $-$28.28 & 20.71 & $-$1.37 & .17 & .61 \\
  & & Mod$\times$Edu (NON-AI:grad)   &  $-$9.53 & 20.71 & $-$0.46 & .65 & .86 \\
  \cmidrule{2-9}
  & \multirow{3}{*}{LF/HF Ratio}
    & Modality (AI)                       &  $-$0.18 &  0.25 & $-$0.72 & .47 & .47 & \multirow{3}{*}{$<$.01\,/\,.12} \\
  & & Mod$\times$Edu (AI:grad)       &     0.01 &  0.25 &  0.03 & .97 & .97 \\
  & & Mod$\times$Edu (NON-AI:grad)   &  $-$0.25 &  0.25 & $-$1.03 & .30 & .61 \\
\midrule
\multirow{6}{*}{\shortstack[l]{EEG\\\scriptsize($N{=}60$\\\scriptsize UNIBA+ITU)}}
  & \multirow{3}{*}{Beta Power}
    & Modality (AI)                       &     3.00 &  3.27 &  0.92 & .36 & .36 & \multirow{3}{*}{.04\,/\,.80} \\
  & & Mod$\times$Edu (AI:grad)       &  $-$2.21 &  3.21 & $-$0.69 & .49 & .49 \\
  & & Mod$\times$Edu (NON-AI:grad)   &  $-$2.15 &  3.14 & $-$0.69 & .49 & .49 \\
  \cmidrule{2-9}
  & \multirow{3}{*}{\textbf{$\theta/\alpha$ Ratio}}
    & \textbf{Modality}              & \textbf{$-$2.30} & \textbf{0.71} & \textbf{$-$3.26} & \textbf{$<$.01} & \textbf{$<$.01}$^*$ & \multirow{3}{*}{\textbf{.07\,/\,.42}} \\
  & & Mod$\times$Edu (AI:grad)       &     1.15 &  0.69 &  1.66 & .10 & .28 \\
  & & Mod$\times$Edu (NON-AI:grad)   &  $-$1.00 &  0.68 & $-$1.47 & .14 & .28 \\
\midrule
\multirow{9}{*}{\shortstack[l]{Gaze\\\scriptsize($N{=}60$\\\scriptsize UNIBA+ITU)}}
  & \multirow{3}{*}{Fixation Count}
    & Modality (AI)                       &  $-$9.79 & 14.31 & $-$0.68 & .49 & .72 & \multirow{3}{*}{.03\,/\,.77} \\
  & & Mod$\times$Edu (AI:grad)       &    10.66 & 14.04 &  0.76 & .45 & .70 \\
  & & Mod$\times$Edu (NON-AI:grad)   & $-$15.75 & 13.75 & $-$1.15 & .25 & .70 \\
  \cmidrule{2-9}
  & \multirow{3}{*}{Mean Fix.\ Duration}
    & Modality (AI)                       &     0.01 &  0.02 &  0.72 & .47 & .72 & \multirow{3}{*}{.01\,/\,.64} \\
  & & Mod$\times$Edu (AI:grad)       &  $-$0.01 &  0.02 & $-$0.62 & .53 & .70 \\
  & & Mod$\times$Edu (NON-AI:grad)   &     0.01 &  0.02 &  0.38 & .70 & .70 \\
  \cmidrule{2-9}
  & \multirow{3}{*}{Blink Rate}
    & Modality (AI)                       &     0.27 &  0.76 &  0.35 & .72 & .72 & \multirow{3}{*}{.01\,/\,.47} \\
  & & Mod$\times$Edu (AI:grad)       &     0.33 &  0.75 &  0.45 & .65 & .70 \\
  & & Mod$\times$Edu (NON-AI:grad)   &     0.59 &  0.73 &  0.81 & .42 & .70 \\
\bottomrule
\end{tabular}
\begin{tablenotes}\footnotesize
\item[$^*$] BH-FDR correction applied within each signal family (EDA, HRV, EEG, Gaze) and effect type (Modality, Mod$\times$Edu)
\item Reference levels: Modality = NON-AI; Education = undergraduate
\item Mod$\times$Edu (AI:grad) and Mod$\times$Edu (NON-AI:grad) denote the effect of graduate (vs.\ undergraduate) education within the AI and NON-AI modality, respectively. 
\end{tablenotes}
\end{threeparttable}
\vspace{-5mm}
}
\end{table}

\begin{table}[tb]
\centering
\caption{Fixed-effect estimates for H2 (second task only); stratified version of Eq.~\ref{eq:lmm-h2}, with \texttt{Task\_order} terms omitted.}
\label{tab:lmm-task2-education}
\resizebox{\textwidth}{!}{%
\begin{threeparttable}
\footnotesize
\setlength{\tabcolsep}{4pt}
\renewcommand{\arraystretch}{0.92}
\begin{tabular}{ll l rrrrr r}
\toprule
\textbf{Family} & \textbf{Metric} & \textbf{Effect} & \textbf{Coef.} & \textbf{SE} & \textbf{$z$} & \textbf{$p$} & \textbf{$p_{\text{BH}}$} & \textbf{$R^2_m / R^2_c$} \\
\midrule
\multirow{9}{*}{\shortstack[l]{EDA\\\scriptsize($N{=}34$\\\scriptsize UNIBA)}}
  & \multirow{3}{*}{Tonic Mean}
    & Modality (AI)                       &    10.26 & 10.38 &  0.99 & .32 & .36 & \multirow{3}{*}{.05\,/\,.80} \\
  & & Mod$\times$Edu (AI:grad)       &     4.22 & 10.38 &  0.41 & .68 & .70 \\
  & & Mod$\times$Edu (NON-AI:grad)   &    13.38 & 10.38 &  1.29 & .20 & .51 \\
  \cmidrule{2-9}
  & \multirow{3}{*}{SCR Amplitude}
    & Modality (AI)                       &  $-$9.72 & 10.69 & $-$0.91 & .36 & .36 & \multirow{3}{*}{.05\,/\,.80} \\
  & & Mod$\times$Edu (AI:grad)       &  $-$4.18 & 10.69 & $-$0.39 & .70 & .70 \\
  & & Mod$\times$Edu (NON-AI:grad)   & $-$12.20 & 10.69 & $-$1.14 & .25 & .51 \\
  \cmidrule{2-9}
  & \multirow{3}{*}{SCR Peaks}
    & Modality (AI)                       &  $-$5.27 &  5.36 & $-$0.98 & .33 & .36 & \multirow{3}{*}{.06\,/\,.82} \\
  & & Mod$\times$Edu (AI:grad)       &  $-$2.76 &  5.36 & $-$0.51 & .61 & .70 \\
  & & Mod$\times$Edu (NON-AI:grad)   &  $-$7.08 &  5.36 & $-$1.32 & .19 & .51 \\
\midrule
\multirow{6}{*}{\shortstack[l]{HRV\\\scriptsize($N{=}34$\\\scriptsize UNIBA)}}
  & \multirow{3}{*}{RMSSD}
    & Modality (AI)                       &    10.95 & 22.40 &  0.49 & .63 & .63 & \multirow{3}{*}{.01\,/\,.79} \\
  & & Mod$\times$Edu (AI:grad)       &  $-$9.36 & 22.39 & $-$0.42 & .68 & .87 \\
  & & Mod$\times$Edu (NON-AI:grad)   &     3.67 & 22.40 &  0.16 & .87 & .87 \\
  \cmidrule{2-9}
  & \multirow{3}{*}{LF/HF Ratio}
    & Modality (AI)                       &     0.18 &  0.29 &  0.63 & .53 & .63 & \multirow{3}{*}{$<$.01\,/\,.14} \\
  & & Mod$\times$Edu (AI:grad)       &  $-$0.09 &  0.29 & $-$0.32 & .75 & .87 \\
  & & Mod$\times$Edu (NON-AI:grad)   &  $-$0.06 &  0.29 & $-$0.22 & .83 & .87 \\
\midrule
\multirow{6}{*}{\shortstack[l]{EEG\\\scriptsize($N{=}60$\\\scriptsize UNIBA+ITU)}}
  & \multirow{3}{*}{Beta Power}
    & Modality (AI)                       &  $-$3.77 &  3.79 & $-$0.99 & .32 & .32 & \multirow{3}{*}{.01\,/\,.80} \\
  & & Mod$\times$Edu (AI:grad)       &     1.82 &  3.64 &  0.50 & .62 & .67 \\
  & & Mod$\times$Edu (NON-AI:grad)   &  $-$1.56 &  3.72 & $-$0.42 & .67 & .67 \\
  \cmidrule{2-9}
  & \multirow{3}{*}{$\theta/\alpha$ Ratio}
    & Modality (AI)                       &     1.11 &  0.81 &  1.38 & .17 & .32 & \multirow{3}{*}{.02\,/\,.39} \\
  & & Mod$\times$Edu (AI:grad)       &  $-$0.87 &  0.77 & $-$1.12 & .26 & .53 \\
  & & Mod$\times$Edu (NON-AI:grad)   &     0.96 &  0.79 &  1.22 & .22 & .53 \\
\midrule
\multirow{9}{*}{\shortstack[l]{Gaze\\\scriptsize($N{=}60$\\\scriptsize UNIBA+ITU)}}
  & \multirow{3}{*}{Fixation Count}
    & Modality (AI)                       &     4.20 & 13.47 &  0.31 & .76 & .76 & \multirow{3}{*}{.08\,/\,.77} \\
  & & Mod$\times$Edu (AI:grad)       & $-$26.65 & 12.94 & $-$2.06 & .04 & .24 \\
  & & Mod$\times$Edu (NON-AI:grad)   &     5.62 & 13.21 &  0.43 & .67 & .80 \\
  \cmidrule{2-9}
  & \multirow{3}{*}{Mean Fix.\ Duration}
    & Modality (AI)                       &  $-$0.01 &  0.02 & $-$0.71 & .48 & .72 & \multirow{3}{*}{.03\,/\,.67} \\
  & & Mod$\times$Edu (AI:grad)       &     0.03 &  0.02 &  1.64 & .10 & .31 \\
  & & Mod$\times$Edu (NON-AI:grad)   &     0.00 &  0.02 &  0.24 & .81 & .81 \\
  \cmidrule{2-9}
  & \multirow{3}{*}{Blink Rate}
    & Modality (AI)                       &     1.41 &  0.82 &  1.73 & .08 & .25 & \multirow{3}{*}{.09\,/\,.51} \\
  & & Mod$\times$Edu (AI:grad)       &     0.98 &  0.78 &  1.25 & .21 & .43 \\
  & & Mod$\times$Edu (NON-AI:grad)   &     0.46 &  0.80 &  0.58 & .56 & .80 \\
\bottomrule
\end{tabular}
\begin{tablenotes}\footnotesize
\item[$^*$] BH-FDR correction applied within each signal family (EDA, HRV, EEG, Gaze) and effect type (Modality, Mod$\times$Edu)
\item Reference levels: Modality = NON-AI; Education = undergraduate
\item Mod$\times$Edu (AI:grad) and Mod$\times$Edu (NON-AI:grad) denote the effect of graduate (vs.\ undergraduate) education within the AI and NON-AI modality, respectively. 
\end{tablenotes}
\end{threeparttable}
}
\vspace{-8mm}
\end{table}

\begin{takeaway}
\textbf{H2: Not supported.} The interaction between modality and developer experience does not reach significance for any physiological metric, neither in the full model nor in the task-stratified analyses. The modality difference in the $\theta/\alpha$ ratio identified under H1 persists once developer experience is included, but is not modulated by experience itself. We therefore find no evidence that developer experience, modeled as academic seniority (i.e., undergraduate vs.\ graduate), is associated with differences in the modality--physiology relationship.
\end{takeaway}

\subsection{H\stageTwo{3}: Physiological measures correlate with performance}
\label{sec:res:h3}

\stageTwo{
Following the analysis plan described in Section~\ref{sec:hypothesis-test}, the H3 results are reported across the three temporal periods of each task coding session (early, mid, and late), with a separate mixed-effects model fitted within each period.

During the early phase of the task session (Table~\ref{tab:glmm_early}), significant main effects of physiological predictors and \texttt{Modality}\,$\times$\,\texttt{Physio} interactions emerge across three signal families.

\texttt{SCR amplitude} is negatively associated with performance ($\beta = -0.076$, $p_{\text{BH}} = .007$): higher SCR amplitude is associated with lower odds of high performance. The interaction with \texttt{Modality} is also significant ($\beta = 0.089$, $p_{\text{BH}} = .015$): the condition-specific refit shows that the negative association between SCR amplitude and performance is present under NON-AI ($\beta = -0.116$, $p = .030$) but absent under AI.
\texttt{EDA tonic mean} is positively associated with performance ($\beta = 0.077$, $p_{\text{BH}} = .011$) and the interaction with \texttt{Modality} is also significant ($\beta = -0.123$, $p_{\text{BH}} = .006$). The condition specific refit shows a positive association between EDA tonic mean and performance under NON-AI condition ($\beta = 0.143$, $p = .016$), with no corresponding association under AI.
For the \texttt{EEG $\theta/\alpha$ ratio}, the modality interaction is also significant ($\beta = 0.469$, $p_{\text{BH}} = .024$), although neither condition-specific refit reaches significance.
Finally, \texttt{blink rate} shows both a significant modality main effect ($\beta = 1.143$, $p_{\text{BH}} = .008$) and a significant interaction ($\beta = -0.338$, $p_{\text{BH}} < .001$); as for the EEG $\theta/\alpha$ ratio, the condition-specific refit does not yield significance in either modality.
}

\begin{table}[t]
\centering
\caption{Bayesian binomial GLMM estimates for H3, early task period ([2{:}00, 5{:}00]~min); model in Eq.~\ref{eq:glmm-h3-full}, with condition-specific refit in Eq.~\ref{eq:glmm-h3-strat}.}
\label{tab:glmm_early}
\resizebox{\textwidth}{!}{
\begin{threeparttable}
\footnotesize
\setlength{\tabcolsep}{4pt}
\renewcommand{\arraystretch}{0.92}

\begin{tabular}{lllrrrrrl}
\toprule
\textbf{Family} & \textbf{Metric} & \textbf{Effect} & \textbf{Coef.} & \textbf{SE} & \textbf{$z$} & \textbf{$p$} & \textbf{$p_{\textsc{bh}}$} & \textbf{OR [95\% CI]} \\

\midrule
\multirow{14}{*}{\shortstack[l]{EDA\\\scriptsize($N{=}34$\\\scriptsize UNIBA)}}
 & \multirow{4}{*}{SCR Amplitude}
   & Modality (AI)  & $-0.950$ & $0.496$ & $-1.92$ & $0.055$ & $0.083$ & $0.39\ [0.15,\ 1.02]$ \\
 & & Task\_order (2)  & $\phantom{-}0.225$ & $0.484$ & $\phantom{-}0.46$ & $0.642$ & $0.642$ & $1.25\ [0.49,\ 3.23]$ \\
 & & \textbf{Physio} & $\boldsymbol{-0.076}$ & $\boldsymbol{0.025}$ & $\boldsymbol{-3.06}$ & $\boldsymbol{0.002}$ & $\boldsymbol{0.007^{**}}$ & $\boldsymbol{0.93\ [0.88,\ 0.97]}$ \\
 & & \textbf{Modality $\times$ Physio} & $\boldsymbol{\phantom{-}0.089}$ & $\boldsymbol{0.035}$ & $\boldsymbol{\phantom{-}2.58}$ & $\boldsymbol{0.010}$ & $\boldsymbol{0.015^{*}}$ & $\boldsymbol{1.09\ [1.02,\ 1.17]}$ \\
\cmidrule(lr){3-9}
 & & \quad\textit{Physio$\mid$AI}     & $-0.007$ & $0.051$ & $-0.14$ & $0.888$ & --- & $0.99\ [0.90,\ 1.10]$ \\
 & & \quad\textbf{\textit{Physio$\mid$NON-AI}} & $\boldsymbol{-0.116}$ & $\boldsymbol{0.054}$ & $\boldsymbol{-2.16}$ & $\boldsymbol{0.030}$ & --- & $\boldsymbol{0.89\ [0.80,\ 0.99]}$ \\
\cmidrule(l){2-9}
 & \multirow{4}{*}{SCR Peaks}
   & Modality (AI)  & $-0.448$ & $0.481$ & $-0.93$ & $0.352$ & $0.352$ & $0.64\ [0.25,\ 1.64]$ \\
 & & Task\_order (2)  & $\phantom{-}0.384$ & $0.475$ & $\phantom{-}0.81$ & $0.420$ & $0.630$ & $1.47\ [0.58,\ 3.73]$ \\
 & & Physio & $-0.019$ & $0.015$ & $-1.28$ & $0.201$ & $0.201$ & $0.98\ [0.95,\ 1.01]$ \\
 & & Modality $\times$ Physio  & $\phantom{-}0.020$ & $0.022$ & $\phantom{-}0.90$ & $0.369$ & $0.369$ & $1.02\ [0.98,\ 1.07]$ \\
\cmidrule(l){2-9}
 & \multirow{4}{*}{Tonic Mean}
   & Modality (AI)  & $-1.162$ & $0.494$ & $-2.35$ & $0.019$ & $0.056$ & $0.31\ [0.12,\ 0.82]$ \\
 & & Task\_order (2)  & $\phantom{-}0.431$ & $0.487$ & $\phantom{-}0.89$ & $0.376$ & $0.630$ & $1.54\ [0.59,\ 3.99]$ \\
 & & \textbf{Physio} & $\boldsymbol{\phantom{-}0.077}$ & $\boldsymbol{0.029}$ & $\boldsymbol{\phantom{-}2.69}$ & $\boldsymbol{0.007}$ & $\boldsymbol{0.011^{*}}$ & $\boldsymbol{1.08\ [1.02,\ 1.14]}$ \\
 & & \textbf{Modality $\times$ Physio} & $\boldsymbol{-0.123}$ & $\boldsymbol{0.040}$ & $\boldsymbol{-3.09}$ & $\boldsymbol{0.002}$ & $\boldsymbol{0.006^{**}}$ & $\boldsymbol{0.88\ [0.82,\ 0.96]}$ \\
\cmidrule(lr){3-9}
 & & \quad\textit{Physio$\mid$AI}     & $-0.025$ & $0.054$ & $-0.46$ & $0.645$ & --- & $0.98\ [0.88,\ 1.08]$ \\
 & & \quad\textbf{\textit{Physio$\mid$NON-AI}} & $\boldsymbol{\phantom{-}0.143}$ & $\boldsymbol{0.059}$ & $\boldsymbol{\phantom{-}2.40}$ & $\boldsymbol{0.016}$ & --- & $\boldsymbol{1.15\ [1.03,\ 1.30]}$ \\
\midrule
\multirow{10}{*}{\shortstack[l]{HRV\\\scriptsize($N{=}34$\\\scriptsize UNIBA)}}
 & \multirow{4}{*}{RMSSD}
   & Modality (AI)  & $-0.243$ & $0.501$ & $-0.49$ & $0.628$ & $0.746$ & $0.78\ [0.29,\ 2.09]$ \\
 & & Task\_order (2)  & $\phantom{-}0.553$ & $0.487$ & $\phantom{-}1.14$ & $0.256$ & $0.395$ & $1.74\ [0.67,\ 4.51]$ \\
 & & Physio & $\phantom{-}0.010$ & $0.007$ & $\phantom{-}1.48$ & $0.140$ & $0.187$ & $1.01\ [1.00,\ 1.02]$ \\
 & & Modality $\times$ Physio  & $-0.016$ & $0.011$ & $-1.49$ & $0.136$ & $0.271$ & $0.98\ [0.96,\ 1.00]$ \\
\cmidrule(l){2-9}
 & \multirow{4}{*}{LF/HF Ratio}
   & Modality (AI)  & $\phantom{-}0.157$ & $0.483$ & $\phantom{-}0.32$ & $0.746$ & $0.746$ & $1.17\ [0.45,\ 3.01]$ \\
 & & Task\_order (2)  & $\phantom{-}0.405$ & $0.476$ & $\phantom{-}0.85$ & $0.395$ & $0.395$ & $1.50\ [0.59,\ 3.81]$ \\
 & & Physio & $\phantom{-}0.187$ & $0.142$ & $\phantom{-}1.32$ & $0.187$ & $0.187$ & $1.21\ [0.91,\ 1.59]$ \\
 & & Modality $\times$ Physio  & $-0.104$ & $0.201$ & $-0.51$ & $0.607$ & $0.607$ & $0.90\ [0.61,\ 1.34]$ \\
\midrule
\multirow{10}{*}{\shortstack[l]{EEG\\\scriptsize($N{=}60$\\\scriptsize UNIBA+ITU)}}
 & \multirow{4}{*}{Beta Power}
   & Modality (AI)  & $\phantom{-}0.189$ & $0.367$ & $\phantom{-}0.51$ & $0.607$ & $0.607$ & $1.21\ [0.59,\ 2.48]$ \\
 & & Task\_order (2)  & $\phantom{-}0.207$ & $0.365$ & $\phantom{-}0.57$ & $0.571$ & $0.571$ & $1.23\ [0.60,\ 2.52]$ \\
 & & Physio & $-0.024$ & $0.026$ & $-0.92$ & $0.360$ & $0.360$ & $0.98\ [0.93,\ 1.03]$ \\
 & & Modality $\times$ Physio  & $-0.041$ & $0.036$ & $-1.13$ & $0.259$ & $0.259$ & $0.96\ [0.89,\ 1.03]$ \\
\cmidrule(l){2-9}
 & \multirow{4}{*}{Theta/Alpha Ratio}
   & Modality (AI)  & $-0.365$ & $0.395$ & $-0.92$ & $0.356$ & $0.607$ & $0.69\ [0.32,\ 1.51]$ \\
 & & Task\_order (2)  & $\phantom{-}0.266$ & $0.395$ & $\phantom{-}0.67$ & $0.501$ & $0.571$ & $1.30\ [0.60,\ 2.83]$ \\
 & & Physio & $\phantom{-}0.132$ & $0.128$ & $\phantom{-}1.03$ & $0.305$ & $0.360$ & $1.14\ [0.89,\ 1.47]$ \\
 & & \textbf{Modality $\times$ Physio} & $\boldsymbol{\phantom{-}0.469}$ & $\boldsymbol{0.187}$ & $\boldsymbol{\phantom{-}2.51}$ & $\boldsymbol{0.012}$ & $\boldsymbol{0.024^{*}}$ & $\boldsymbol{1.60\ [1.11,\ 2.31]}$ \\
\cmidrule(lr){3-9}
 & & \quad\textit{Physio$\mid$AI}     & $\phantom{-}0.489$ & $0.249$ & $\phantom{-}1.96$ & $0.050$ & --- & $1.63\ [1.00,\ 2.66]$ \\
 & & \quad\textit{Physio$\mid$NON-AI} & $\phantom{-}0.369$ & $0.222$ & $\phantom{-}1.66$ & $0.097$ & --- & $1.45\ [0.94,\ 2.23]$ \\
\midrule
\multirow{14}{*}{\shortstack[l]{Gaze\\\scriptsize($N{=}60$\\\scriptsize UNIBA+ITU)}}
 & \multirow{4}{*}{Fixation Count\tnote{a}}
   & Modality (AI)  & --- & --- & --- & --- & --- & n.c. \\
 & & Task\_order (2)  & --- & --- & --- & --- & --- & n.c. \\
 & & Physio & --- & --- & --- & --- & --- & n.c. \\
 & & Modality $\times$ Physio  & --- & --- & --- & --- & --- & n.c. \\
\cmidrule(l){2-9}
 & \multirow{4}{*}{Mean Fixation Dur.}
   & Modality (AI)  & $-0.239$ & $0.364$ & $-0.66$ & $0.512$ & $0.768$ & $0.79\ [0.39,\ 1.61]$ \\
 & & Task\_order (2)  & $\phantom{-}0.247$ & $0.362$ & $\phantom{-}0.68$ & $0.494$ & $0.741$ & $1.28\ [0.63,\ 2.60]$ \\
 & & Physio & $\phantom{-}1.485$ & $0.841$ & $\phantom{-}1.77$ & $0.077$ & $0.232$ & $4.42\ [0.85,\ 22.96]$ \\
 & & Modality $\times$ Physio  & $\phantom{-}0.768$ & $1.143$ & $\phantom{-}0.67$ & $0.502$ & $0.753$ & $2.16\ [0.23,\ 20.27]$ \\
\cmidrule(l){2-9}
 & \multirow{4}{*}{Blink Rate}
   & \textbf{Modality (AI)}  & $\boldsymbol{\phantom{-}1.143}$ & $\boldsymbol{0.383}$ & $\boldsymbol{\phantom{-}2.99}$ & $\boldsymbol{0.003}$ & $\boldsymbol{0.008^{**}}$ & $\boldsymbol{3.14\ [1.48,\ 6.64]}$ \\
 & & Task\_order (2)  & $\phantom{-}0.443$ & $0.381$ & $\phantom{-}1.16$ & $0.245$ & $0.735$ & $1.56\ [0.74,\ 3.28]$ \\
 & & Physio & $\phantom{-}0.089$ & $0.065$ & $\phantom{-}1.37$ & $0.171$ & $0.257$ & $1.09\ [0.96,\ 1.24]$ \\
 & & \textbf{Modality $\times$ Physio} & $\boldsymbol{-0.338}$ & $\boldsymbol{0.082}$ & $\boldsymbol{-4.13}$ & $\boldsymbol{<\!0.001}$ & $\boldsymbol{<\!0.001^{***}}$ & $\boldsymbol{0.71\ [0.61,\ 0.84]}$ \\
\cmidrule(lr){3-9}
 & & \quad\textit{Physio$\mid$AI}     & $-0.127$ & $0.106$ & $-1.20$ & $0.231$ & --- & $0.88\ [0.72,\ 1.08]$ \\
 & & \quad\textit{Physio$\mid$NON-AI} & $\phantom{-}0.162$ & $0.144$ & $\phantom{-}1.12$ & $0.262$ & --- & $1.18\ [0.89,\ 1.56]$ \\
\bottomrule
\end{tabular}
\begin{tablenotes}\footnotesize
\item Reference levels: Modality = NON-AI; Task\_order = first; Outcome: P(High performance); low performance = reference."
\item Significance codes refer to $p_{\textsc{bh}}$: $^{*}\,q<0.05$; $^{**}\,q<0.01$; $^{***}\,q<0.001$. Rows in \textbf{bold} indicate effects with $p_{\textsc{bh}} < 0.05$. BH correction computed within (period $\times$ family $\times$ effect type).
\item Indented rows below a significant interaction report the condition-specific slope of the physiological predictor refitted within each \textsc{Modality} level (Eq.~\ref{eq:glmm-h3-strat}).
\item[a] Fixation Count GLMM did not converge (n.c.); estimates are not interpretable.
\end{tablenotes}
\end{threeparttable}
}
\vspace{-5mm}
\end{table}

\stageTwo{
During the mid phase of the task session (see Table~\ref{tab:glmm_mid}), significant \texttt{Modality} $\times$ \texttt{Physio} interactions emerge across all four signal families. For \texttt{SCR amplitude}, the interaction reaches significance ($\beta = 0.068$, $p_{\text{BH}} = .041$), but the condition-specific refit does not yield significant slopes in either modality. For \texttt{EDA tonic mean} the interaction reach significance ($\beta = -0.083$, $p_{\text{BH}} = .041$) and the condition specific refit shows a positive association between EDA tonic mean and performance under the NON-AI condition ($\beta = 0.086$, $p = .028$). In the AI condition no significant association was found. 
For \texttt{RMSSD}, both the physiological predictor ($\beta = 0.017$, $p_{\text{BH}} = .029$) and its interaction with \texttt{Modality} ($\beta = -0.032$, $p_{\text{BH}} = .001$) are significant, but the condition-specific refit does not yield significant slopes in either modality.
}
\stageTwo{
For the \texttt{$\theta/\alpha$ ratio}, the interaction with \texttt{Modality} is significant ($\beta = 0.535$, $p_{\text{BH}} = .017$); as for SCR amplitude and RMSSD, the condition specific refit does not yield significant slopes in either modality.
Finally, \texttt{blink rate} shows significance on all three terms: \texttt{Modality} ($\beta = 1.158$, $p_{\text{BH}} = .007$); the physiological predictor ($\beta = 0.243$, $p_{\text{BH}} = .006$), positively associated with high performance; and the \texttt{Modality} $\times$ \texttt{Physio} interaction ($\beta = -0.469$, $p_{\text{BH}} < .001$). The condition-specific refit reveals slopes of opposite sign across conditions: negative under AI ($\beta = -0.281$, $p = .028$), and positive under NON-AI ($\beta = 0.369$, $p = .039$).
}

\stageTwo{
During the late phase of the task coding session (Table~\ref{tab:glmm_late}), the remaining significant effects concentrates in the EDA and HRV families. Two EDA metrics show direct associations with performance: \texttt{SCR amplitude} ($\beta = -0.033$, $p_{\text{BH}} = .040$) and \texttt{SCR peaks} ($\beta = -0.047$, $p_{\text{BH}} = .002$), both negatively. In the HRV family, the modality interaction is significant for \texttt{RMSSD} ($\beta = -0.049$, $p_{\text{BH}} = .001$), although neither condition-specific slope reaches significance individually. A similar pattern emerges for the \texttt{LF/HF Ratio}, where the modality interaction is also significant ($\beta = -0.390$, $p_{\text{BH}} = .032$) but no condition-specific slope reaches significance. 
}

\begin{takeaway}
\textbf{H3: Partially supported.} Several physiological measures show significant interactions with modality across all three task periods, but the within-condition picture is heterogeneous. EDA-derived measures (SCR amplitude, tonic mean) correlate with performance only under NON-AI. Blink rate correlates with performance in both conditions, but with opposite signs. The remaining significant interactions (EEG $\theta/\alpha$ in the early period; HRV RMSSD and LF/HF in the late period) do not produce robust within-condition slopes. The expected difference in physiological--performance coupling between conditions therefore applies primarily to EDA-derived measures.
\end{takeaway}

\begin{table}[t]
\centering
\caption{Bayesian binomial GLMM estimates for H3, mid task period (3-min interval centred on the task midpoint); same model as Table~\ref{tab:glmm_early}.}
\label{tab:glmm_mid}
\resizebox{\textwidth}{!}{
\begin{threeparttable}
\footnotesize
\setlength{\tabcolsep}{4pt}
\renewcommand{\arraystretch}{0.92}
\begin{tabular}{lllrrrrrl}
\toprule
\textbf{Family} & \textbf{Metric} & \textbf{Effect} & \textbf{Coef.} & \textbf{SE} & \textbf{$z$} & \textbf{$p$} & \textbf{$p_{\textsc{bh}}$} & \textbf{OR [95\% CI]} \\
\midrule
\multirow{14}{*}{\shortstack[l]{EDA\\\scriptsize($N{=}34$\\\scriptsize UNIBA)}}
 & \multirow{4}{*}{SCR Amplitude}
   & Modality (AI)  & $-0.812$ & $0.486$ & $-1.67$ & $0.095$ & $0.143$ & $0.44\ [0.17,\ 1.15]$ \\
 & & Task\_order (2)  & $\phantom{-}0.317$ & $0.490$ & $\phantom{-}0.65$ & $0.518$ & $0.518$ & $1.37\ [0.53,\ 3.58]$ \\
 & & Physio & $-0.040$ & $0.019$ & $-2.05$ & $0.040$ & $0.060$ & $0.96\ [0.92,\ 1.00]$ \\
 & & \textbf{Modality $\times$ Physio} & $\boldsymbol{\phantom{-}0.068}$ & $\boldsymbol{0.031}$ & $\boldsymbol{\phantom{-}2.21}$ & $\boldsymbol{0.027}$ & $\boldsymbol{0.041^{*}}$ & $\boldsymbol{1.07\ [1.01,\ 1.14]}$ \\
\cmidrule(lr){3-9}
 & & \quad\textit{Physio$\mid$AI}     & $\phantom{-}0.017$ & $0.045$ & $\phantom{-}0.39$ & $0.699$ & --- & $1.02\ [0.93,\ 1.11]$ \\
 & & \quad\textit{Physio$\mid$NON-AI} & $-0.064$ & $0.035$ & $-1.80$ & $0.071$ & --- & $0.94\ [0.88,\ 1.01]$ \\
\cmidrule(l){2-9}
 & \multirow{4}{*}{SCR Peaks}
   & Modality (AI)  & $-0.423$ & $0.481$ & $-0.88$ & $0.379$ & $0.379$ & $0.66\ [0.26,\ 1.68]$ \\
 & & Task\_order (2)  & $\phantom{-}0.363$ & $0.476$ & $\phantom{-}0.76$ & $0.446$ & $0.518$ & $1.44\ [0.57,\ 3.66]$ \\
 & & Physio & $-0.023$ & $0.014$ & $-1.69$ & $0.091$ & $0.091$ & $0.98\ [0.95,\ 1.00]$ \\
 & & Modality $\times$ Physio  & $\phantom{-}0.016$ & $0.020$ & $\phantom{-}0.81$ & $0.417$ & $0.417$ & $1.02\ [0.98,\ 1.06]$ \\
\cmidrule(l){2-9}
 & \multirow{4}{*}{Tonic Mean}
   & Modality (AI)  & $-0.965$ & $0.495$ & $-1.95$ & $0.051$ & $0.143$ & $0.38\ [0.14,\ 1.01]$ \\
 & & Task\_order (2)  & $\phantom{-}0.387$ & $0.489$ & $\phantom{-}0.79$ & $0.429$ & $0.518$ & $1.47\ [0.56,\ 3.84]$ \\
 & & Physio & $\phantom{-}0.051$ & $0.022$ & $\phantom{-}2.29$ & $0.022$ & $0.060$ & $1.05\ [1.01,\ 1.10]$ \\
 & & \textbf{Modality $\times$ Physio} & $\boldsymbol{-0.083}$ & $\boldsymbol{0.035}$ & $\boldsymbol{-2.38}$ & $\boldsymbol{0.017}$ & $\boldsymbol{0.041^{*}}$ & $\boldsymbol{0.92\ [0.86,\ 0.99]}$ \\
\cmidrule(lr){3-9}
 & & \quad\textit{Physio$\mid$AI}     & $-0.032$ & $0.048$ & $-0.66$ & $0.508$ & --- & $0.97\ [0.88,\ 1.06]$ \\
 & & \quad\textbf{\textit{Physio$\mid$NON-AI}} & $\boldsymbol{\phantom{-}0.086}$ & $\boldsymbol{0.039}$ & $\boldsymbol{\phantom{-}2.20}$ & $\boldsymbol{0.028}$ & --- & $\boldsymbol{1.09\ [1.01,\ 1.18]}$ \\
\midrule
\multirow{10}{*}{\shortstack[l]{HRV\\\scriptsize($N{=}34$\\\scriptsize UNIBA)}}
 & \multirow{4}{*}{RMSSD}
   & Modality (AI)  & $-0.470$ & $0.511$ & $-0.92$ & $0.358$ & $0.358$ & $0.63\ [0.23,\ 1.70]$ \\
 & & Task\_order (2)  & $\phantom{-}0.373$ & $0.501$ & $\phantom{-}0.75$ & $0.456$ & $0.456$ & $1.45\ [0.54,\ 3.88]$ \\
 & & \textbf{Physio} & $\boldsymbol{\phantom{-}0.017}$ & $\boldsymbol{0.007}$ & $\boldsymbol{\phantom{-}2.44}$ & $\boldsymbol{0.015}$ & $\boldsymbol{0.029^{*}}$ & $\boldsymbol{1.02\ [1.00,\ 1.03]}$ \\
 & & \textbf{Modality $\times$ Physio} & $\boldsymbol{-0.032}$ & $\boldsymbol{0.009}$ & $\boldsymbol{-3.39}$ & $\boldsymbol{<\!0.001}$ & $\boldsymbol{0.001^{**}}$ & $\boldsymbol{0.97\ [0.95,\ 0.99]}$ \\
\cmidrule(lr){3-9}
 & & \quad\textit{Physio$\mid$AI}     & $-0.012$ & $0.013$ & $-0.91$ & $0.363$ & --- & $0.99\ [0.96,\ 1.01]$ \\
 & & \quad\textit{Physio$\mid$NON-AI} & $\phantom{-}0.021$ & $0.016$ & $\phantom{-}1.30$ & $0.194$ & --- & $1.02\ [0.99,\ 1.05]$ \\
\cmidrule(l){2-9}
 & \multirow{4}{*}{LF/HF Ratio}
   & Modality (AI)  & $\phantom{-}0.733$ & $0.487$ & $\phantom{-}1.51$ & $0.132$ & $0.265$ & $2.08\ [0.80,\ 5.41]$ \\
 & & Task\_order (2)  & $\phantom{-}0.476$ & $0.483$ & $\phantom{-}0.99$ & $0.324$ & $0.456$ & $1.61\ [0.62,\ 4.15]$ \\
 & & Physio & $-0.000$ & $0.139$ & $-0.00$ & $0.999$ & $0.999$ & $1.00\ [0.76,\ 1.31]$ \\
 & & Modality $\times$ Physio  & $-0.362$ & $0.202$ & $-1.79$ & $0.074$ & $0.074$ & $0.70\ [0.47,\ 1.04]$ \\
\midrule
\multirow{10}{*}{\shortstack[l]{EEG\\\scriptsize($N{=}60$\\\scriptsize UNIBA+ITU)}}
 & \multirow{4}{*}{Beta Power}
   & Modality (AI)  & $\phantom{-}0.182$ & $0.375$ & $\phantom{-}0.49$ & $0.627$ & $0.627$ & $1.20\ [0.58,\ 2.50]$ \\
 & & Task\_order (2)  & $\phantom{-}0.262$ & $0.372$ & $\phantom{-}0.70$ & $0.482$ & $0.482$ & $1.30\ [0.63,\ 2.69]$ \\
 & & Physio & $-0.059$ & $0.027$ & $-2.17$ & $0.030$ & $0.060$ & $0.94\ [0.89,\ 0.99]$ \\
 & & Modality $\times$ Physio  & $-0.039$ & $0.038$ & $-1.03$ & $0.303$ & $0.303$ & $0.96\ [0.89,\ 1.04]$ \\
\cmidrule(l){2-9}
 & \multirow{4}{*}{Theta/Alpha Ratio}
   & Modality (AI)  & $-0.399$ & $0.401$ & $-1.00$ & $0.319$ & $0.627$ & $0.67\ [0.31,\ 1.47]$ \\
 & & Task\_order (2)  & $\phantom{-}0.413$ & $0.400$ & $\phantom{-}1.03$ & $0.302$ & $0.482$ & $1.51\ [0.69,\ 3.31]$ \\
 & & Physio & $\phantom{-}0.138$ & $0.098$ & $\phantom{-}1.41$ & $0.160$ & $0.160$ & $1.15\ [0.95,\ 1.39]$ \\
 & & \textbf{Modality $\times$ Physio} & $\boldsymbol{\phantom{-}0.535}$ & $\boldsymbol{0.204}$ & $\boldsymbol{\phantom{-}2.62}$ & $\boldsymbol{0.009}$ & $\boldsymbol{0.017^{*}}$ & $\boldsymbol{1.71\ [1.15,\ 2.55]}$ \\
\cmidrule(lr){3-9}
 & & \quad\textit{Physio$\mid$AI}     & $\phantom{-}0.454$ & $0.265$ & $\phantom{-}1.71$ & $0.087$ & --- & $1.57\ [0.94,\ 2.64]$ \\
 & & \quad\textit{Physio$\mid$NON-AI} & $\phantom{-}0.098$ & $0.139$ & $\phantom{-}0.71$ & $0.481$ & --- & $1.10\ [0.84,\ 1.45]$ \\
\midrule
\multirow{14}{*}{\shortstack[l]{Gaze\\\scriptsize($N{=}60$\\\scriptsize UNIBA+ITU)}}
 & \multirow{4}{*}{Fixation Count\tnote{a}}
   & Modality (AI)  & --- & --- & --- & --- & --- & n.c. \\
 & & Task\_order (2)  & --- & --- & --- & --- & --- & n.c. \\
 & & Physio & --- & --- & --- & --- & --- & n.c. \\
 & & Modality $\times$ Physio  & --- & --- & --- & --- & --- & n.c. \\
\cmidrule(l){2-9}
 & \multirow{4}{*}{Mean Fixation Dur.}
   & Modality (AI)  & $\phantom{-}0.203$ & $0.366$ & $\phantom{-}0.55$ & $0.579$ & $0.868$ & $1.22\ [0.60,\ 2.51]$ \\
 & & Task\_order (2)  & $\phantom{-}0.261$ & $0.364$ & $\phantom{-}0.72$ & $0.474$ & $0.711$ & $1.30\ [0.64,\ 2.65]$ \\
 & & Physio & $\phantom{-}0.366$ & $0.838$ & $\phantom{-}0.44$ & $0.662$ & $0.993$ & $1.44\ [0.28,\ 7.45]$ \\
 & & Modality $\times$ Physio  & $-0.759$ & $1.150$ & $-0.66$ & $0.510$ & $0.764$ & $0.47\ [0.05,\ 4.46]$ \\
\cmidrule(l){2-9}
 & \multirow{4}{*}{Blink Rate}
   & \textbf{Modality (AI)}  & $\boldsymbol{\phantom{-}1.158}$ & $\boldsymbol{0.378}$ & $\boldsymbol{\phantom{-}3.06}$ & $\boldsymbol{0.002}$ & $\boldsymbol{0.007^{**}}$ & $\boldsymbol{3.18\ [1.52,\ 6.68]}$ \\
 & & Task\_order (2)  & $\phantom{-}0.332$ & $0.373$ & $\phantom{-}0.89$ & $0.374$ & $0.711$ & $1.39\ [0.67,\ 2.90]$ \\
 & & \textbf{Physio} & $\boldsymbol{\phantom{-}0.243}$ & $\boldsymbol{0.079}$ & $\boldsymbol{\phantom{-}3.09}$ & $\boldsymbol{0.002}$ & $\boldsymbol{0.006^{**}}$ & $\boldsymbol{1.27\ [1.09,\ 1.49]}$ \\
 & & \textbf{Modality $\times$ Physio} & $\boldsymbol{-0.469}$ & $\boldsymbol{0.097}$ & $\boldsymbol{-4.85}$ & $\boldsymbol{<\!0.001}$ & $\boldsymbol{<\!0.001^{***}}$ & $\boldsymbol{0.63\ [0.52,\ 0.76]}$ \\
\cmidrule(lr){3-9}
 & & \quad\textbf{\textit{Physio$\mid$AI}}     & $\boldsymbol{-0.281}$ & $\boldsymbol{0.128}$ & $\boldsymbol{-2.20}$ & $\boldsymbol{0.028}$ & --- & $\boldsymbol{0.76\ [0.59,\ 0.97]}$ \\
 & & \quad\textbf{\textit{Physio$\mid$NON-AI}} & $\boldsymbol{\phantom{-}0.369}$ & $\boldsymbol{0.178}$ & $\boldsymbol{\phantom{-}2.07}$ & $\boldsymbol{0.039}$ & --- & $\boldsymbol{1.45\ [1.02,\ 2.05]}$ \\
\bottomrule
\end{tabular}
\begin{tablenotes}\footnotesize
\item See Table~\ref{tab:glmm_early} for column and effect-label definitions, reference levels, significance codes, and the convention on indented condition-specific rows.
\item[a] Fixation Count GLMM did not converge (n.c.); estimates are not interpretable.
\end{tablenotes}
\end{threeparttable}
}
\vspace{-5mm}
\end{table}

\begin{table}[tb]
\centering
\caption{Bayesian binomial GLMM estimates for H3, late task period (final 3~min); same model as Table~\ref{tab:glmm_early}.}
\label{tab:glmm_late}
\resizebox{\textwidth}{!}{
\begin{threeparttable}
\footnotesize
\setlength{\tabcolsep}{4pt}
\renewcommand{\arraystretch}{0.92}
\begin{tabular}{lllrrrrrl}
\toprule
\textbf{Family} & \textbf{Metric} & \textbf{Effect} & \textbf{Coef.} & \textbf{SE} & \textbf{$z$} & \textbf{$p$} & \textbf{$p_{\textsc{bh}}$} & \textbf{OR [95\% CI]} \\
\midrule
\multirow{12}{*}{\shortstack[l]{EDA\\\scriptsize($N{=}34$\\\scriptsize UNIBA)}}
 & \multirow{4}{*}{SCR Amplitude}
   & Modality (AI)  & $-0.298$ & $0.523$ & $-0.57$ & $0.568$ & $0.568$ & $0.74\ [0.27,\ 2.07]$ \\
 & & Task\_order (2)  & $\phantom{-}0.053$ & $0.509$ & $\phantom{-}0.10$ & $0.917$ & $0.917$ & $1.05\ [0.39,\ 2.86]$ \\
 & & \textbf{Physio} & $\boldsymbol{-0.033}$ & $\boldsymbol{0.015}$ & $\boldsymbol{-2.22}$ & $\boldsymbol{0.027}$ & $\boldsymbol{0.040^{*}}$ & $\boldsymbol{0.97\ [0.94,\ 1.00]}$ \\
 & & Modality $\times$ Physio  & $\phantom{-}0.031$ & $0.023$ & $\phantom{-}1.36$ & $0.174$ & $0.262$ & $1.03\ [0.99,\ 1.08]$ \\
\cmidrule(l){2-9}
 & \multirow{4}{*}{SCR Peaks}
   & Modality (AI)  & $-0.542$ & $0.483$ & $-1.12$ & $0.262$ & $0.392$ & $0.58\ [0.23,\ 1.50]$ \\
 & & Task\_order (2)  & $\phantom{-}0.315$ & $0.479$ & $\phantom{-}0.66$ & $0.511$ & $0.766$ & $1.37\ [0.54,\ 3.50]$ \\
 & & \textbf{Physio} & $\boldsymbol{-0.047}$ & $\boldsymbol{0.014}$ & $\boldsymbol{-3.37}$ & $\boldsymbol{<\!0.001}$ & $\boldsymbol{0.002^{**}}$ & $\boldsymbol{0.95\ [0.93,\ 0.98]}$ \\
 & & Modality $\times$ Physio  & $\phantom{-}0.018$ & $0.021$ & $\phantom{-}0.84$ & $0.401$ & $0.401$ & $1.02\ [0.98,\ 1.06]$ \\
\cmidrule(l){2-9}
 & \multirow{4}{*}{Tonic Mean}
   & Modality (AI)  & $-0.680$ & $0.496$ & $-1.37$ & $0.170$ & $0.392$ & $0.51\ [0.19,\ 1.34]$ \\
 & & Task\_order (2)  & $\phantom{-}0.394$ & $0.488$ & $\phantom{-}0.81$ & $0.420$ & $0.766$ & $1.48\ [0.57,\ 3.86]$ \\
 & & Physio & $\phantom{-}0.026$ & $0.014$ & $\phantom{-}1.81$ & $0.071$ & $0.071$ & $1.03\ [1.00,\ 1.06]$ \\
 & & Modality $\times$ Physio  & $-0.043$ & $0.024$ & $-1.81$ & $0.070$ & $0.209$ & $0.96\ [0.91,\ 1.00]$ \\
\midrule
\multirow{10}{*}{\shortstack[l]{HRV\\\scriptsize($N{=}34$\\\scriptsize UNIBA)}}
 & \multirow{4}{*}{RMSSD}
   & Modality (AI)  & $-0.720$ & $0.535$ & $-1.35$ & $0.178$ & $0.178$ & $0.49\ [0.17,\ 1.39]$ \\
 & & Task\_order (2)  & $\phantom{-}0.442$ & $0.520$ & $\phantom{-}0.85$ & $0.395$ & $0.395$ & $1.56\ [0.56,\ 4.31]$ \\
 & & Physio & $\phantom{-}0.018$ & $0.008$ & $\phantom{-}2.11$ & $0.035$ & $0.069$ & $1.02\ [1.00,\ 1.04]$ \\
 & & \textbf{Modality $\times$ Physio} & $\boldsymbol{-0.049}$ & $\boldsymbol{0.014}$ & $\boldsymbol{-3.43}$ & $\boldsymbol{<\!0.001}$ & $\boldsymbol{0.001^{**}}$ & $\boldsymbol{0.95\ [0.93,\ 0.98]}$ \\
\cmidrule(lr){3-9}
 & & \quad\textit{Physio$\mid$AI}     & $-0.030$ & $0.018$ & $-1.62$ & $0.106$ & --- & $0.97\ [0.94,\ 1.01]$ \\
 & & \quad\textit{Physio$\mid$NON-AI} & $\phantom{-}0.012$ & $0.015$ & $\phantom{-}0.83$ & $0.409$ & --- & $1.01\ [0.98,\ 1.04]$ \\
\cmidrule(l){2-9}
 & \multirow{4}{*}{LF/HF Ratio}
   & Modality (AI)  & $\phantom{-}0.888$ & $0.489$ & $\phantom{-}1.82$ & $0.069$ & $0.139$ & $2.43\ [0.93,\ 6.34]$ \\
 & & Task\_order (2)  & $\phantom{-}0.425$ & $0.483$ & $\phantom{-}0.88$ & $0.379$ & $0.395$ & $1.53\ [0.59,\ 3.94]$ \\
 & & Physio & $\phantom{-}0.094$ & $0.124$ & $\phantom{-}0.76$ & $0.449$ & $0.449$ & $1.10\ [0.86,\ 1.40]$ \\
 & & \textbf{Modality $\times$ Physio} & $\boldsymbol{-0.390}$ & $\boldsymbol{0.182}$ & $\boldsymbol{-2.15}$ & $\boldsymbol{0.032}$ & $\boldsymbol{0.032^{*}}$ & $\boldsymbol{0.68\ [0.47,\ 0.97]}$ \\
\cmidrule(lr){3-9}
 & & \quad\textit{Physio$\mid$AI}     & $-0.337$ & $0.268$ & $-1.26$ & $0.208$ & --- & $0.71\ [0.42,\ 1.21]$ \\
 & & \quad\textit{Physio$\mid$NON-AI} & $-0.403$ & $0.270$ & $-1.49$ & $0.136$ & --- & $0.67\ [0.39,\ 1.13]$ \\
\midrule
\multirow{8}{*}{\shortstack[l]{EEG\\\scriptsize($N{=}60$\\\scriptsize UNIBA+ITU)}}
 & \multirow{4}{*}{Beta Power}
   & Modality (AI)  & $-0.041$ & $0.377$ & $-0.11$ & $0.913$ & $0.913$ & $0.96\ [0.46,\ 2.01]$ \\
 & & Task\_order (2)  & $\phantom{-}0.159$ & $0.378$ & $\phantom{-}0.42$ & $0.674$ & $0.676$ & $1.17\ [0.56,\ 2.46]$ \\
 & & Physio & $-0.054$ & $0.026$ & $-2.03$ & $0.042$ & $0.084$ & $0.95\ [0.90,\ 1.00]$ \\
 & & Modality $\times$ Physio  & $\phantom{-}0.004$ & $0.036$ & $\phantom{-}0.12$ & $0.904$ & $0.904$ & $1.00\ [0.94,\ 1.08]$ \\
\cmidrule(l){2-9}
 & \multirow{4}{*}{Theta/Alpha Ratio}
   & Modality (AI)  & $-0.163$ & $0.382$ & $-0.43$ & $0.668$ & $0.913$ & $0.85\ [0.40,\ 1.79]$ \\
 & & Task\_order (2)  & $\phantom{-}0.157$ & $0.376$ & $\phantom{-}0.42$ & $0.676$ & $0.676$ & $1.17\ [0.56,\ 2.44]$ \\
 & & Physio & $\phantom{-}0.076$ & $0.103$ & $\phantom{-}0.74$ & $0.457$ & $0.457$ & $1.08\ [0.88,\ 1.32]$ \\
 & & Modality $\times$ Physio  & $\phantom{-}0.345$ & $0.162$ & $\phantom{-}2.13$ & $0.033$ & $0.067$ & $1.41\ [1.03,\ 1.94]$ \\
\midrule
\multirow{12}{*}{\shortstack[l]{Gaze\\\scriptsize($N{=}60$\\\scriptsize UNIBA+ITU)}}
 & \multirow{4}{*}{Fixation Count\tnote{a}}
   & Modality (AI)  & --- & --- & --- & --- & --- & n.c. \\
 & & Task\_order (2)  & --- & --- & --- & --- & --- & n.c. \\
 & & Physio & --- & --- & --- & --- & --- & n.c. \\
 & & Modality $\times$ Physio  & --- & --- & --- & --- & --- & n.c. \\
\cmidrule(l){2-9}
 & \multirow{4}{*}{Mean Fixation Dur.}
   & Modality (AI)  & $-0.005$ & $0.366$ & $-0.01$ & $0.990$ & $1.000$ & $1.00\ [0.49,\ 2.04]$ \\
 & & Task\_order (2)  & $\phantom{-}0.260$ & $0.364$ & $\phantom{-}0.71$ & $0.475$ & $0.713$ & $1.30\ [0.64,\ 2.65]$ \\
 & & Physio & $\phantom{-}0.625$ & $0.828$ & $\phantom{-}0.75$ & $0.451$ & $0.676$ & $1.87\ [0.37,\ 9.48]$ \\
 & & Modality $\times$ Physio  & $-0.023$ & $1.142$ & $-0.02$ & $0.984$ & $1.000$ & $0.98\ [0.10,\ 9.16]$ \\
\cmidrule(l){2-9}
 & \multirow{4}{*}{Blink Rate}
   & Modality (AI)  & $\phantom{-}0.493$ & $0.373$ & $\phantom{-}1.32$ & $0.186$ & $0.557$ & $1.64\ [0.79,\ 3.40]$ \\
 & & Task\_order (2)  & $\phantom{-}0.358$ & $0.371$ & $\phantom{-}0.97$ & $0.334$ & $0.713$ & $1.43\ [0.69,\ 2.96]$ \\
 & & Physio & $\phantom{-}0.078$ & $0.068$ & $\phantom{-}1.15$ & $0.252$ & $0.676$ & $1.08\ [0.95,\ 1.23]$ \\
 & & Modality $\times$ Physio  & $-0.172$ & $0.091$ & $-1.89$ & $0.059$ & $0.176$ & $0.84\ [0.70,\ 1.01]$ \\
\bottomrule
\end{tabular}
\begin{tablenotes}\footnotesize
\item See Table~\ref{tab:glmm_early} for column and effect-label definitions, reference levels, significance codes, and the convention on indented condition-specific rows.
\item[a] Fixation Count GLMM did not converge (n.c.); estimates are not interpretable.
\end{tablenotes}
\end{threeparttable}
}
\end{table}

\subsection{H\stageTwo{4}: Subjective--objective alignment differs across conditions}
\label{sec:res:h4}


\stageTwo{As reported in Table~\ref{tab:glmm_workload_interaction}, after BH correction across the seven NASA-TLX models within each effect type, only the \texttt{Physical} dimension shows a significant \texttt{Modality} $\times$ \texttt{Dimension} interaction ($\beta = 2.096$, $p_{\text{BH}} < .001$). The condition-specific analysis clarifies the source of the interaction: within the NON-AI condition, high physical demand is associated with lower performance ($\beta = -2.380$, $p < .001$), whereas within the AI condition this association is absent ($\beta = 0.049$, $p = .909$). The main effect of \texttt{Physical} on objective performance also survives BH correction ($\beta = -2.613$, $p_{\text{BH}} < .001$), negatively associated with performance. Two additional main effects survive BH correction independently of modality: self-reported \texttt{Performance} (NASA-TLX) ($\beta = 1.135$, $p_{\text{BH}} < .01$), positively associated with objective performance, and \texttt{Effort} ($\beta = -0.998$, $p_{\text{BH}} < .01$), negatively associated with it. No \texttt{Modality} $\times$ \texttt{Dimension} interaction emerges for the \texttt{Mental}, \texttt{Temporal}, \texttt{Frustration}, or aggregated \texttt{Mean} dimensions, indicating that the modality-dependent alignment between subjective workload and objective performance is restricted to the physical demand dimension.}

\begin{table}[htb]
\centering
\caption{Bayesian binomial GLMM estimates for H4: \texttt{Modality}~$\times$~\texttt{NASA-TLX}-dimension interactions on binary task performance; model in Eq.~\ref{eq:glmm-h4}.}
\label{tab:glmm_workload_interaction}
\resizebox{\textwidth}{!}{
\begin{threeparttable}
\footnotesize
\setlength{\tabcolsep}{4pt}
\renewcommand{\arraystretch}{0.92}
\begin{tabular}{llrrrrrr}
\toprule
\textbf{Dimension} & \textbf{Effect} & \textbf{Coef.} & \textbf{SE} & \textbf{$z$} & \textbf{$p$} & \textbf{$p_{\mathrm{BH}}$} & \textbf{OR [95\% CI]} \\
\midrule
\multirow{5}{*}{Mental}
  & Intercept                                & $-0.049$ & $0.251$ & $-0.19$ & $0.846$ & & $0.95\ [0.58,\ 1.56]$ \\
  & Modality (AI)                            & $-0.112$ & $0.354$ & $-0.32$ & $0.751$ & $0.751$ & $0.89\ [0.45,\ 1.79]$ \\
  & Mental (high)                       & $-0.151$ & $0.397$ & $-0.38$ & $0.703$ & $0.703$ & $0.86\ [0.40,\ 1.87]$ \\
  & Task\_order (2)                          & $\phantom{-}0.270$ & $0.353$ & $\phantom{-}0.77$ & $0.444$ & $0.680$ & $1.31\ [0.66,\ 2.62]$ \\
  & Modality $\times$ Mental          & $\phantom{-}0.253$ & $0.678$ & $\phantom{-}0.37$ & $0.709$ & $0.709$ & $1.29\ [0.34,\ 4.87]$ \\
\midrule
\multirow{7}{*}{Physical}
  & \textbf{Intercept}                       & $\boldsymbol{\phantom{-}0.875}$ & $\boldsymbol{0.255}$ & $\boldsymbol{\phantom{-}3.43}$ & $\boldsymbol{<\!0.001}$ & & $\boldsymbol{2.40\ [1.46,\ 3.95]}$ \\
  & Modality (AI)                            & $-0.925$ & $0.347$ & $-2.67$ & $0.008$ & $0.053$ & $0.40\ [0.20,\ 0.78]$ \\
  & \textbf{Physical (high)}            & $\boldsymbol{-2.613}$ & $\boldsymbol{0.369}$ & $\boldsymbol{-7.09}$ & $\boldsymbol{<\!0.001}$ & $\boldsymbol{<\!0.001^{***}}$ & $\boldsymbol{0.07\ [0.04,\ 0.15]}$ \\
  & Task\_order (2)                          & $\phantom{-}0.617$ & $0.364$ & $\phantom{-}1.69$ & $0.090$ & $0.631$ & $1.85\ [0.91,\ 3.78]$ \\
  & \textbf{Modality $\times$ Physical\_cat} & $\boldsymbol{\phantom{-}2.096}$ & $\boldsymbol{0.497}$ & $\boldsymbol{\phantom{-}4.22}$ & $\boldsymbol{<\!0.001}$ & $\boldsymbol{<\!0.001^{***}}$ & $\boldsymbol{8.14\ [3.07,\ 21.56]}$ \\
\cmidrule(lr){2-8}
  & \quad\textit{Physical$\mid$AI}\tnote{a}      & $\phantom{-}0.049$ & $0.426$ & $\phantom{-}0.11$ & $0.909$ & & $1.05\ [0.46,\ 2.42]$ \\
  & \quad\textbf{\textit{Physical$\mid$NON-AI}}\tnote{a} & $\boldsymbol{-2.380}$ & $\boldsymbol{0.450}$ & $\boldsymbol{-5.29}$ & $\boldsymbol{<\!0.001^{***}}$ & & $\boldsymbol{0.09\ [0.04,\ 0.22]}$ \\
\midrule
\multirow{5}{*}{Temporal}
  & Intercept                                & $-0.338$ & $0.250$ & $-1.35$ & $0.177$ & & $0.71\ [0.44,\ 1.16]$ \\
  & Modality (AI)                            & $\phantom{-}0.441$ & $0.352$ & $\phantom{-}1.25$ & $0.210$ & $0.553$ & $1.55\ [0.78,\ 3.10]$ \\
  & Temporal (high)                     & $\phantom{-}0.647$ & $0.378$ & $\phantom{-}1.71$ & $0.087$ & $0.087$ & $1.91\ [0.91,\ 4.01]$ \\
  & Task\_order (2)                          & $\phantom{-}0.109$ & $0.353$ & $\phantom{-}0.31$ & $0.758$ & $0.758$ & $1.11\ [0.56,\ 2.23]$ \\
  & Modality $\times$ Temporal          & $-1.120$ & $0.575$ & $-1.95$ & $0.051$ & $0.051$ & $0.33\ [0.11,\ 1.01]$ \\
\midrule
\multirow{5}{*}{Performance}
  & Intercept                                & $-0.520$ & $0.245$ & $-2.12$ & $0.034$ & & $0.59\ [0.37,\ 0.96]$ \\
  & Modality (AI)                            & $\phantom{-}0.233$ & $0.341$ & $\phantom{-}0.68$ & $0.494$ & $0.734$ & $1.26\ [0.65,\ 2.46]$ \\
  & \textbf{Performance (high)}         & $\boldsymbol{\phantom{-}1.135}$ & $\boldsymbol{0.351}$ & $\boldsymbol{\phantom{-}3.24}$ & $\boldsymbol{0.001}$ & $\boldsymbol{0.001^{**}}$ & $\boldsymbol{3.11\ [1.56,\ 6.19]}$ \\
  & Task\_order (2)                          & $\phantom{-}0.190$ & $0.345$ & $\phantom{-}0.55$ & $0.582$ & $0.680$ & $1.21\ [0.61,\ 2.38]$ \\
  & Modality $\times$ Performance       & $-0.789$ & $0.449$ & $-1.76$ & $0.079$ & $0.079$ & $0.45\ [0.19,\ 1.10]$ \\
\midrule
\multirow{5}{*}{Effort}
  & Intercept                                & $\phantom{-}0.404$ & $0.251$ & $\phantom{-}1.61$ & $0.108$ & & $1.50\ [0.92,\ 2.45]$ \\
  & Modality (AI)                            & $-0.227$ & $0.356$ & $-0.64$ & $0.524$ & $0.734$ & $0.80\ [0.40,\ 1.60]$ \\
  & \textbf{Effort (high)}              & $\boldsymbol{-0.998}$ & $\boldsymbol{0.362}$ & $\boldsymbol{-2.76}$ & $\boldsymbol{0.006}$ & $\boldsymbol{0.006^{**}}$ & $\boldsymbol{0.37\ [0.18,\ 0.75]}$ \\
  & Task\_order (2)                          & $\phantom{-}0.384$ & $0.355$ & $\phantom{-}1.08$ & $0.279$ & $0.680$ & $1.47\ [0.73,\ 2.95]$ \\
  & Modality $\times$ Effort            & $-0.265$ & $0.618$ & $-0.43$ & $0.669$ & $0.669$ & $0.77\ [0.23,\ 2.58]$ \\
\midrule
\multirow{5}{*}{Frustration}
  & Intercept                                & $\phantom{-}0.178$ & $0.251$ & $\phantom{-}0.71$ & $0.479$ & & $1.19\ [0.73,\ 1.96]$ \\
  & Modality (AI)                            & $-0.418$ & $0.353$ & $-1.18$ & $0.237$ & $0.553$ & $0.66\ [0.33,\ 1.32]$ \\
  & Frustration (high)                  & $-0.564$ & $0.352$ & $-1.60$ & $0.109$ & $0.109$ & $0.57\ [0.29,\ 1.13]$ \\
  & Task\_order (2)                          & $\phantom{-}0.246$ & $0.353$ & $\phantom{-}0.70$ & $0.486$ & $0.680$ & $1.28\ [0.64,\ 2.55]$ \\
  & Modality $\times$ Frustration       & $\phantom{-}0.852$ & $0.526$ & $\phantom{-}1.62$ & $0.105$ & $0.105$ & $2.34\ [0.84,\ 6.58]$ \\
\midrule
\multirow{5}{*}{Mean (aggregated)\tnote{b}}
  & Intercept                                & $-0.059$ & $0.252$ & $-0.23$ & $0.816$ & & $0.94\ [0.58,\ 1.54]$ \\
  & Modality (AI)                            & $-0.149$ & $0.355$ & $-0.42$ & $0.675$ & $0.751$ & $0.86\ [0.43,\ 1.73]$ \\
  & Mean (high)                         & $-0.096$ & $0.348$ & $-0.28$ & $0.783$ & $0.783$ & $0.91\ [0.46,\ 1.80]$ \\
  & Task\_order (2)                          & $\phantom{-}0.243$ & $0.354$ & $\phantom{-}0.69$ & $0.493$ & $0.680$ & $1.28\ [0.64,\ 2.55]$ \\
  & Modality $\times$ Mean             & $\phantom{-}0.301$ & $0.523$ & $\phantom{-}0.57$ & $0.565$ & $0.565$ & $1.35\ [0.48,\ 3.77]$ \\
\bottomrule
\end{tabular}
\begin{tablenotes}\footnotesize
\item Reference levels: Modality = NON-AI; NASA-TLX-Dimension = low; Task\_order = first.
\item $p_{\mathrm{BH}}$: Benjamini–Hochberg FDR-corrected $p$-values, computed per effect type across the seven NASA-TLX models. Significance codes (applied to $p_{\mathrm{BH}}$): $^{*}\,p<0.05$; $^{**}\,p<0.01$; $^{***}\,p<0.001$. Rows in \textbf{bold} indicate effects significant after BH correction. Intercept and split-analysis rows are not part of the correction family and report uncorrected $p$ only.
\item[a] Indented rows below a significant interaction report the effect of the NASA-TLX dimension fitted within each \textsc{Modality} level (model: \textsc{performance} $\sim$ \textsc{dimension} $+$ \textsc{Task\_order} $+ (1 \mid \text{Participant})$); these are not BH-corrected.
\item[b] Aggregated NASA-TLX score: arithmetic mean of the six $z$-score-normalized sub-dimensions, then median-split.
\end{tablenotes}
\end{threeparttable}
}
\end{table}

\begin{takeaway}
\textbf{H4: Partially supported.} Of the six NASA-TLX sub-dimensions, only the Physical dimension interacts significantly with modality: high self-reported physical demand is associated with lower objective performance under NON-AI but not under AI. The Performance and Effort dimensions show modality-independent associations with objective performance (positive and negative, respectively), while Mental, Temporal, Frustration, and the aggregated overall workload show no significant interaction with modality. The alignment between subjective workload and objective performance therefore differs across conditions only for a subset of the NASA-TLX dimensions.
\end{takeaway}

\section{Discussion}\label{sec:discussion}

\stageTwo{Our four hypotheses received heterogeneous support. We organize the discussion below around the four research questions and close with the theoretical and practical implications that emerge across them.}

\subsection{\stageTwo{RQ1: Physiological differences between AI- and non-AI-assisted programming}}\label{sec:disc:rq1}

\stageTwo{\textbf{Short answer.} Physiological measures differ between AI- and non-AI-assisted programming, but the difference is limited: it concentrates on brain activity (lower $\theta/\alpha$ under AI on the first task) and in oculomotor marker (higher blink rate under AI on the second task) both indicators of cognitive workload. No corresponding differences emerge in HRV or EDA. H1 is therefore partially supported.}

\stageTwo{Of the ten physiological metrics tested under H1, only two differentiate the AI and non-AI conditions, and the two appear in different tasks: the EEG $\theta/\alpha$ ratio on the first task, and the gaze blink rate on the second. The $\theta/\alpha$ ratio is significantly lower under AI than under non-AI both across all participants ($\beta = -1.11$, $p_{\mathrm{BH}} = .01$; Table~\ref{tab:lmm-full-combined}) and when the analysis is restricted to the first task ($\beta = -1.16$, $p_{\mathrm{BH}} = .04$; Table~\ref{tab:lmm-task1}). None of the HRV, EDA and oculomotor metrics reach significance.} 

\stageTwo{The $\theta/\alpha$ ratio is a well-established EEG marker of cognitive load that increases with mental effort~\citep{borghini2014measuring,PUMA:theta:alpha:cognitiveLoad}. The lower ratio observed under AI, therefore, points to reduced working-memory engagement, consistent with developers offloading part of the generative effort to the model~\citep{grinschgl2021offloading,lee2025critical}. This direction is consistent with \citet{kosmyna2025brain}, who reported lower alpha-band connectivity during LLM-assisted essay writing relative to a brain-only baseline, and converges with the broader programming-neuroimaging literature, in which program comprehension activates working-memory regions~\citep{siegmund2014,siegmund2017,peitek2021}.}

\stageTwo{On the second task, only the gaze blink rate differentiates the two conditions ($\beta = 1.76$, $p_{\mathrm{BH}} < .01$; Table~\ref{tab:lmm-task2}), with higher rates under AI. Because cognitive demand is known to suppress blinking during visually engaging tasks~\citep{rosenfield2015cognitive,lenskiy2016blink}, the lower blink rate observed under non-AI is again interpretable as evidence of greater cognitive load when the solution must be produced unaided. Both tasks therefore point to the same conclusion: non-AI is associated with greater cognitive engagement, shown by higher $\theta/\alpha$ on the first task and lower blink rate on the second.}

\stageTwo{A final caveat concerns effect magnitude. Marginal $R^2$ values remain below $0.10$ across the H1 models. Although the AI/non-AI modality has a statistically significant effect on the $\theta/\alpha$ ratio, the fixed effects explain only $R^2_m = 0.02$ of its variance, and including between-participant variability via the random intercept yields $R^2_c = 0.24$. The effect is therefore statistically reliable but modest in size relative to individual differences.}

\subsection{\stageTwo{RQ2: Moderating role of developer experience}}\label{sec:disc:rq2}

\stageTwo{\textbf{Short answer.} The physiological response to AI assistance was the same for undergraduate and graduate students. H2 is therefore not supported.}

\stageTwo{None of the \texttt{Modality}\,$\times$\,\texttt{Education} interaction terms reach significance for any of the ten physiological metrics, either across all participants (Table~\ref{tab:lmm-full-education}) or in the task-stratified analyses (Tables~\ref{tab:lmm-task1-education} and~\ref{tab:lmm-task2-education}). At the same time, the H1 main effect of modality on the $\theta/\alpha$ ratio remains significant once developer experience is added to the model ($\beta = -1.16$, $p_{\mathrm{BH}} < .01$; Table~\ref{tab:lmm-full-education}). The cognitive-engagement difference between AI and non-AI is therefore observed in undergraduate and graduate students alike, and the H1 finding does not depend on academic seniority.}

\stageTwo{Three factors may explain the absence of moderation. First, our developer-experience contrast is narrower than the one studied in prior work. \citet{friedmann2024pair} observed different interaction patterns between experienced and novice programmers (acceleration of familiar tasks among the experienced, exploratory use among the novices), whereas our undergraduate-versus-graduate contrast covers a much smaller range of programming experience. Second, ChatGPT has become widely familiar among CS students at both academic levels, which may have leveled out any baseline difference in how undergraduates and graduates approach an AI assistant. Third, with 60 participants split across modality and education levels, our design has limited statistical power to detect interaction effects of the size observed in our data.}

\stageTwo{Undergraduate and graduate students showed similar responses to AI assistance in our data. Whether a wider range of programming experience, such as the experienced vs.\ novice comparison studied by \citet{friedmann2024pair}, would show a different pattern remains an open question that calls for replication in industry settings.}

\subsection{\stageTwo{RQ3: Physiological correlates of performance across modalities}}\label{sec:disc:rq3}

\stageTwo{\textbf{Short answer.} Physiological measures correlate with performance, but the coupling depends on modality: the EDA--performance association is present under the non-AI modality and absent under AI. EEG and HRV interactions reach significance overall, but within each condition the physiology--performance association is at most marginal. H3 is therefore partially supported.}

\stageTwo{The clearest finding is the EDA pattern. In the early task period, both SCR amplitude (\texttt{Modality}\,$\times$\,\texttt{Physiological}: $\beta = 0.089$, $p_{\mathrm{BH}} = .015$; Table~\ref{tab:glmm_early}) and tonic EDA ($\beta = -0.123$, $p_{\mathrm{BH}} = .006$) show significant interactions with modality, and the stratified analyses indicate that the association between EDA and performance is present under non-AI (SCR amplitude: $\beta = -0.116$, $p = .030$; tonic EDA: $\beta = 0.143$, $p = .016$) but absent under AI. This is consistent with prior work showing that EDA tracks programming task difficulty~\citep{fritz2014icse} and reflects stress-related arousal~\citep{westerink2020deriving}, and aligns with the cognitive-offloading interpretation~\citep{grinschgl2021offloading,lee2025critical} discussed under H1: when developers offload part of the generative effort to the model, the EDA--performance association observed under the non-AI modality is no longer present.}

\stageTwo{The pattern evolves across task periods. Because we partitioned each task into early, mid, and late periods to capture how cognitive demands evolve, we can also observe how the modality--physiology relationship changes across them. The \texttt{Modality}\,$\times$\,\texttt{Physiological} interaction is significant in three signal families in the early period (EDA, EEG, and gaze; Table~\ref{tab:glmm_early}) and in all four families in the mid period (Table~\ref{tab:glmm_mid}). In the late period, the pattern narrows: HRV interactions remain significant, while SCR amplitude and SCR peaks correlate with performance regardless of modality (Table~\ref{tab:glmm_late}). One interpretation is that the AI/non-AI difference in physiology--performance coupling is strongest across the early and mid periods, when developers devise and refine the solution. In the mid period, the blink-rate--performance association differs in sign between conditions: under the AI modality, a higher blink rate is associated with lower performance ($\beta = -0.281$, $p = .028$); under non-AI, with higher performance ($\beta = 0.369$, $p = .039$).}

\stageTwo{For several signals, the \texttt{Modality}\,$\times$\,\texttt{Physiological} interaction is significant overall, but when we fit the model separately to AI and non-AI, neither condition shows a significant physiology--performance association. This applies to the EEG $\theta/\alpha$ ratio and the gaze blink rate in the early period (interactions $\beta = 0.469$, $p_{\mathrm{BH}} = .024$ and $\beta = -0.338$, $p_{\mathrm{BH}} < .001$; Table~\ref{tab:glmm_early}) and to HRV RMSSD and LF/HF in the late period (interactions $\beta = -0.049$, $p_{\mathrm{BH}} = .001$ and $\beta = -0.390$, $p_{\mathrm{BH}} = .032$; Table~\ref{tab:glmm_late}). The expected physiology--performance coupling difference between modalities therefore applies most cleanly to EDA in the early and mid periods and to mid-period blink rate, and only weakly elsewhere.}

\subsection{\stageTwo{RQ4: Subjective--objective alignment across modalities}}\label{sec:disc:rq4}

\stageTwo{\textbf{Short answer.} The alignment between perceived, self-reported workload and objective performance differs across modalities for only one (Physical demand) of the six NASA-TLX dimensions. H4 is therefore partially supported.}

\stageTwo{Of the six NASA-TLX dimensions, only Physical demand shows a significant interaction with modality ($\beta = 2.096$, $p_{\mathrm{BH}} < .001$; Table~\ref{tab:glmm_workload_interaction}). The stratified analysis shows that under the non-AI condition high self-reported physical demand is associated with lower objective performance ($\beta = -2.380$, $p < .001$), whereas under AI no such association is observed.
}

\stageTwo{Self-reported Performance and Effort dimensions show modality-in\-de\-pen\-dent associations with objective performance, in opposite directions ($\beta = 1.135$ and $-0.998$, respectively; $p_{\mathrm{BH}} < .01$). Developers can therefore still accurately assess their performance and effort when using AI.

Taken together, these findings refine the perception--reality gap reported by~\citet{becker2025metr} in their experiment, where experienced open-source developers expected a speedup from AI tools but instead experienced a slowdown. The gap we observe is not general but dimension-specific.}

\subsection{\stageTwo{Implications}}\label{sec:disc:implications}

\stageTwo{Our findings carry theoretical implications for how AI-assisted programming is studied and practical implications for the design of AI-assisted development tools. We discuss each in turn.}

\subsubsection{\stageTwo{Theoretical Implications}}

\stageTwo{\textbf{AI is not faster solo programming.} The $\theta/\alpha$ ratio is an EEG marker of cortical engagement that increases with mental effort, so the lower values observed under AI in H1 are consistent with reduced cortical engagement rather than a faster version of the same cognitive process~\citep{grinschgl2021offloading,kosmyna2025brain}; a modality-dependent $\theta/\alpha$--performance coupling also appears in H3. Effort and cognitive-load models calibrated on solo coding therefore do not transfer directly to AI-assisted conditions: a developer who completes a task in less time with AI is not simply running the same internal process more efficiently, and frameworks that treat AI-assisted programming as merely a faster version of solo programming may misrepresent cognitive cost.}


\stageTwo{\textbf{EDA-based effort measures need re-validation.} The EDA--per\-for\-mance association observed under non-AI in the early and mid periods of H3 is no longer observed under AI. Because EDA has long been associated with task difficulty in software-engineering studies~\citep{fritz2014icse} and with stress-related arousal in broader psychophysiology work~\citep{westerink2020deriving}, this modality-dependent pattern suggests that EDA-based effort measures validated against non-AI baselines should be re-validated before being applied to AI-assisted conditions.}

\stageTwo{\textbf{Workload by dimension, not aggregate.} Subjective workload should be analyzed at the dimension level, not as a single aggregate score, in AI-coding research. Among the six NASA-TLX dimensions, only Physical demand shows a modality-dependent association with objective performance, so studies that report only the aggregated workload score might mask the one place where AI assistance shifts the perception--performance relationship. The perception--reality gap reported by~\citet{becker2025metr} is local rather than global and instruments used to measure developer experience with AI should preserve dimension-level resolution.}

\subsubsection{\stageTwo{Practical Implications}}

\stageTwo{\textbf{In-tool feedback mechanisms.} The lower cortical engagement observed under AI is consistent with cognitive offloading during AI-assisted programming. AI assistants could therefore benefit from features that encourage critical review of generated code, such as explicit review prompts, side-by-side comparisons of alternative implementations, or step-by-step inspection workflows that surface what the model produced and why, to counter risks of overreliance and skill atrophy reported in adjacent work~\citep{lee2025critical,storey2026cognitive}.}

\stageTwo{\textbf{Signal choice for biometric monitoring.} Two physiological markers discriminated AI from non-AI in our 60-participant lab sample, each on one task: the $\theta/\alpha$ ratio on the first task and the gaze blink rate on the second task. The remaining metrics (EDA, HRV, EEG beta power, fixation count, and mean fixation duration) did not discriminate AI from non-AI. Designers integrating biometric sensing should weigh signal quality against developer acceptability: blink rate is less obtrusive than EEG and can be captured by a webcam, but its discriminative power in our data was task-specific. No single channel was reliable across both tasks; biometric monitoring of AI-assisted programming should therefore combine multiple signals.}

\section{Threats to Validity}\label{sec:limitations}

\textit{Internal validity.} First, the 2$\times$2 crossover design carries the risk of carryover, sequence, and treatment~$\times$~period effects, which are inherently difficult to disentangle in programming contexts where problem-solving strategies and acquired knowledge cannot be `washed out' between tasks~\citep{vegas2016tse}. \stageTwo{To mitigate this risk, \texttt{Task\_order} was included as a fixed effect and \texttt{Modality}\,$\times$\,\texttt{Task\_order} as an interaction term when testing H1 and H2. In H2, a significant main effect of \texttt{Task\_order} on the EEG $\theta/\alpha$ ratio was observed, without a corresponding \texttt{Modality}\,$\times$\,\texttt{Task\_order} interaction. Following~\citet{vegas2016tse}, residual carryover effects cannot be fully ruled out.}
Second, pilot testing minimized the potential impact of task design differences.
In addition, we performed opportunistic sampling to enable stratification of participants by developer experience, operationalized as academic seniority. \stageTwo{This yielded an approximately balanced composition across the pooled sample (see Table~\ref{tab:participants}), enabling the H2 test on the pooled dataset as registered.}
Third, to reduce Hawthorne effects from biometric monitoring, we ensured comfortable conditions with acclimation time and between-task breaks. 
Finally, as site-specific analyses have intrinsically lower statistical power, our multisite approach served as an internal replication mechanism that enabled pooled analyses to address this limitation.

\textit{External validity.} Our two-site study design enables generalization across two diverse physical and cultural contexts. \stageTwo{However, two deviations from the registered protocol (see~\ref{appendix:deviations}) limit what can be claimed about cross-site generalizability: site was not modeled as a fixed or random effect in our analyses, so cross-site equivalence was not formally tested; EDA and HRV could similarly be collected only at Uniba, so the findings based on these two signals rest on a single-site subsample.}
Limitations include our university student sample, which may not fully represent professional developers, although there is evidence suggesting that the differences may not be as large as expected~\citep{salman2015icse}. \stageTwo{We addressed this limitation by stratifying by developer experience (operationalized as academic seniority) and acknowledging this during interpretation; we note, however, that this contrast does not capture the full range of professional programming experience.} The laboratory setting, 
may also differ from natural programming environments. To mitigate this threat, we designed the lab environment to feel as natural and unobtrusive as possible and enhance ecological validity by collecting feedback on the difficulty of the tasks from the participants recruited for the study.

\textit{Construct validity.} While EEG, eye tracking, EDA, and HRV provide useful proxies for cognitive states, they may not perfectly reflect the constructs of interest. To strengthen validity, we relied on multiple converging measures, including self-report, and on established preprocessing pipelines.
Another threat concerns participants' prior experience with AI tools, which may influence their performance and physiological responses. \stageTwo{AI familiarity was measured during screening (see Table~\ref{tab:participants}), but deviating from the registered protocol (see~\ref{appendix:deviations}), the composite score was not entered as a covariate in our primary models. Between-participant variance, including AI familiarity, is partly absorbed by the participant random intercept used in our mixed-effects models.}

\stageTwo{\textit{Conclusion validity.} The marginal $R^2$ values across the H1 and H2 models are below 0.10. The AI/non-AI modality and developer experience therefore account for only a small share of the variability in physiological responses compared to the individual differences between participants. 
In addition, the Fixation Count GLMM did not converge in any of the three H3 time periods (early, mid, late), so the \texttt{Modality}~$\times$~\texttt{Fixation\_Count} interaction has no interpretable estimates and no conclusion can be drawn for that metric.} 

\section{Conclusion}\label{sec:conclusion}
\stageTwo{This paper reported a multisite, preregistered biometric study comparing AI-assisted and non-assisted development. We collected EEG, eye tracking, electrodermal activity, and heart rate variability from 60 computer science students at two universities, alongside a rubric-based performance score and self-reported workload.

Two physiological signals---the EEG $\theta/\alpha$ ratio and the gaze blink rate---differed between AI and non-AI development, both consistent with reduced cognitive engagement when developers offload generative effort to the model. This pattern did not differ between undergraduate and graduate students. Electrodermal activity correlated with performance under the non-AI condition but not under AI, and among the six NASA-TLX dimensions only Physical demand showed the same asymmetric pattern. The AI/non-AI difference therefore concentrates on specific eye-tracking and EEG markers rather than spanning the heart rate and electrodermal measures tested. Likewise the perception--performance gap reported in prior work appears in only one of the six workload dimensions rather than across all of them.

These findings inform the design of AI assistants and the choice of biometric signals for monitoring developer state. Replication in industry settings and across a wider range of professional experience remains an open avenue for future work.}

\section*{Data Availability}\label{sec:data-availability}
\stageTwo{To support reproducibility, all artifacts produced for this study are publicly available as a replication package.\footnote{\url{10.5281/zenodo.20120212}}
The package includes: (i)~raw multimodal recordings in BIDS format per participant and session, 
including EEG, EDA-HRV physiological signals (Uniba only), GAZE, behavioral 
logs (key presses, mouse activity, GPT interactions), screen recordings, 
per-participant code submissions with their rubric-based evaluations, 
and self-reported workload scores (NASA-TLX);
(ii)~aggregated per-site and combined datasets of extracted physiological features (EEG, EDA-HRV, and gaze metrics), self-reported 
workload scores (NASA-TLX), code quality scores, and 
participant metadata for each participant and task;
(iii)~Python notebooks reproducing the preprocessing pipelines (EEG, EDA, HRV, GAZE) and the statistical analyses for H1--H4 reported in Section~\ref{sec:hypothesis-test};
(iv)~the rubric template used to assess coding tasks correctness, and the questionnaires 
administered during the study (demographic-technological and post-experiment 
debriefing), provided in both English and Italian;
(v)~the full experiment infrastructure in both English and Italian versions, 
including the oTree application managing experiment flow, the Dockerized 
coding environment presented to participants, and the Java programming 
tasks used in the study.
}

\stageTwo{In compliance with GDPR and the ethics protocols approved at both sites, raw video and screen recordings, raw biometric traces that may contain identifying information (e.g., individual EEG segments aligned with screen activity), and any free-text responses that could re-identify a participant are \emph{not} released. Aggregated, windowed, and feature-level data are released after de-identification.}


\section*{Acknowledgements}
We thank all the students who volunteered to take part in the experiments.
We are grateful to Nadja Brix Koch, Theis Helth Stensgaard, and Guillaume Andrea Desaphy for helping with the protocol registrations and pilot study execution.
This research was co-funded by the NRRP Initiative, Mission 4, Component 2, Investment 1.3 - Partnerships extended to universities, research centers, companies, and research D.D. MUR n. 341, 15.03.2022 – Next Generation EU (``FAIR - Future Artificial Intelligence Research", code PE00000013, CUP H97G22000210007), the Complementary National Plan PNC-I.1 - Research initiatives for innovative technologies and pathways in the health and welfare sector - D.D. 931 of 06/06/2022 (``DARE - DigitAl lifelong pRevEntion initiative", code PNC0000002, CUP B53C22006420001), and by the European Union - NextGenerationEU through the Italian Ministry of University and Research, Projects PRIN 2022 (``QualAI: Continuous Quality Improvement of AI-based Systems'', grant n. 2022B3BP5S, CUP: H53D23003510006).

\appendix
\renewcommand{\thesection}{Appendix~\Alph{section}}

\section{Protocol Deviations}\label{appendix:deviations}

\stageTwo{This appendix documents all deviations from the preregistered Stage~1 protocol~\citep{stage1report}, grouped by data acquisition, measurement, statistical modeling, and presentation.}

\stageTwo{\textit{EDA and HRV collection at ITU}.
The registered protocol specified the collection of electrodermal activity (EDA) and heart rate variability (HRV) at both sites.
However, no wristband device could be acquired at the ITU site.
As a result, EDA and HRV data were collected only at the Uniba site.
Analyses involving these measures are therefore restricted to the Uniba subsample, and we account for this as a limitation.}

\stageTwo{\textit{Performance operationalization}.
The registered protocol committed to three performance indicators: time to completion, functional correctness (percentage of passed tests in a comprehensive unit-test suite), and a task-correctness score obtained through manual verification of code. Of these, only the task-correctness score was used in the analyses reported here. Functional correctness was dropped because no participant produced a solution that compiled and ran successfully against the prepared test suite. Time-to-completion was dropped because every participant used the full time slot granted for each task, leaving no between-participant variation. The task-correctness score was operationalized through a structured assessment rubric, with design and application procedure described in Section~\ref{sec:variables-measurements}.}

\stageTwo{\textit{Eye-tracking metric set}. The executed analysis used fixation count, mean fixation duration, and \emph{blink rate} as eye-tracking metrics. lthough not registered, Blink rate added as an established marker of cognitive load in code-reading paradigms~\citep{lenskiy2016blink, rosenfield2015cognitive}. The registered protocol committed to scan path length and to exploring AOI-based measures such as dwell time and inter-AOI transitions; these, however, were dropped in fa
vor of a smaller set focused on cognitive-load interpretation~\citep{sharafi2020practical}. A}

\stageTwo{\textit{Statistical model for H1 and H2}.
The registered protocol specified a mixed ANOVA with programming condition as a within-subjects factor and task order as a between-subjects factor. We instead fitted linear mixed-effects models (LMMs) with by-participant random intercepts. The change was motivated by the windowed feature-extraction pipeline. Physiological signals were segmented into overlapping 60-second windows, producing multiple correlated observations per participant, which violates the independence assumption of the mixed ANOVA. The LMM specification preserves the fixed-effects structure intended by the original test (\texttt{Modality}, \texttt{Task\_order}, and their interaction for H1; \texttt{Modality}\,$\times$\,\texttt{Education} added for H2), so the hypotheses tested are unchanged.}



\stageTwo{\textit{Site as a random effect}.
The registered protocol planned to include site as a random effect in our mixed-effects models. 
We did not include site in our regression analysis because each participant belongs to a single site, therefore any baseline difference between the two sites is already captured by the per-participant terms that every model includes.}

\stageTwo{\textit{AI experience score not used as a covariate}.
The registered protocol committed to a composite 12-item AI experience score as a covariate in all primary analyses. We computed the score from the screening questionnaire but did not include it in the models. The items showed strong ceiling effects in our sample: 81.7\% of participants reported using AI tools \textit{Daily}, 81.7\% had first used AI coding tools before this year, and 90.0\% rated themselves in the top two AI-literacy levels (Table~\ref{tab:participants}). The score therefore had too little variation to be used as a covariate; we report its components in Table~\ref{tab:participants} for transparency.}

\stageTwo{\textit{Hypothesis renumbering}.
In Stage~2, the hypothesis on the moderating role of developer experience moved from H4 to H2, shifting the physiology--performance correlation from H2 to H3 and subjective--objective alignment from H3 to H4. H1 (physiological differences) is unchanged, and the substantive content of each hypothesis is preserved.}


\bibliographystyle{spbasic}
\bibliography{references}

\end{document}